%% file: efficient.tex
\documentclass[acmsmall]{acmart}
\pdfoutput=1 
\newcommand\tomACMVersion0

\newcommand\ifTomACMVersion[2]{\ifthenelse{\equal\tomACMVersion1}{#1}{#2}}
\makeatletter
\newcommand\ifACMAnonymous[2]{\if@ACM@anonymous #1 \else #2 \fi}
\makeatother

\startPage{1}

\acmJournal{PACMPL}
\acmVolume{1}
\acmNumber{XXXX} % CONF = POPL or ICFP or OOPSLA
\acmArticle{1}
\acmYear{2023}
\acmMonth{1}
\acmDOI{} % \acmDOI{10.1145/nnnnnnn.nnnnnnn}

\usepackage{mathtools}
\usepackage{booktabs}
\usepackage{subcaption}
\usepackage[capitalise]{cleveref}
\usepackage{float}
\usepackage{stmaryrd}
\usepackage{xifthen}
\usepackage{tikz}
\usepackage{makecell}
\usepackage{xspace}
\usepackage{wrapfig}
\usepackage[final]{pdfpages}
\usepackage{cleveref}
\usepackage{xcolor}
\usepackage{mleftright}
\usepackage{mathpartir}
\usepackage[normalem]{ulem}
\usepackage{multirow}
\usepackage[frozencache]{minted}

\usetikzlibrary{arrows.meta}

\input{defs}

\newcommand\agdagithublink{\url{https://github.com/tomsmeding/efficient-chad-agda}}

\allowdisplaybreaks

\begin{document}

\title{Efficient CHAD}

\author{Tom J.\ Smeding}
\orcid{0000-0002-4986-6820}
\affiliation{%
    \department{Department of Information and Computing Sciences}
    \institution{Utrecht University}
    % \streetaddress{Heidelberglaan 8}
    \city{Utrecht}
    % \postcode{2584 CS}
    \country{The Netherlands}
}
\email{t.j.smeding@uu.nl}

\author{Matthijs I.\:L.\:V\'ak\'ar}
\orcid{0000-0003-4603-0523}
\affiliation{%
    \department{Department of Information and Computing Sciences}
    \institution{Utrecht University}
    % \streetaddress{Heidelberglaan 8}
    \city{Utrecht}
    % \postcode{2584 CS}
    \country{The Netherlands}
}
\email{m.i.l.vakar@uu.nl}

%%
%% By default, the full list of authors will be used in the page
%% headers. Often, this list is too long, and will overlap
%% other information printed in the page headers. This command allows
%% the author to define a more concise list
%% of authors' names for this purpose.
% \renewcommand{\shortauthors}{Trovato and Tobin, et al.}

%%
%% The abstract is a short summary of the work to be presented in the
%% article.
\begin{abstract}
We show how the basic Combinatory Homomorphic Automatic Differentiation (CHAD) 
algorithm can be optimised, using well-known methods, 
to yield a simple, composable, and generally applicable reverse-mode automatic differentiation (AD)
technique that has the correct computational complexity that we would expect of 
reverse-mode AD.
Specifically, we show that the standard optimisations of sparse vectors and state-passing style 
code (as well as defunctionalisation/closure conversion, for higher-order languages) 
give us a purely functional algorithm that is most of the way to the correct complexity, with (functional) mutable updates taking care of 
the final $\log$-factors.
We provide an Agda formalisation of our complexity proof.
Finally, we discuss how the techniques apply to differentiating parallel functional array programs:
the key observations are 1) that all required mutability is (commutative, associative) accumulation, which lets us preserve task-parallelism and 2) that we can write down data-parallel derivatives for most data-parallel array primitives.
\end{abstract}

%%
%% The code below is generated by the tool at http://dl.acm.org/ccs.cfm.
%% Please copy and paste the code instead of the example below.
%%
\begin{CCSXML}
<ccs2012>
 <concept>
  <concept_id>10002950.10003714.10003715.10003748</concept_id>
  <concept_desc>Mathematics of computing~Automatic differentiation</concept_desc>
  <concept_significance>500</concept_significance>
 </concept>
 <concept>
  <concept_id>10011007.10011006.10011008.10011009.10011012</concept_id>
  <concept_desc>Software and its engineering~Functional languages</concept_desc>
  <concept_significance>300</concept_significance>
 </concept>
 <concept>
  <concept_id>10011007.10010940.10010992.10010993</concept_id>
  <concept_desc>Software and its engineering~Correctness</concept_desc>
  <concept_significance>300</concept_significance>
 </concept>
</ccs2012>
\end{CCSXML}

\ccsdesc[500]{Mathematics of computing~Automatic differentiation}
\ccsdesc[300]{Software and its engineering~Functional languages}
\ccsdesc[300]{Software and its engineering~Correctness}

\keywords{automatic differentiation, source transformation, functional programming}

\maketitle

\section{Introduction}
\label{sec:intro}
Automatic differentiation (AD) is a widely-used method for computing derivatives of functions 
$f:\R^n\to\R^m$ represented as code \cite{adbook-2008-griewank-walther,ad-2018-survey-automatic-differentiation,ad-2018-survey-ad-implementation}.
The key idea is to transform the original source code to efficiently and compositionally compute its derivative, using the chain-rule
% while optimising shared subcomputations.
while preventing duplication of computation.
Forward-mode AD composes  derivatives of program fragments in the original execution order.
By contrast, reverse-mode AD composes transposed derivatives of program fragments in reverse.
Reverse-mode AD is preferred when $n\gg m$, for example in machine learning applications that require gradient computation for scalar-valued functions on high-dimensional spaces.

The motivation of this paper is to develop a compositional, purely functional, generally applicable, correct, efficient reverse AD technique for the functional, parallel array programming paradigm.
This paradigm is gaining popularity in frameworks such as XLA/HLO \cite{ad-2017-xla}, Accelerate \cite{acc-2011-cuda,acc-2013-optim,accelerate-docs} and Futhark \cite{futhark-2017-thesis,futhark-2017-pldi} as a high-level interface for writing code that makes efficient use of modern parallel hardware.
By developing such an AD technique, we hope to bridge some of the gap between theory and practice in the programming language literature about AD.

Our starting point is the CHAD automatic differentiation technique \cite{vakar-2022-chad,nunes-2022-chad-expressive,vakar-2021-higher-order-reverse-ad,adfp-2018-categories-ad},
which has our desired features of compositionality, pure functionality, general applicability and verified correctness, but is not efficient. Further, it has not yet been extended to parallel array operations.

The main contributions of this paper to the existing body of literature (see \cref{sec:related-work}) are:
\begin{itemize}
\item the observation that the CHAD algorithm for first-order functional programs (\cref{sec:basic-chad})~has efficiency problems (\cref{sec:efficiency-problems}) that are solvable, up to some log-factors, by using a sparse implementation of the required data types (\cref{sec:sparsity}) and state-passing style (\cref{sec:monadic-lifting});
\item a further observation that the remaining log-factors can be optimised away by making the vectors mutable, breaking pure functionality (\cref{sec:log-factors});
\item a \emph{mechanically formalised complexity proof}:  the resulting implementation achieves the correct computational complexity expected of reverse AD algorithms (\cref{sec:informal-amortisation,sec:complexity-proof});
\item a demonstration that the same techniques enable efficient AD of array programs (\cref{sec:efficient-chad-on-array-types});
\item the insight that all required state is only written to using (commutative, associative) accumulation operations that can easily be parallelised (\cref{sec:parallelism});
\item the finding that the naive CHAD algorithm for higher-order functions is fundamentally inefficient, but that efficiency can be restored through the use of defunctionalisation or closure conversion (\cref{sec:efficient-chad-on-function-types}).
\end{itemize}
This paper aims to provide a comprehensive exploration of the proposed AD algorithm while addressing its efficiency and applicability to a wide range of programming contexts.

\section{Basic Reverse-Mode CHAD}
\label{sec:basic-chad}
We summarise the basic reverse-mode CHAD algorithm described and proved correct in \cite{vakar-2022-chad,nunes-2022-chad-expressive,vakar-2021-higher-order-reverse-ad,adfp-2018-categories-ad, ad-2019-vytiniotisdifferentiable}, before proceeding to explain how its implementation can be made efficient.
Consider programs written in a standard first-order functional language with the types and programs of \cref{fig:source-language-1}.
The types of this language are built using a primitive type $\R$ of real numbers, tuple types $\Unit$ and $\Pair{\ty}{\ty[2]}$, and sum types $\Sum{\ty}{\ty[2]}$.
This language allows us to build complex programs by combining primitive (partial) functions $\op:\R^n\to\R$, where $\R^n\defeq \Pair{\R}{\Pair}{\cdots}{\R}$, and $\tSign{}:\R\to\B$, which can be used to perform case distinctions on real numbers.\footnote{The choice of including $\tSign{}$ is arbitrary; we simply include it to show that conditionals are fully supported.}
We postpone the treatment of array and function types until \cref{sec:efficient-chad-on-array-types} and \cref{sec:efficient-chad-on-function-types}, respectively.
% This sentence is formulated precisely so that the two long definitions do not get split over lines.
We use syntactic sugar $\tTuple{\trm_1,\ldots,\trm_n}\defeq \tPair{\tPair{\trm_1}{\trm_2},\cdots }{\trm_n}$, as well as destructuring assignments: $\letin{\tPair{\var}{\var[2]}}{\trm}{\trm[2]}\defeq \letin{\var[3]}{\trm}{\letin{\var}{\tFst\var[3]}{\letin{\var[2]}{\tSnd\var[3]}{\trm[2]}}}$ and $\funabs{\tPair{\var}{\var[2]}}{\trm}\defeq \funabs{\var[3]}{\letin{\tPair{\var}{\var[2]}}{\var[3]}}{\trm}$.

Reverse-mode CHAD associates to every type $\ty$ two types:
\begin{itemize}
\item a type $\Dsyn{\ty}_1$ of primal values of type $\ty$; here $\Dsyn{\R}_1\defeq\R$,
$\Dsyn{\Unit}_1\defeq \Unit$, and $\Dsyn{\Pair{\ty}{\ty[2]}}_1\defeq \Pair{\Dsyn{\ty}_1}{\Dsyn{\ty[2]}_1}$; note that, for now, $\Dsyn{\ty}_1=\ty$ for all types $\ty$ (this will change in \cref{sec:efficient-chad-on-function-types});
\item a type $\Dsyn{\ty}_2$ of cotangent values of type $\ty$, which has the structure of a commutative monoid.
\end{itemize}
Given a well-typed program $\Gamma\vdash \trm:\ty$ of type $\ty$ in typing context $\Gamma=\var_1:\ty_1,\ldots,\var_n:\ty_n$, reverse-mode CHAD associates 
to that a program $\Dsyn[\Gamma]{\trm}$ that computes the pair of the (primal) value of $\trm$ and as well as its transposed derivative:
\begin{equation}
  \Dsyn{\Gamma}_1\vdash \Dsyn[\Gamma]{\trm}:\Pair{\Dsyn{\ty}_1}{(\Function{\Dsyn{\ty}_2}{\Env{\Dsyn{\Gamma}_2}})}.
  \label{eq:chad-trans-type}
\end{equation}
Here, we write $\Dsyn{\var_1:\ty_1,\ldots,\var_n:\ty_n}_i\defeq \var_1:\Dsyn{\ty_1}_i,\ldots,\var_n:\Dsyn{\ty_n}_i$ for $i=1,2$.
If $\Gamma$ is an environment, say $\var_1 : \ty_1, \ldots, \var_n : \ty_n$, then $\Env{\Gamma}$ (for ``environment vector'') is an abstract type containing one value for each type $\ty_i$ in $\Gamma$.
For now, we could naively implement $\Env{\Gamma}\defeq \Pair{\Pair{\ty_1}{\cdots}}{\ty_n}$, associating $\Pair{}{}$ to the left;\footnote{In this paper we consider environments ordered lists.} with this definition, the codomain of the backpropagator in \cref{eq:chad-trans-type} expands to $\Pair{\Pair{\Dsyn{\ty_1}_2}{\cdots}}{\Dsyn{\ty_n}_2}$.
We will later consider more efficient implementations of $\Env{\Gamma}$.

\begin{figure}
  \figheading{Types: \vspace{-13pt}}
  \begin{align*}
    \begin{array}{l@{\ \ }c@{\ \ }l}
    \ty, \ty[2] &\coloneqqq&
      \R \mid \Unit \mid \Pair{\ty}{\ty[2]} \mid \Sum{\ty}{\ty[2]}
    \end{array}
  \end{align*}
  \figheading{Terms:  \vspace{-13.7pt}}
  \begin{align*}
    \qquad\begin{array}{l@{\ \ }c@{\ \ }ll}
    \trm, \trm[2], \trm[3] &\coloneqqq&
    \var \mid \mathrlap{\letin{\var : \ty} {\trm}{\trm[2]}\mid \tUnit \mid \tPair{\trm}{\trm[2]}\mid \tFst \trm \mid \tSnd \trm 
    } \\
      &\mid& \mathrlap{\tInl\trm\mid \tInr\trm\mid \tSMatch{\trm}{\var}{\trm[2]}{\var[2]}{\trm[3]}}\\
      &\mid& r & \text{(literal $\R$ values)} \\
      &\mid& \tSign \trm & \text{(sign function $\R\ra\B$, where $\B\defeq \Sum{\Unit}{\Unit}$)} \\
      &\mid& \tOpApp{\trm_1,\ldots,\trm_n} & \text{($\op \in \Op_n$, possibly partial primitive operation $\R^n \rightharpoonup \R$)}
    \end{array}
  \end{align*}
  \caption{\label{fig:source-language-1}
    The source language of this paper's reverse AD transformation.
  }
\end{figure}

\begin{figure}
  \begin{align*}
    \Dsyn{\R}_2  \defeq \LR\qquad
    \Dsyn{\Unit}_2\defeq \LUnit\qquad
    \Dsyn{\Pair{\ty}{\ty[2]}}_2  \defeq \LPair{\Dsyn{\ty}_2}{\Dsyn{\ty[2]}_2}\qquad
    \Dsyn{\Sum{\ty}{\ty[2]}}_2 \defeq\LSum{\Dsyn{\ty}_2}{\Dsyn{\ty[2]}_2}
  \end{align*}
  \begin{align*}
  \Dsyn[\Gamma]{\var : \ty}&\defeq \tPair{\var : \Dsyn{\ty}_2}{\funabs{d}{\tOne\var{\Dsyn{\ty}_2}{\Dsyn{\Gamma}_2}\ d}}\\
  \Dsyn[\Gamma]{\letin{\var:\ty}{\trm}{\trm[2]}}&\defeq
  \begin{array}[t]{@{}l@{}}
    \letin{\tPair{\var}{\var'}}{\Dsyn[\Gamma]{\trm}}{
      \letin{\tPair{\var[2]}{\var[2]'}}{\Dsyn[\Gamma,\var:\ty]{\trm[2]}}{}}\\
    \tPair{\var[2]}{\funabs{d}{\letin{\tPair{d_1}{d_2}}{\tSplit{\Dsyn{\Gamma}_2}{\Dsyn{\ty}_2} (\var[2]'\ d)}{\tPlus{d_1}{x'\ d_2}}}}
  \end{array}
  \\
  \Dsyn[\Gamma]{\tUnit}&\defeq \tPair{\tUnit}{\funabs{d}{\tZero}}\\
  \Dsyn[\Gamma]{\tPair{\trm}{\trm[2]}}&\defeq
  \begin{array}[t]{@{}l@{}}
  \letin{\tPair{\var}{\var'}}{\Dsyn[\Gamma]{\trm}}{
    \letin{\tPair{\var[2]}{\var[2]'}}{\Dsyn[\Gamma]{\trm[2]}}{}}\\
  \tPair{\tPair{\var}{\var[2]}}{\funabs{d}{\tPlus{\var'\, (\tlFst{d})}{\var[2]'\, (\tlSnd{d})}}}\end{array}\\
  \Dsyn[\Gamma]{\tFst{\trm}}&\defeq \letin{\tPair{\var}{\var'}}{\Dsyn[\Gamma]{\trm}}{\tPair{\tFst\var}{\funabs{d}{\var'\,\tlPair{d}{\tZero}}}}\\
  \Dsyn[\Gamma]{\tSnd{\trm}}&\defeq \letin{\tPair{\var}{\var'}}{\Dsyn[\Gamma]{\trm}}{\tPair{\tSnd\var}{\funabs{d}{\var'\,\tlPair{\tZero}{d}}}}\\
  \Dsyn[\Gamma]{\tInl{\trm}}&\defeq \letin{\tPair{\var}{\var'}}{\Dsyn[\Gamma]{\trm}}{\tPair{\tInl{\var}}{\funabs{d}{\var'\,(\tlCast d)}}}\\
  \Dsyn[\Gamma]{\tInr{\trm}}&\defeq \letin{\tPair{\var}{\var'}}{\Dsyn[\Gamma]{\trm}}{\tPair{\tInr{\var}}{\funabs{d}{\var'\,(\trCast d)}}}\\
  \DsynCaseSumHang & \defeq
  \begin{array}[t]{@{}l@{}}
  \letin{\tPair{\var[3]}{\var[3]'}}{\Dsyn[\Gamma]{\trm}}{} \\
  \tSAMatch{\var[3]}{\var}{
    \begin{array}[t]{@{}l@{}}\letin{\tPair{\var_1}{\var_2}}{\Dsyn[\Gamma,\var:\ty]{\trm[2]}}{}\\
      \tPairOpen{\var_1}\tPairSep{\funabs{d}{
        \begin{array}[t]{@{}l@{}}
          \letin{\tPair{v_1}{v_2}}{\tSplit{\Dsyn{\Gamma}_2}{\Dsyn{\ty}_2}\,(\var_2\ d)}{}\\
          \tPlus{v_1}{\var[3]'\,(\tlInl v_2)}\tPairClose
        \end{array}
          }}
    \end{array}
    }{\var[2]}
  {
    \begin{array}[t]{@{}l@{}}\letin{\tPair{\var[2]_1}{\var[2]_2}}{\Dsyn[{\Gamma,\var[2]:\ty[2]}]{\trm[3]}}{}\\
      \tPairOpen{\var[2]_1}\tPairSep{\funabs{d}{
        \begin{array}[t]{@{}l@{}}
          \letin{\tPair{w_1}{w_2}}{\tSplit{\Dsyn{\Gamma}_2}{\Dsyn{\ty[2]}_2}\,(\var[2]_2\ d)}{}\\
          \tPlus{w_1}{\var[3]'\,(\tlInr w_2)}\tPairClose
        \end{array}
          }}
    \end{array}  
  }
  \end{array}
  \\
  \Dsyn[\Gamma]{r}&\defeq \tPair{r}{\funabs{d}{\tZero}}\\
  \Dsyn[\Gamma]{\tSign\trm}&\defeq \tPair{\tSign\trm}{\funabs{d}{\tZero}}\\
  \Dsyn[\Gamma]{\tOpApp{\trm_1,\ldots,\trm_n}}&\defeq  \begin{array}[t]{@{}l@{}}
  \letin{\tPair{\var_1}{\var_1'}}{\Dsyn[\Gamma]{\trm_1}}{}
  % \\ \midletvdots \\
  \ldots\;
  \letin{\tPair{\var_n}{\var_n'}}{\Dsyn[\Gamma]{\trm_n}}{}\\
  \tPairOpen{\tOpApp{\var_1,\ldots,\var_n}}\tPairSep{\funabs{d}
  \begin{array}[t]{@{}l@{}}
  \letin{\tTuple{d_1,\ldots,d_n}}{\tDOpApp{\var_1,\ldots,\var_n}{d}}{}\\\tPlus{\var_1'\ d_1}{\tPlus{\cdots}{\var_n'\ d_n}}\tPairClose
  \end{array}}
  \end{array}
  \end{align*}
  \caption{\label{fig:basic-chad} The basic reverse-mode CHAD definitions for transforming types and programs.}
\end{figure}

For the special case where the input term is scalar-valued, i.e.\ $\Gamma\vdash \trm:\R$, the transformed term
$\Gamma\vdash\Dsyn[\Gamma]{\trm}:\Pair{\R}{(\LR\to \Env{\Dsyn{\Gamma}_2})}$
computes, with sharing, the (primal) function value of $\trm$ 
as well as a function that computes its \emph{gradient} if we feed it $1$ as an input.
$\Env{\Dsyn{\Gamma}_2}$ then stores a representation of this gradient.

The basic reverse-mode CHAD algorithm is defined in \cref{fig:basic-chad}.
The transformation generates code in an extension of the source language
with function types and, displayed in \textcolor{separate}{blue}, linear types\footnote{
Linear types in the sense that they come equipped with the algebraic structure of a commutative monoid
and that the functions we consider between these types are monoid homomorphisms.
Note, however, that we combine these types using the (bi)additive conjunction (the biproduct)
rather than the multiplicative conjunction (the tensor product).
In that sense, we can freely use the structural rules of dereliction and contraction.
See \cite{vakar-2022-chad} for details.
}:
\[
\ty,\ty[2]\coloneqqq \cdots\mid \Function{\ty}{\ty[2]}\textcolor{separate}{{} \mid \LR\mid \LUnit\mid \LPair{\ty}{\ty[2]}\mid \LSum{\ty}{\ty[2]}}.
\]
For these linear types, we consider the following programs:
\[
\trm,\trm[2] \coloneqqq \funabs{\var:\ty}\trm{} \textcolor{separate}{{} \mid \trm\,\trm[2]\mid \tZero\mid \tPlus{\trm}{\trm[2]}
\mid \tlFst{} \mid \tlSnd{} \mid \tlPair{\trm}{\trm[2]}\mid \tlCast{} \mid \trCast{} \mid \tlInl{} \mid \tlInr{} \mid \tOne{}{}{}\mid \tSplit{}{}\mid  D\op^t }
\]
with the typing of \cref{fig:linear-types-api}.
In particular, all linear types are commutative monoids.

\begin{figure}
  \[
  \begin{array}{ll}
    \begin{array}[t]{@{}l@{}}
    \tOne\var\ty\Gamma : \ty \to \Env{\Gamma} \\
    \tSplit{\Gamma}{\ty}:\Env{(\Gamma,\var:\ty)}\to \Pair{\Env{\Gamma}}{\ty}
    \end{array}
    &
    \begin{array}[t]{@{}l@{}}
    D\op^t:\R^n\rightharpoonup \LR \to \LR^n\\
    (\text{for $\op\in\Op^n$})
    \end{array}
    \\
    \begin{array}[t]{@{}l@{}}
  \tlFst{}:\LPair{\ty}{\ty[2]}\to \ty\\
  \tlSnd{}:\LPair{\ty}{\ty[2]}\to \ty[2]\\
  \tlPair{-}{-}:\ty\to\ty[2]\to\LPair{\ty}{\ty[2]}\\
    \end{array}
  &
  \begin{array}[t]{@{}l@{}}
  \tlCast{}:\LSum{\ty}{\ty[2]}\to \ty\\
  \trCast{}:\LSum{\ty}{\ty[2]}\to \ty[2]\\
  \tlInl{}:\ty\to \LSum{\ty}{\ty[2]}\\
  \tlInr{}:\ty[2]\to\LSum{\ty}{\ty[2]}\\
  \end{array}
  \\[-12pt]
  \tZero[\ty]:\ty\qquad
  (\tPlus[\ty]{}{}):\ty\to\ty\to\ty
  \end{array}
  \]
  \caption{\label{fig:linear-types-api} The API for the linear types, which we view as abstract types that have multiple implementations.}
\end{figure}

For now, the reader may keep in mind the naive implementation of the linear types of Fig. \ref{fig:naive-api}, which we will later replace with a more optimised one.
\begin{figure}
\[
\begin{array}{@{}l@{}}
\begin{array}{llll}\LR\defeq \R& \LUnit\defeq \Unit & \LPair{\ty}{\ty[2]}\defeq \Pair{\ty}{\ty[2]}& \LSum{\ty}{\ty[2]}\defeq \Sum{\Unit}{(\Sum{\ty}{\ty[2]})}\\
\tZero[\LR]\defeq 0& \tZero[\LUnit]\defeq \tUnit& \tZero[\LPair{\ty}{\ty[2]}]\defeq \tPair{\tZero[\ty]}{\tZero[{\ty[2]}]}& \tZero[\LSum{\ty}{\ty[2]}]\defeq\tInl\tUnit\\
\begin{array}[t]{@{}l@{}}
  \tPlus[\LR]{\trm}{\trm[2]}\defeq\qquad\;\\
  \;\; \trm+\trm[2]\\
  \\
  \mathrlap{
  \tlPair{-}{-}\defeq\funabs\var\funabs{\var[2]}\tPair\var{\var[2]}}\\
  \mathrlap{
    \tlFst{}\defeq \funabs\var\tFst\var
  }\\
  \mathrlap{
    \tlSnd{}\defeq \funabs\var\tSnd\var
  }\\
  \mathrlap{
    \tlInl{}\defeq\funabs\var\tInr(\tInl\var)
  }\\
\mathrlap{
  \tlInr{}\defeq\funabs\var\tInr(\tInr\var)
}
\end{array} & 
\begin{array}[t]{@{}l@{}}\tPlus[\LUnit]{\trm}{\trm[2]}\defeq\qquad\;\\
  \;\; \tUnit
\end{array} & 
\begin{array}[t]{@{}l@{}}
\tPlus[\LPair{\ty}{\ty[2]}]{\trm}{\trm[2]}\defeq \\
\;\;\letin{\tPair{\var_1}{\var_2}}{\trm}{}\\
\;\;\letin{\tPair{\var[2]_1}{\var[2]_2}}{\trm[2]}{}\\
\;\;\tPair{\tPlus[\ty]{\var_1}{\var[2]_1}}{\tPlus[{\ty[2]}]{\var_2}{\var[2]_2}}
\end{array}& 
\begin{array}[t]{@{}l@{}}
\tPlus[\LSum{\ty}{\ty[2]}]{\trm}{\trm[2]}\defeq\qquad\;\\
 \;\;\tSAMatch{\trm}{\_}{\trm[2]}{\var}{
  \tSAMatch{\var}{\var'}{\tPlus[\ty]{\var'}{\tlCast\trm[2]}}{\var''}{\tPlus[{\ty[2]}]{\var''}{\trCast\trm[2]}}
    }
    % \text{(see footnote\footnotemark)}
\end{array}
\end{array}\\
\begin{array}{l}
\tlCast{}\defeq\funabs\var \tSMatch{\var}{\_}{\tZero}{\var'}{\tSMatch{\var'}{\var''}{\var''}{\_}{\tError}}\\
\trCast{}\defeq\funabs\var \tSMatch{\var}{\_}{\tZero}{\var'}{\tSMatch{\var'}{\_}{\tError}{\var''}{\var''}}\\
\tOne{\var_i}{\ty_i}{\Gamma}\defeq\funabs{\var_i}{\tTuple{\tZero[\ty_1],\cdots,\tZero[\ty_{i-1}],\var_i,\tZero[\ty_{i+1}],\cdots,\tZero[\ty_n]}}\qquad\qquad
\tSplit{\Gamma}{\ty} \defeq \funabs{\var[2]}{\var[2]}
\end{array}
\end{array}
\]
\caption{\label{fig:naive-api}
  The naive implementation of the API required for our linear types that store cotangents.
  `$\protect\tOne{}{}{}$' produces a \emph{one-hot} vector: an all-zero structure except for one element.
}
\end{figure}
The linear operations $D\op^t:\R^n \rightharpoonup \R\to \R^n $ are assumed to be implementations of the transposed derivatives of the operations $\op$.\footnote{We choose this typing, including the incoming cotangent, over the perhaps more traditional $D\op^t : \R^n \rightharpoonup \R^n$ to generalise cleanly to primitive operations returning larger structures than a single scalar.}

A note on scoping: we assume that all variable names are unique in the input program, and implicitly consider all variables on the right-hand side of `$\defeq$' in \cref{fig:basic-chad} fresh unless they coincide with a source program variable (such as $\var$ in $\Dsyn[\Gamma]{\textbf{let} \ldots}$).
In effect, we consider variable names to be syntactic sugar for De Bruijn indices.

Finally, note that our formulation of CHAD does not use dependent types, in contrast to earlier work on the algorithm\footnote{\cite{nunes-2022-chad-expressive} presents the same basic algorithm but with very fine-grained linear dependent types \cite{krishnaswami2015integrating,DBLP:conf/fossacs/Vakar15} to guide the correctness proof.}: even if this paper is about asymptotic complexity, we are interested in implementation efficiency and typical high-performance array languages do \emph{not} support dependent types in their input language.
The $\tError$ cases in \cref{fig:naive-api} arise from the conversion to simple types; the well-typedness of the algorithm with dependent types proves that the $\tError$ cases are unreachable.

% \footnotetext{$  \tPlus[\LSum{\ty}{\ty[2]}]{\trm}{\trm[2]}\defeq\tSMatch{\trm}{\_}{\trm[2]}{\var}{
%   \tSMatch{\var}{\var'}{\tPlus[\ty]{\var'}{\tlCast\trm[2]}}{\var''}{\tPlus[{\ty[2]}]{\var''}{\trCast\trm[2]}}
%     }$}

\section{Key Ideas}\label{sec:key-ideas}

\paragraph{Key Complexity Criterion}
As far as we are aware, the expected asymptotic complexity of reverse AD was first studied by
\cite{linnainmaa1976taylor}.
In its basic form (without e.g.\ checkpointing), reverse AD should compute the 
gradient of a function $f:\R^n\to \R$ in time proportional to the time required to compute the original function.
Put more precisely, there exist some (relatively small) constants $c$ and $c'$ such that for all functions $f:\R^n\to \R$ that are expressible in some programming language, the cost of computing the gradient $\nabla f(x)$ is less than $c'$ plus $c$ times the cost of computing $f(x)$, for any input $x$.
It is important that $c$ and $c'$ are uniform: they depend neither on the program $f$ nor on the input $x$.
Formally, we therefore demand that 
\begin{equation}
  % \begin{array}{l}
\!\!\exists c, c' > 0.\ 
\forall (\var : \ty\vdash \trm : \R).\ 
\forall \var : \ty.\ 
\forall d : \LR.\ 
\cost{{\tSnd{\Dsyn[\Gamma]{\trm}}\ d}}{\var=\var, d=d} 
 \leq
  c'
  + c \cdot \cost{\trm}{\var=\var}
% \end{array}
\label{eq:complexity-criterion}
\end{equation}
where $\cost{\trm}{\var_1=v_1,\ldots, \var_n=v_n}$ is the time cost of evaluating the program $\trm$ in the environment where the variables $\var_1,\ldots,\var_n$ are in scope and have values $v_1,\ldots,v_n$.
This is the key complexity property that we prove (in \cref{sec:complexity-proof}) that CHAD satisfies after the optimisations described in this paper.\footnote{To be precise, seeing as one of our optimisations makes CHAD generate monadic code, the complexity criterion will change slightly to include the handler $\tRun{}{}$ for the relevant monad.}
In fact, because of the inductive structure of the proof, we prove a stronger statement that generalises \cref{eq:complexity-criterion} to programs of the form $\var_1:\ty[2]_1,\ldots,\var_n:\ty[2]_n\vdash\trm:\ty$.

\paragraph{Identifying Complexity Challenges}
We want to prove this criterion by induction on the structure of programs. Therefore, \cref{eq:complexity-criterion} (generalised to terms of types other than $\R$) should apply not only to the input program as a whole, but also to all subterms.
This is a very strong requirement!
In particular, it immediately exposes efficiency problems with the naive implementation of the commutative monoids in \cref{fig:naive-api}: when we
\begin{enumerate}
\item\label{item:key-ideas-problems-discard} discard values (e.g.\ with projections such as $\tFst{}$), we need zeros $\tZero[\ty]$ to take constant time;
\item\label{item:key-ideas-problems-share} share values (use let-bound variables multiple times), we need cotangent addition $\tPlus[\ty]{}{}$ to take constant time (seemingly --- see below);
\item\label{item:key-ideas-problems-envadd} have terms with multiple subterms, such as $\tPair{-}{-}$, we need environment addition $\tPlus[\Env\Gamma]{}{}$ to take constant time;
\item\label{item:key-ideas-problems-onehot} reference variables, we need one-hot environment vectors $\tOne\var\ty\Gamma$ to take constant time.
\end{enumerate}

\paragraph{Solving the Problems}
Problems (\ref{item:key-ideas-problems-discard}) and (\ref{item:key-ideas-problems-onehot}) we address in \cref{sec:sparsity} by replacing the naive definitions of $\LPair{\ty}{\ty[2]}$ and $\Env\Gamma$ with sparse versions.
While these fixes keep the code transformation itself as-is, the tree-map we use to implement $\Env\Gamma$ introduces log-factors in the complexity of certain operations.

In \cref{sec:monadic-lifting}, we then address problem (\ref{item:key-ideas-problems-envadd}) by eliminating environment vector additions entirely.
We move from divide-and-conquer style (creating environment vectors in subterms and merging them where they meet in enclosing terms) to state-passing style.
That is, we create a \emph{single} environment vector at the beginning of the computation and update it with individual contributions as we go.
This change puts the code in monadic style, meaning that we can even perform those updates \emph{mutably} and eliminate the log-factors from certain operations again.
At this point, due to the state-passing style, the only place where we still use $\tPlus[\ty]{}{}$ is in the case for variables ($\Dsyn[\Gamma]{\var : \ty}$).

\paragraph{Proving the Complexity Criterion}
It turns that the algorithm now already has the right complexity!
Point (\ref{item:key-ideas-problems-share}) above is not \emph{actually} a requirement, because we can use an \emph{amortisation argument} to discount the additions performed in $\Dsyn[\Gamma]{\var : \ty}$ against the work done to build up the cotangents that are being added.
The argument works by making two observations:
\begin{enumerate}
\item\label{item:key-ideas-prop-affine} CHAD treats cotangents in an affine way (i.e.\ after adding something to a cotangent, the original cotangent is not used any more);
\item\label{item:key-ideas-prop-sum} The addition of our sparsely represented cotangents can be performed in cost that is proportional to the size of their intersection, and the sum is \emph{smaller} in size than the summands.
\end{enumerate}
Property (\ref{item:key-ideas-prop-affine}) means that it is valid to associate a ``computation budget'' with each built-up cotangent value (this budget will not be magically duplicated), and property (\ref{item:key-ideas-prop-sum}) means that $\tPlus[\ty]{}{}$ preserves the invariant that we always have this budget.
\cref{sec:informal-amortisation} explains the details.

We implement this amortisation argument by strengthening the induction hypothesis of the complexity proof to be aware of this computation budget.
The theorem that we prove (\cref{eq:theorem-phi} in \cref{sec:complexity-proof}) is thus slightly different from the original criterion, which fortunately follows as a corollary (\cref{eq:theorem-cor}), yielding the final result.
We have formalised the complexity proof in the Agda proof assistant.
No encoding of big-$O$ theory was necessary, because the only big-$O$ expressions used in our work are of the form $f = O(g)$, which can be expressed simply by an existential constant, as indeed we did in \cref{eq:complexity-criterion}.

\paragraph{Efficient CHAD for Arrays}
To show that CHAD, also in asymptotically-efficient form, extends to parallel array operations, \cref{sec:efficient-chad-on-array-types} gives derivatives for three such operations:
\begin{itemize}
  \item `$\taBuild{\trm[2]}{i}{\trm}$', which constructs an array of length $\trm[2]$ with value $\trm$ at position $i$;
  \item $\taIndex{\trm}{\trm[2]}$, which indexes the array $\trm$ at index $\trm[2]$;
  \item `$\taFold{\var}{\trm}{\trm[2]}$', which reduces the non-empty array $\trm[2] : \Array\ty$ using the element combination function $\Gamma, \var : \Pair\ty\ty \vdash \trm : \ty$.
\end{itemize}
The methodology applies also to further array operations that complete the full set supported by typical array languages.
We have chosen this small subset to illustrate the most important and difficult points.

Arrays behave like product types in the sense that indexing behaves like a generalised projection.
Consequently, we need the same sparsity tricks in $\Dsyn{\Array\ty}_2$ as we used for $\Dsyn{\Pair{\ty}{\ty[2]}}_2$ --- and more, because arrays can have any length and pairs always have length 2.\footnote{The fact that arrays are \emph{variably} sized is not important here.}
The representation that we choose for $\Dsyn{\Array\ty}_2$ is `$\Bag{(\Pair{\Z}{\Dsyn{\ty}_2})}$': a representation of a list of index--value pairs that supports constant-time concatenation.
This data type works for the complexity property\footnote{This we do not formally prove in this work; the formal proof is limited to the language in \cref{fig:source-language-1}.} in a sequential setting; the fact that it parallelises poorly is not fully resolved in this paper, but we discuss some mitigation approaches in \cref{sec:parallelism}.

However, we do argue why parallelisation of our optimised CHAD algorithm is possible in the first place: this may seem unlikely given that the transformation now produces monadic code!
But monads do not inherently prevent parallelism: the left-hand and right-hand side of a bind operation ($\tBind$) are of course sequentialised, but such a bind operation only occurs in full generality in our output programs when there was already a data-dependency in the source program (such as for let-bindings; see \cref{fig:monadic-chad} in \cref{sec:monadic-lifting}).
Such places indeed inherently do not have parallelism, but many others do: this possibility is evidenced by the use of command sequencing $x \tSeq y=x\tBind \lambda \_.\ y$ instead of the general ($\tBind$).

Because we only use \emph{accumulation} instead of fully general mutation in the monad --- that is, we only \emph{add} values to the state --- we can rearrange effects at will (our monad is \emph{commutative}) and thus parallelisation is possible, using atomic operations for the actual implementation.
In particular, we have a parallel implementation of $(\tSeq)$.

\paragraph{Closure Conversion for Efficient CHAD of Lambdas}
Naive CHAD of higher order functions leads to a duplication of work, due to separation of the transposed derivative w.r.t. function arguments and captured context variables.
This inefficiency can be exploited to get an exponential blow-up in complexity.
An obvious solution is to get rid of all captured context variables, i.e. to apply closure conversion before CHAD.
One option for this is to apply full defunctionalisation first, which uses a global program analysis to reduce function types to a combination of finite sum types and tuples, constructs we already know how to differentiate.
Another option, which does retain locality and compositionality of the code transformation, is to explain how to differentiate the infinite sum types or existential types required to do typed closure conversion.
We discuss both in \cref{sec:efficient-chad-on-function-types}.

\section{Finding and Solving Efficiency Problems}\label{sec:efficiency-problems}

The basic algorithm defined in \cref{sec:basic-chad} is relatively simple and pleasant to analyse.
However, it fails to satisfy the complexity criterion in \cref{eq:complexity-criterion}, so there are some complexity issues to address.

\paragraph{Identifying Complexity Problems}
To spot the complexity problems in the algorithm from \cref{fig:basic-chad}, we can try to prove the complexity criterion in \cref{eq:complexity-criterion} inductively.
To do this, we first have to strengthen the statement to also apply to subexpressions with larger contexts and non-$\R$ result~types:
\begin{equation}
\begin{array}{@{}l@{}}
  \exists c, c' > 0.\ 
  \forall (\var_1 : \ty[2]_1, \ldots, \var_n : \ty[2]_n \vdash \trm : \ty).\ 
  \forall \var_1 : \ty[2]_1, \ldots, \var_n : \ty[2]_n.\ 
  \forall d : \Dsyn{\ty}_2. \\
  \qquad
  \cost{{\tSnd{\Dsyn[\Gamma]{\trm}}\ d}}{\var_1=\var_1, \ldots, \var_n=\var_n, d=d} 
   \leq
    c'
    + c \cdot \cost{\trm}{\var_1=\var_1, \ldots, \var_n=\var_n}
\end{array}
  \label{eq:complexity-criterion-generalised}
\end{equation}
Interpreting \cref{eq:complexity-criterion-generalised} intuitively, the criterion states that the cost of evaluating $\Dsyn[\Gamma]{\trm}$, plus the cost of calling the backpropagator in its second component, must be within a constant factor of the cost of evaluating the original expression $\trm$.\footnote{The $c'$ is convenient for the proof, but could technically be removed assuming that all terms have cost $\geq1$.}

For instance, consider $\trm = \tPair{\trm_1}{\trm_2}$.
We can see (in \cref{fig:basic-chad}) that $\Dsyn[\Gamma]{\tPair{\trm_1}{\trm_2}}$ evaluates $\Dsyn[\Gamma]{\trm_1}$ and $\Dsyn[\Gamma]{\trm_2}$, and the body of the backpropagator calls the backpropagators of $\trm_1$ and $\trm_2$.
By the induction hypothesis, these are offset by the evaluation of the original $\trm_1$ and $\trm_2$ on the left-hand side.
Thus, the inductive step of the proof would hold for pair expressions --- were it not for the (single) use of $\tPlus{}{}$ on environment vectors $\Env\Gamma$ in the backpropagator: the fact that $\tPlus[\Env\Gamma]{}{}$ is expensive on the naive representation of $\Env\Gamma$ from \cref{sec:basic-chad} makes the pair case problematic.

\paragraph{Inefficient Monoid Structures on Cotangents}
Similar reasoning elsewhere in the code transformation uncovers some more operations that are required to be efficient and are potentially not.
Naively (but see \cref{sec:informal-amortisation}), all of the operations shown in \cref{fig:constant-time-methods} must be constant-time.

\begin{table}
\begin{center}
\begin{tabular}{l|l|l|l}
\textbf{Type} & \multicolumn{2}{l|}{\textbf{Operation}} & \textbf{Reason for appearance} \\ \hline
\multirow{4}*{$\Env\Gamma$}
  & zero & $\tZero[\Env\Gamma]$ & constants (e.g.\ $\tUnit$) \\
  & plus & $\tPlus[\Env\Gamma]{}{}$ & multiple subterms (e.g.\ $\tPair{\trm}{\trm[2]}$ or $\tOpApp{\trm_1, \ldots, \trm_n}$) \\
  & one-hots & $\tOne\var\ty\Gamma$ & variable references ($\var$) \\
  & splitting & $\tSplit\Gamma\ty$ & scopes (e.g.\ $\mathbf{let}$, $\mathbf{case}$) \\ \hline
\multirow{3}*{Cotangents}
  & zero & $\tZero[\ty]$ & unused variables \\
  & plus & $\tPlus[\ty]{}{}$ & variable sharing; see \cref{sec:informal-amortisation} \\
  & one-hots & e.g.\ $\tlPair{d}{\tZero}$ & product projections (e.g. $\tFst{}$); already constant-time
\end{tabular}
\end{center}
\caption{\label{fig:constant-time-methods}
  Operations required to be constant-time (seemingly, in the case of $\protect\tPlus[\ty]{}{}$).
  \vspace{-0.25cm}
}
\end{table}

Because our source language type system (\cref{fig:source-language-1}) does not have unbounded-size products (i.e.\ $n$-ary tuples), one-hot cotangents are not an issue, because just building the one-hot value from (constant-time) zeros will take bounded cost anyway.
An example of this is the $\tlPair{d}{\tZero}$ in $\Dsyn[\Gamma]{\tFst{\trm}}$ in \cref{fig:basic-chad}.
(Had we included $n$-ary tuples in our language, these one-hots would have been a larger issue, needing a solution similar to that for arrays (\cref{sec:efficient-chad-on-array-types}); after all, arrays differ from (homogeneous) $n$-ary tuples only in that their size is not known statically, which does not matter much from a complexity perspective.)
Thus we can focus on the other operations.

As an example of the need for constant-time zero, consider the program $\trm$:
% \TODO{removable}
\[
  \var[1] : \Pair \R \ty \vdash \tOpApp{\tFst{\var[1]}, \ldots, \tFst{\var[1]}} : \R
\]
for some some $n$-ary operation $\op$.
Because of the $\tZero$ in the rule for $\Dsyn[\Gamma]{\tFst{\trm}}$, the backpropagator of the derivative of $\trm$ --- that is, $\tSnd{\Dsyn[{\var[1]} : \Pair \R \ty]{t}}$ --- will use $\tZero[{\Dsyn{\ty}_2}]$ once for each occurrence of $\tFst{}$ to create $n$ zero values of type $\Dsyn{\ty}_2$.
If $\tZero[{\Dsyn{\ty}_2}]$ takes time $T_0$, then the backpropagator of the derivative of $t$ will take time at least on the order of $n \cdot T_0$, whereas $t$ itself is $O(n)$.
Hence $T_0$ must indeed be constant, i.e.\ not dependent on $\ty$.

In addition to the zeros, the $n - 1$ uses of $\tPlus[{\Dsyn{\ty}_2}]{}{}$ in $\tSnd{\Dsyn[\Gamma]{\tOpApp{\ldots}}}$ also take time, and thus it would seem that $\tPlus{}{}$ needs to be constant-time as well, explaining its presence in the table above.
However, as we explain in \cref{sec:informal-amortisation}, with a smarter analysis we can weaken this requirement.

% This problem is not limited to products: deep nesting of coproducts $\Sum{\ty[1]}{\ty[2]}$ also results in problems when creating zeros
% \footnote{Note that $\tZero[\Dsyn{\Sum{\ty[1]}{\ty[2]}}_2]$ is not actually implementable currently; this will be fixed as a byproduct of \cref{sec:sparsity}.}
% with $\tZero[\Dsyn{\Sum{\ty[1]}{\ty[2]}}_2]$ or adding derivative values using $\tPlus[{\Dsyn{\Sum{\ty[1]}{\ty[2]}}_2}]{}{}$, which will take time on the order of the nesting depth.

Our current, naive implementations of $\Env\Gamma$ and $\Dsyn{\ty}_2$ do not at all reach these constant-time requirements, so we need to do something about this.
Indeed, so far, $\Env{\Gamma}$ is a simple product of the types in $\Gamma$ (as defined in \cref{sec:basic-chad}: $\Env{(x_1 : \ty_1, \ldots, x_n : \ty_n)} = \ty_1 \times \cdots \times \ty_n$):
this representation has expensive zero, plus and one-hot (all at least $O(n)$ work), although it supports efficient $\tSplit\Gamma\ty$.

\paragraph{Solving the Problems with Sparse Representations}
We address the problem of expensive zeros in \cref{sec:sparsity} by choosing \emph{sparse} representations of the monoids
% in \cref{fig:linear-types-api,fig:naive-api}
that are not already sparse: $\Dsyn{\Pair{\ty}{\ty[2]}}_2$ and $\Env\Gamma$.
Then (in \cref{sec:monadic-lifting}) we lift the output of the transformation to monadic code in order to eliminate the need for $\tPlus[\Env\Gamma]{}{}$, as well as to prepare for making $\tOne{}{}{}$ and $\tSplit{}{}$ constant time (which became logarithmic because of the sparse representation) using a mutable array in \cref{sec:log-factors}.
Finally, we only have $\tPlus[\ty]{}{}$ left, which then turns out to not be a problem any more: an amortisation argument (informally in \cref{sec:informal-amortisation}, more formally in \cref{sec:complexity-proof}) shows that we can discount $\tPlus[\Dsyn\ty_2]{}{}$ against building up of cotangent values.

\subsection{Step 1: Sparsity}\label{sec:sparsity}

% solution: clever (sparse) implementation of monoid data structures and environments
% Remaining problem: just log factors (in the sequential setting)

Sparse data structures aim to represent the uninteresting parts of a data structure as compactly as possible, focusing on the interesting (usually the \emph{non-zero}) parts.
Such representations not only conserve memory but also serve to avoid computing with many useless (zero) values, since the result is typically as uninteresting as the input: zero.

% Some of our complexity problems arise from the need to fully represent (large) zeros in our derivative monoids: the cost of $\tZero$ and  $\tOne{\var}{\ty}{\Gamma}$.
% Although it is less clear how to solve the problems with $\tPlus{}{}$, a sparse representation will end up being helpful there too.

\paragraph{Fixing Cotangent Zeros}
Given the tree structure of our types (which are built out of products and sums), a natural first attempt for a sparse representation is to add an explicit, redundant value representing ``zero'' for products.
($\tZero[\LUnit]$, $\tZero[\LR]$ and $\tZero[\LSum{\ty}{\ty[2]}]$ are already constant-time.)
That is, we change the implementation of $\LPair{\ty}{\ty[2]}$ from $\Pair{\ty}{\ty[2]}$ to $\Sum{\Unit}{(\Pair{\ty}{\ty[2]})}$.
The implementation of its API then changes to:
% \[
% \begin{array}{ll}
% \tZero[\LPair{\ty}{\ty[2]}]\defeq \tInl\tUnit&
% \tPlus[\LPair{\ty}{\ty[2]}]{\trm}{\trm[2]}\defeq
% \begin{array}[t]{@{}l@{}}
% \tSAMatch{\trm}{\_}{\trm[2]}{\var}{\tSAMatch{\trm[2]}{\_}{\tInr\var}{\var[2]}{
%   \begin{array}[t]{@{}l@{}}
%     \letin{\tPair{\var_1}{\var_2}}{\var}{}\\
%     \letin{\tPair{\var[2]_1}{\var[2]_2}}{\var[2]}{}\\
%     \tInr\tPair{\tPlus[\ty]{\var_1}{\var[2]_1}}{\tPlus[{\ty[2]}]{\var_2}{\var[2]_2}}
%   \end{array}
% }}
% \end{array}\\
% \tlPair{-}{-}\defeq\funabs{\var}\funabs{\var[2]}\tInr{\tPair{\var}{\var[2]}}\qquad\;&
% \tlFst{}\defeq \funabs{\var}{\tSMatch{\var}{\_}{\tZero}{\var'}{\tFst\var'}}\\
% &
% \tlSnd{}\defeq \funabs{\var}{\tSMatch{\var}{\_}{\tZero}{\var'}{\tSnd\var'}}
% \end{array}
% \]
\[
  \begin{array}{l@{\qquad}l}
  \tZero[\LPair{\ty}{\ty[2]}] \defeq \tInl\tUnit &
  \tPlus[\LPair{\ty}{\ty[2]}]{\tInl{\_}}{\var[2]} \defeq \var[2] \\
  \tlPair{\var}{\var[2]} \defeq \tInr{\tPair{\var}{\var[2]}} &
  \tPlus[\LPair{\ty}{\ty[2]}]{\var}{\tInl{\_}} \defeq \var \\
  & \tPlus[\LPair{\ty}{\ty[2]}]{\tInr{\tPair{\var_1}{\var_2}}}{\tInr{\tPair{\var[2]_1}{\var[2]_2}}} \defeq \tInr{\tPair{\tPlus{\var_1}{\var[2]_1}}{\tPlus{\var_2}{\var[2]_2}}} \\[0.2em]
  \multicolumn{2}{l}{\tlFst{\var} \defeq \tSMatch{\var}{\_}{\tZero}{\var'}{\tFst\var'}} \\
  \multicolumn{2}{l}{\tlSnd{\var} \defeq \tSMatch{\var}{\_}{\tZero}{\var'}{\tSnd\var'}}
  \end{array}
\]
The result is that $\tZero[\Dsyn{\ty}_2]$ is now constant time for all types $\ty$ in our source-language type system (\cref{fig:source-language-1}).

\paragraph{Fixing Environment Zeros and One-Hots}
Since we implemented $\Env{(x_1 : \ty_1, \ldots, x_n : \ty_n)}$ as $\ty_1 \times \cdots \times \ty_n$ in \cref{sec:basic-chad}, the $\tZero[\Env{\Gamma}]$ in $\Dsyn[\Gamma]{\tUnit}$ and the $\tOne{\var}{\Dsyn{\ty}_2}{\Dsyn{\Gamma}_2}$ in $\Dsyn[\Gamma]{\var : \ty}$ still have to create a big tuple filled with zeros.
A standard way to make a sequence (a total map from bounded integers to values) sparse is to use an associative array,
often implemented as a balanced binary tree in programming languages\footnote{And in purely functional programming languages, almost always, because there a \emph{mutable} hashmap does not work well.} (which is essentially a \emph{partial} map from keys to values).
Since~our keys are slightly odd --- they are pointers into the environment, and the type of the associated~value depends on the index --- we will explicitly specify the interface of such a map data structure instead of directly instantiating the standard data structure. However, it should be clear that it can be implemented with a standard immutable tree-map, modulo type-safety.
This API is shown in \cref{fig:env-map-interface}.

\begin{figure}
\begin{align*}
&\textbf{data}\ \EnvMap{\Gamma} \qquad \textit{--- all types in $\Gamma$ must be monoids.} \\
&\begin{array}{@{}l@{\ }l@{\quad}l@{}}
  \mEmpty &: \EnvMap{\Gamma} & \textit{--- $O(1)$} \\
  \mGet\var\tau\Gamma{} &: \EnvMap{\Gamma} \ra \ty & \textit{--- $O(\log(\text{map size}))$} \\
  \mPush{} &: \EnvMap{\Gamma} \ra \EnvMap{(\Gamma, \var : \ty)} & \textit{--- only a type change; $O(1)$} \\
  \mPopx{} &: \EnvMap{(\Gamma, \var : \ty)} \ra \EnvMap\Gamma & \textit{--- $O(\log(\text{map size}))$; map delete operation} \\
  \mModify\var\ty\Gamma{}{} &: (\ty \ra \ty) \ra \EnvMap\Gamma \ra \EnvMap\Gamma & \textit{--- $O(\log(\text{map size})) + (\text{one call to the function})$} \\
  \mUnion{}{} &: \EnvMap\Gamma \ra \EnvMap \Gamma \ra \EnvMap \Gamma & \textit{--- [see caption]; uses $\tPlus{}{}$ on types in $\Gamma$} \\
  \mPop{} &:  \EnvMap{(\Gamma, \var : \ty)} \ra \Pair{\EnvMap\Gamma}{\ty} & \textit{--- derived: $\mPop e \defeq \letin{\var[2]}{e}\tPair{\mPopx \var[2]}{\mGet \var {} {} \var[2]}$}
\end{array}
\end{align*}
\caption{\label{fig:env-map-interface}
  The interface of the environment map, used in \cref{sec:sparsity}.
  Here, the notation ``$\var : \ty \in \Gamma$'' means a pointer into $\Gamma$, i.e.\ an integer from 1 to the length of $\Gamma$.
  Ignoring type safety, this interface, including the noted time complexities, can be implemented with a standard immutable tree map.
  Regarding $\protect\mUnion{}{}$: if passed two maps of size $m$ and $n$, with $m \leq n$, its runtime is $O(m \log(\frac nm + 1)) + (\textit{the required $\tPlus{}{}$ calls})$, and this complexity is optimal.~\cite{algorithms-1979-merging-tarjan}
  The first term is $O(n)$ when $m \approx n$ and $O(\log(n))$ when $m$ is small.
}
\end{figure}

In the interface, $\mEmpty$ creates an empty map, and $\mGet\var\tau\Gamma e$ gets the entry for $\var$ from the map $e$, returning $\tZero[\tau]$ if there is no entry there yet.
$\mPush{}$ and $\mPopx{}$, respectively, extend and reduce the environment, where $\mPopx{}$ performs actual work because it removes the now-extraneous~entry from the underlying map structure; $\mPush{}$ simply changes the type to make an additional key allowable, but does not need to add the key yet.
(Conceptually, $\mPush{}$ pushes a $\tZero$, but zeros are~elided due to sparsity.)
$\mModify\var\tau\Gamma f e$ applies the function $f$ to the value in the map at the variable $\var$, pre-initialising with $\tZero[\tau]$ if necessary.
$\mUnion{}{}$ takes the union of the two maps, adding overlapping values using the addition operation $\tPlus[\ty]{}{}$ of the monoids $\ty$ contained in $\Gamma$. 
Finally, $\mPop{}$ is a derived operation.

Using this interface, we can implement $\Env\Gamma$ as $\EnvMap\Gamma$ and instantiate its methods as follows:
\[
  \begin{array}{@{}l@{\ \defeq\ }l@{\qquad}l@{\ \defeq\ }l@{}}
    \tZero[\Env\Gamma] & \mEmpty
    & \tOne{\var}{\ty}{\Gamma}\ v & \mModify\var\ty\Gamma{(\funabs{\var[2]}{\tPlus{v}{\var[2]}})}{\mEmpty} \\
    \tPlus[\Env\Gamma]{}{} & \mUnion{}{}
    & \tSplit\Gamma\ty & \mPop{}.
  \end{array}
\]
At this point, zero cotangents (due to our new sparse $\LPair{\ty}{\ty[2]}$) as well as zero and one-hot environment vectors can all be constructed in constant time as required.
Note that the program transformation of \cref{fig:basic-chad} has not changed: simply the implementation of some of the types has changed.

Thus the remaining points of care in the derivative program are \emph{addition} of cotangent values and environment cotangents, as well as $\tSplit\Gamma\ty$, which has inadvertently become log-time with this change in representation of $\Env\Gamma$; we address these in the following two sections.

\subsection{Step 2: Monadic Lifting}\label{sec:monadic-lifting}

Let us first focus on the addition operation on environment cotangents ($\Env{\Dsyn{\Gamma}_2}$), which occurs in the differentiated program whenever evaluation of the corresponding source program term involves evaluating more than one subterm.\footnote{In \cref{fig:basic-chad}, this happens for let-bindings, pair constructors, case elimination and primitive operations (of arity $\geq 2$).}
Usually, however, these source program terms do not take time on the order of the size of the environment, and hence do not ``pay'' (in the sense of a direct inductive proof of the complexity criterion \cref{eq:complexity-criterion-generalised} as explained at the start of \cref{sec:efficiency-problems}) for $\mUnion{}{}$, which is at least logarithmic in the size of its arguments.\footnote{An amortisation argument in the style of \cref{sec:informal-amortisation} will not work here; \ifTomACMVersion{\cite[Appendix A]{efficient-chad-arxiv} gives a counterexample}{a counterexample is given in \cref{app:union-not-efficient}}.}

Fortunately, we have some yet-unexploited flexibility: because the environment cotangent is only ever modified by \emph{adding} contributions to it (and since addition is commutative and associative), we can rearrange these additions at will.
For example, instead of returning the environment cotangent contributions from each subprogram and merging the results using $\mUnion{}{}$, we can also pass a growing accumulator around in state-passing style, adding individual $\tOne\var\ty\Gamma$ contributions to it as the derivative program executes.
Using a slightly modified definition of $\tOne{}{}{}$:
\[
  \tOnep\var\ty\Gamma\ v\ e \defeq
    \mModify\var\ty\Gamma{(\funabs{\var[2]}{\tPlus{v}{\var[2]}})}{e}
\]
we can add the single contribution (here $v$) from $\Dsyn[\Gamma]{\var : \ty}$ in time $O(\log(\text{map size})) + (\text{cost of `$\tPlus{}{}$'})$ to the passed state $e$ of type $\Env\Gamma$.

To use this $\tOnep{}{}{}$, we have to change the structure of the derivative program somewhat: instead of passing environment contributions upwards, merging them as control flows meet, we pass around a \emph{single} environment cotangent in state-passing style, that we modify with $\tOnep{}{}{}$ each time we encounter a variable reference.
The result is that the derivative program then lives inside a variant of a \emph{local state monad} \cite{fp-2002-notions-computation-monads,fp-2010-local-state-staton}: it is a state monad\footnote{Due to the absence of a `get' operation, it can also be seen as a (CPS-style) Writer monad. This is also connected to its parallelisability (\cref{sec:parallelism}). To more explicitly describe its implementation, however, we will call it a state monad.} (that we will call $\EnvM{}{}$, for \underline{e}nvironment \underline{v}ector \underline{m}onad) whose state is divided up into components, one for each entry in the environment.
$\tZero[\Env\Gamma]$ becomes $\tReturn\tUnit$, $\tPlus[\Env\Gamma]{}{}$ becomes $\tSeq$, $\tOne\var\ty\Gamma$ becomes $\tOnep\var\ty\Gamma$, and usages of $\tSplit\Gamma\ty$ become usages of $\tScope\Gamma\ty$ (see below) instead, using $\tBind$ to extract the results:
\[
  \begin{array}{@{}l@{}}
    \EnvM\Gamma\ty \defeq \EnvMap\Gamma \ra \Pair\ty{\EnvMap\Gamma} \qquad \textit{--- again all types in $\Gamma$ must be monoids.} \\
    \begin{array}{@{}l@{\ :\ }l}
    \tOne\var\ty\Gamma & \ty \ra \EnvM\Gamma\Unit \hspace{2.35cm} \textit{--- implemented in terms of $\tOnep\var\ty\Gamma$ from above} \\
    \tScope\Gamma\ty & \EnvM{(\Gamma, \var : \ty)}{\ty[2]} \ra \EnvM\Gamma{(\Pair{\ty[2]}\ty)} \qquad \textit{--- initial $\ty$ value is $\tZero[\ty]$} \\
    \tRun{}{} & \EnvM{\Gamma}{\ty} \ra \ET\Gamma \ra \Pair{\ty}{\ET\Gamma}
    \end{array}
  \end{array}
\]
% $\EnvM{}{}$ is the new local state monad, with methods as shown.

\paragraph{About the Methods}
Notable here is that the role of $\tSplit{}{}$ is now fulfilled by $\tScope{}{}$.
In principle, because we are now state-passing, we must not only \emph{pop} (split off) the outermost variable of the environment cotangent vector returned by the derivative of a subcomputation with an extended scope (e.g.\ the body of a \textbf{let}), but also \emph{push} an empty entry on the incoming vector before being able to pass it on to that same subcomputation in the first place.
Had we chosen to use pure state-passing style ($\Env\Gamma = \EnvMap\Gamma \ra \EnvMap\Gamma$), we would indeed have used
% $\mPush{} : \EnvMap{\Gamma} \ra \EnvMap{(\Gamma, \var : \ty)}$
$\mPush{}$
from \cref{fig:env-map-interface}; however, here in monadic style, such separate $\textbf{push}$ and $\tSplit{}{}$ would not typecheck.
Thus, $\tScope{}{}$ combines both.

Fortunately, $\Dsyn[\Gamma]{}$ is compositional, meaning that $\Dsyn[\Gamma']{\trm[2]}$ is a subterm of $\Dsyn[\Gamma]{\trm}$ whenever $\trm[2]$ is a subterm of $\trm$.
(And $\Dsyn[\Gamma]{\trm}$ does not depend in any other way on the structure of $\trm[2]$.)
Therefore, we can \emph{scope} the usage of an extended environment to the monadic subcomputation that handles the subterm with that extended environment in the style of a local state monad. 
This scoping is done by the updated $\tScope\Gamma\ty$ above: conceptually, it first extends the $\EnvMap\Gamma$ in the state to an $\EnvMap{(\Gamma, \var : \ty)}$ (the push step --- semantically storing $\tZero[\ty]$ in the new cell but operationally, because of sparsity, just changing the monad type), then runs the subcomputation of type $\EnvM{(\Gamma, \var : \ty)}{\ty[2]}$ with that extended state, and finally pops off the extra value of type $\ty$ and returns it along with the return value of the subcomputation (of type $\ty[2]$).

Finally, $\tRun{}{}$ is the handler (see \cite{fp-2013-effect-handlers}) of the monad, where for an environment $\Gamma = \var_1 : \ty_1, \ldots, \var_n : \ty_n$ we define $\ET\Gamma \defeq \Pair{\Pair{\Pair{\ty_1}{\ty_2}}{\ldots}}{\ty_n}$.

\paragraph{What Did We Achieve?}

\begin{figure}
  \[
    \Dsyn{\Gamma}_1
      \vdash \Dsyn[\Gamma]{\trm} : \Pair{\Dsyn{\ty}_1}{(\Function{\Dsyn{\ty}_2}{\EnvM{\Dsyn{\Gamma}_2}\Unit})}
  \] 
  \begin{align*}
  \Dsyn[\Gamma]{\var : \ty}
    &\defeq \tPair{\var : \Dsyn{\ty}_2}{\funabs{d}{\tOne\var{\Dsyn{\ty}_2}{\Dsyn{\Gamma}_2}\ d}} \\
  \Dsyn[\Gamma]{\letin{\var : \ty}{\trm}{\trm[2]}}
    &\defeq
    \begin{array}[t]{@{}l@{}}
      \letin{\tPair{\var}{\var'}}{\Dsyn[\Gamma]{\trm}}{
        \letin{\tPair{\var[2]}{\var[2]'}}{\Dsyn[\Gamma,\var:\ty]{\trm[2]}}{}} \\
        \tPair{\var[2]}
             {\funabs{d}{\tDo{\tPair{\tUnit}{\textit{d\var}} \leftarrow \tScope{\Dsyn{\Gamma}_2}{\Dsyn{\ty}_2}\ (\var[2]'\ d) ; \var'\ \textit{d\var}}}}
    \end{array} \\
  \Dsyn[\Gamma]{\tUnit}
    &\defeq \tPair{\tUnit}{\funabs{d}{\tReturn{\tUnit}}} \\
  \Dsyn[\Gamma]{\tPair{\trm}{\trm[2]}}
    &\defeq
    \begin{array}[t]{@{}l@{}}
      \letin{\tPair{\var}{\var'}}{\Dsyn[\Gamma]{\trm}}{
        \letin{\tPair{\var[2]}{\var[2]'}}{\Dsyn[\Gamma]{\trm[2]}}{}} \\
      \tPair{\tPair{\var}{\var[2]}}{\funabs{d}{\tDo{\var'\, (\tlFst{d}) ; \var[2]'\, (\tlSnd{d})}}}
    \end{array} \\
  \Dsyn[\Gamma]{\tFst{\trm}}
    & \defeq
    \letin{\tPair{\var}{\var'}}{\Dsyn[\Gamma]{\trm}}
          {\tPair{\tFst\var}{\funabs{d}{\var'\, \tlPair{d}{\tZero}}}}
          \\
  \ifTomACMVersion{}{
    \Dsyn[\Gamma]{\tSnd{\trm}}
    & \defeq
    \letin{\tPair{\var}{\var'}}{\Dsyn[\Gamma]{\trm}}
          {\tPair{\tSnd\var}{\funabs{d}{\var'\, \tlPair{\tZero}{d}}}}
          \\
    \Dsyn[\Gamma]{\tInl{\trm}}&\defeq \letin{\tPair{\var}{\var'}}{\Dsyn[\Gamma]{\trm}}{\tPair{\tInl{\var}}{\funabs{d}{\var'\,(\tlCast d)}}}\\
    \Dsyn[\Gamma]{\tInr{\trm}}&\defeq \letin{\tPair{\var}{\var'}}{\Dsyn[\Gamma]{\trm}}{\tPair{\tInr{\var}}{\funabs{d}{\var'\,(\trCast d)}}}\\
    \DsynCaseSumHang&\defeq
    \begin{array}[t]{@{}l@{}}
    \letin{\tPair{\var[3]}{\var[3]'}}{\Dsyn[\Gamma]{\trm}}{}\\
    \tSAMatch{\var[3]}{\var}{
      \begin{array}[t]{@{}l@{}}\letin{\tPair{\var_1}{\var_2}}{\Dsyn[\Gamma,\var:\ty]{\trm[2]}}{}\\
        \tPairOpen{\var_1}\tPairSep{\funabs{d}{
          \begin{array}[t]{@{}l@{}}
            \tDo{\tPair{\tUnit}{dz}}\leftarrow{\tScope{\Dsyn{\Gamma}_2}{\Dsyn{\ty}_2}\,(\var_2\ d)}{}\\
            \phantom{\tDo{}}\var[3]'\,(\tlInl dz)\tPairClose
          \end{array}
            }}
      \end{array}
      }{\var[2]}
    {
      \begin{array}[t]{@{}l@{}}\letin{\tPair{\var[2]_1}{\var[2]_2}}{\Dsyn[{\Gamma,\var[2]:\ty[2]}]{\trm[3]}}{}\\
        \tPairOpen{\var[2]_1}\tPairSep{\funabs{d}{
          \begin{array}[t]{@{}l@{}}
            \tDo{\tPair{\tUnit}{dz}}\leftarrow {\tScope{\Dsyn{\Gamma}_2}{\Dsyn{\ty[2]}_2}\,(\var[2]_2\ d)}{}\\
            \phantom{\tDo{}}\var[3]'\,(\tlInr dz)\tPairClose
          \end{array}
            }}
      \end{array}  
    }
    \end{array}
    \\
    \Dsyn[\Gamma]{r}&\defeq \tPair{r}{\funabs{d}{\tReturn\tUnit}}\\
    \Dsyn[\Gamma]{\tSign\trm}&\defeq \tPair{\tSign\trm}{\funabs{d}{\tReturn\tUnit}}\\
  }
  \Dsyn[\Gamma]{\tOpApp{\trm_1,\ldots,\trm_n}}&\defeq  \begin{array}[t]{@{}l@{}}
  \letin{\tPair{\var_1}{\var_1'}}{\Dsyn[\Gamma]{\trm_1}}{}
  % \\ \midletvdots \\
  \ldots\;
  \letin{\tPair{\var_n}{\var_n'}}{\Dsyn[\Gamma]{\trm_n}}{}\\
  \tPairOpen{\tOpApp{\var_1,\ldots,\var_n}}\tPairSep{\funabs{d}
  \begin{array}[t]{@{}l@{}}
  \letin{\tTuple{d_1,\ldots,d_n}}{\tDOpApp{\var_1,\ldots,\var_n}{d}}{}\\{\tDo{}\var_1'\ d_1;\ }{{\cdots;\ }{\var_n'\ d_n}}\tPairClose
  \end{array}}
  \end{array}
  \end{align*}
\caption{\label{fig:monadic-chad}
  \protect\ifTomACMVersion
    {Some representative rules for the}
    {The}
  CHAD definitions updated after \cref{sec:monadic-lifting}.
}
\end{figure}

In contrast to the changes in \cref{sec:sparsity}, we now have to change the code transformation to produce monadic code;
\ifTomACMVersion
  {some representative rules of the updated code transformation are}
  {the updated code transformation is}
shown in \cref{fig:monadic-chad}.
\ifTomACMVersion
  {(For the full rules, see the \cite[Fig.\ 6]{efficient-chad-arxiv}.)}
  {}
It is helpful to compare the new typing of the code transformation (at the top of \cref{fig:monadic-chad}) with the original typing in \cref{eq:chad-trans-type}, and to compare the new rules with \cref{fig:basic-chad}.
Note how the structure is the same.

Let us look back at the operations in \cref{fig:constant-time-methods} and make up the balance.
All zeros are now constant time and $\tPlus[\Env\Gamma]{}{}$ is gone.
On the other hand, $\tOne\var\ty\Gamma\ d$ adds (using $\tPlus[\ty]{}{}$) the cotangent value $d$ to the entry in the environment cotangent (in the monad state) corresponding to the variable $x$.
This takes time logarithmic in the size of the environment cotangent (thus usually $O(\log |\Gamma|)$, unless most variables are unused), plus the time required to invoke $\tPlus[\ty]{}{}$ on $d$ and the running total.
Furthermore, $\tSplit\Gamma\ty$ is also logarithmic in the size of the environment cotangent.

In \cref{sec:log-factors}, we will replace the logarithmic-time operations with constant-time ones by using a mutable array instead of a persistent tree map ($\EnvMap{}$); this will modify only the implementation of $\EnvM{}{}$ and its methods, keeping the code transformation itself completely as-is.\footnote{We can do this because $\EnvM{}{}$ is used as a black-box monad.}
Then, the only remaining potential problem is the cotangent addition ($\tPlus[\ty]{}{}$) in $\tOne{}{}{}$.
However, as we discuss below in \cref{sec:informal-amortisation}, we can amortise the cost of these additions against the \emph{creation} of the cotangent values being added, meaning that the code transformation as it is now in \cref{fig:monadic-chad} --- with the efficient implementation of $\EnvM{}{}$ from \cref{sec:log-factors} --- is actually \emph{already finished and asymptotically efficient}.

\subsection{Step 3: There Is No Step 3 (The Amortisation Argument, Informally)}\label{sec:informal-amortisation}

\paragraph{Affine Use of Cotangents}
To see how we are going to argue that the use of $\tPlus[\Dsyn{\ty}_2]{}{}$ in $\Dsyn[\Gamma]{\var:\ty}$ is already efficient after the monadic lifting of \cref{sec:monadic-lifting}, first observe that in \cref{fig:monadic-chad} (and already in \cref{fig:basic-chad}), cotangent values are used in an \emph{affine}\footnote{That is: linear, but dropping is allowed.} manner: they are mostly not duplicated, and when they are (in $\Dsyn[\Gamma]{\tPair{\trm}{\trm[2]}}$), the structure is split using $\tlFst{}$ and $\tlSnd{}$ before using the constituent parts affinely again.
(We could encode this affine usage with a resource-aware type system, but this does not really bring benefit for our presentation; the observation is simply useful to explain why the amortisation argument will go through, and in a way it is proved by the amortisation argument itself.)
Because of this affine usage, once a cotangent value has been added to another, only the sum will again be used elsewhere; the values used to build this sum will never be used again.

\paragraph{Addition of Sparse Structures}
The second ingredient that we need is an observation about the cost of $\tPlus[\ty]{}{}$.
The addition of two sparse cotangent values $\var$ and $\var[2]$ of type $\Dsyn{\ty}_2$, i.e.\ $\tPlus[\Dsyn{\ty}_2]{\var}{\var[2]}$, is essentially a zip of the two (sparse) structures.
In this zip, we assume that the two cotangents, as far as they are defined (i.e.\ not omitted due to sparsity), have equal structure: all values of type $\LPair{\ty}{\ty[2]}$ have the form $\tlPair{\var}{\var[2]}$ for further structures $\var$ and $\var[2]$, and although a value of type $\LSum{\ty}{\ty[2]}$ can be both $\tlInl{\var}$ and $\tlInr{\var}$, we raise a runtime error if the two do not correspond (see \cref{fig:naive-api} and the $\tError$ calls inside $\tlCast{}$ and $\trCast{}$ --- this will not occur in practice because CHAD has been proved correct~\cite{vakar-2022-chad}).

Hence, this zipping operation visits precisely the common ``prefix'', or rather the \emph{intersection},~of both structures --- no more, no less.
Subtrees that occur in only one of the two arguments to $\tPlus{}{}$~are simply returned as-is, which is constant-time.
An example is shown in \cref{fig:plus-sparse-cost}.
The~circled nodes are the nodes that the two inputs share, i.e.\ neither has omitted.
The implementation of $\tPlus{}{}$ does not have to recurse into the left subtree of $\var$ (blue), nor into some of the branches of $\var[2]$~(red).

\begin{figure}
\includegraphics[width=9cm]{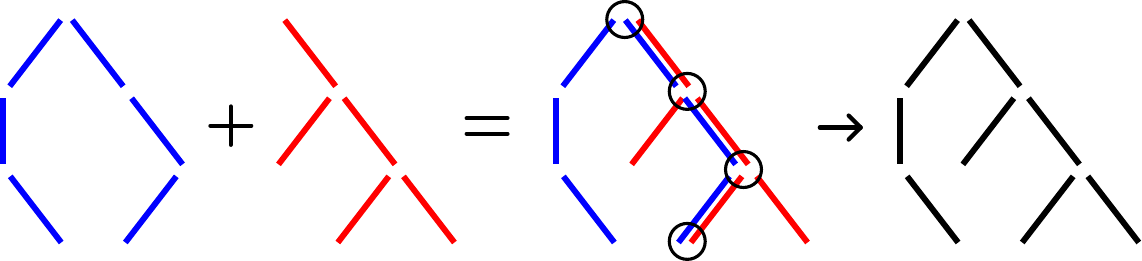}
\caption{\label{fig:plus-sparse-cost}
  Pictured are two sparse structures, $\var$ in blue and $\var[2]$ in red, together with their overlay.
  Leaves represent $\LR$, nodes with a vertical child indicate (one branch of) $\LSum{\ty}{\ty[2]}$, and nodes with diagonal children indicate $\LPair{\ty}{\ty[2]}$.
  Omitted lines are children omitted due to sparsity.
  Computing the sum of $\var$ and $\var[2]$ involves work on the overlap of the two structures; in this case, at 4 nodes.
  The structure of the sum is shown in black on the right.
}
\end{figure}

More formally, define the size of a cotangent value as follows:
\[
  \begin{array}{l@{\qquad}l}
    \multicolumn{2}{l}{\size[\ty]{} : \Dsyn{\ty}_2 \ra \Z_{>0}} \\
    \size[\Unit]{\tUnit} = 1 &
    \size[\Sum{\ty[2]}{\ty}]{(\tInl\tUnit)} = 1 \\
    \size[\R]{x} = 1 &
    \size[\Sum{\ty[2]}{\ty}]{(\tInr(\tInl\var))} = 1 + \size[{\ty[2]}]\var \\
    \size[\Pair{\ty[2]}{\ty}]{(\tInl\tUnit)} = 1 &
    \size[\Sum{\ty[2]}{\ty}]{(\tInr(\tInr{\var[2]}))} = 1 + \size[\ty]{\var[2]} \\
    \size[\Pair{\ty[2]}{\ty}]{(\tInr{\tPair{\var}{\var[2]}})} = 1 + \size[{\ty[2]}]\var + \size[\ty]{\var[2]}
  \end{array}
\]
Then, formalising the observation of \cref{fig:plus-sparse-cost}, we have the following inequality:
\[
  \forall \ty.~
  \forall a, b : \Dsyn{\ty}_2.~
    \cost{\tPlus{\var}{\var[2]}}{\enventry{\var}{a}{\Dsyn{\ty}_2}, \enventry{\var[2]}{b}{\Dsyn{\ty}_2}}
      \leq \cpoten \cdot (\size[\ty]{a} + \size[\ty]{b} - \size[\ty]{(\tPlus{a}{b})})
\]
% where $\cost{\trm}{\mathcal V}$ is the cost, in our (call-by-value) cost model (see \cref{sec:complexity-proof}), of evaluating the term $\trm$ in the environment valuation $\mathcal V$.
Here, $\cpoten$ is the number of computation steps (in our cost model) that $\tPlus{}{}$ at most requires to fully handle one node in a cotangent value.
Note that because the structure of $\tPlus{a}{b}$ is the union of the structures of $a$ and $b$, the expression $(\size[\ty]{a} + \size[\ty]{b} - \size[\ty]{(\tPlus{a}{b})})$ is the size of the intersection of $a$ and $b$.
Now we define our \emph{potential function} as $\poten[\ty]{d} \defeq \cpoten \cdot \size[\ty]{d}$, measuring the number of computation steps that we budget in a node and can still use when consuming the cotangent value later.
Then, rearranging terms, the inequality rewrites as follows:
\begin{equation}
  \forall \ty.~
  \forall a, b : \Dsyn{\ty}_2.~
    \cost{\tPlus{\var}{\var[2]}}{\enventry{\var}{a}{\Dsyn{\ty}_2}, \enventry{\var[2]}{b}{\Dsyn{\ty}_2}}
      - \poten[\ty]{a} - \poten[\ty]{b}
      + \poten[\ty]{(\tPlus{a}{b})}
      \leq 0
  \label{eq:plus-amortisation-law}
\end{equation}
in other words, after accounting for the potential flowing in (through $a$ and $b$ --- these are computation steps we already counted elsewhere) and flowing out (through $\tPlus{a}{b}$ --- the steps we are budgeting for later), addition is free.

\paragraph{Amortising Sums}
Armed with these two observations, first consider the simplified situation~where a large number of cotangents are successively added to a single, threaded-through accumulator:
\[
  \tPlus{(\tPlus{(\tPlus{(\tPlus{(\tPlus{d}{d_1})}d_2)}{d_3})}{d_4})}{\ldots}
\]
Each of the additions involved takes time at most proportional to the size of the $d_i$ added there --- ``size'' being the number of materialised nodes in memory.
Furthermore, since we do not duplicate structures, constructing $d_i$ itself will also have taken time \emph{at least} on the order of the size of $d_i$.
Thus, this chain of additions (at most) duplicates the work done in constructing the $d_i$, which we had to do anyway; so essentially these additions are free.
We have \emph{amortised} the additions against the construction of their inputs.

Of course, in a general derivative program, the additions will not necessarily be done in such a linear fashion, but a more precise analysis of the above situation will show how the amortisation argument works in general.
Indeed, because of our second observation (\cref{fig:plus-sparse-cost}), the cost of $\tPlus{(\ldots)}{d_i}$ is not really proportional to the whole size of $d_i$: only the \emph{intersection} of $(\ldots)$ and $d_i$ is traversed in $\tPlus{}{}$.
Furthermore, the parts of $d_i$ that are \emph{not} traversed are included as-is in the sum without processing.
We can still amortise against those non-traversed subtrees of the sum!

Indeed, assuming we could still amortise against the entirety of both arguments $d_1$ and $d_2$ of an addition $\tPlus{d_1}{d_2}$, their sum will consist of a number of subtrees that were included unchanged, connected by their intersection that $\tPlus{}{}$ did traverse once.
However, on precisely that intersection, we had \emph{two} input nodes we could amortise against: we can arbitrarily choose to amortise this addition against the overlapping part of $d_1$, and still have the corresponding part of $d_2$ to amortise against later.
So in the end, the entire result of the addition can still be amortised against, preserving the invariant that we can always still amortise against the entirety of a cotangent value.
Seeing that we never copy cotangents, there is no risk of amortising against the same cotangent twice.

% (In the rest of the paper we will assume without loss of generality that $\cpoten = 1$; by doing so, we effectively assume that $\tPlus{}{}$ can process one tree node completely in 1 computation step.
% This is without loss of generality because in the end we will use our cost model up to a constant factor only; we can give any operation cost 1 if we want, as long as that operation is constant-time.)

This affine usage is also the reason we can speak (informally) of potential being \emph{carried by} a cotangent value, and thus flowing into, or out of, a computation together with that value.

It turns out that if we modify the full complexity criterion to account for potential flowing in and out like in \cref{eq:plus-amortisation-law}, we get a statement that easily proves itself by induction; the only thing we still need to do then is implementing $\EnvM{}{}$ efficiently, removing the logarithmic factors in its complexity.
This optimisation is described in \cref{sec:log-factors}, after which \cref{sec:complexity-proof} sketches the full complexity proof.

\section{Getting Rid of Log Factors}\label{sec:log-factors}

The current implementation of $\EnvM{}{}$ is not quite efficient enough for our purposes: $\tOne\var{}\Gamma\ d$ not only adds $d$ (using $\tPlus{}{}$) to the value for $\var$ in the monad state, but also takes $O(\log |\Gamma|)$ time to find and update that value in the $\Map$, a purely functional tree map.
Similarly, $\tScope{}{}$ has logarithmic overhead for updating values in the $\Map$.
These logarithmic computations violate the complexity criterion (because neither variable references nor let-bindings in the source program account for those logarithmic costs), but fortunately we can do better by replacing the $\Map$ with a mutable~array.

Encoding mutability in a functional language can be done in multiple ways, and for the complexity proof it does not matter which we choose --- as long as it can be encapsulated in the $\EnvM{}{}$ API in such a way that its methods have the complexities listed in \cref{fig:envm-api}.\footnote{
  Indeed, our Agda formalisation of the complexity proof is actually generic over all the implementations in this section, because it assumes only the complexities in \cref{fig:envm-api} together with some equations on that API that define the semantics of the methods.
    The implementation is kept black-box; see \cref{sec:agda-formalisation}.
}

\begin{table}
\begin{center}
  \begin{tabular}{l|l}
    \textbf{Method} & \textbf{Cost} \\ \hline
    \makecell[l]{
      $\tOne\var\ty\Gamma : \ty \ra \EnvM\Gamma\Unit$ \\
      $\tOne\var\ty\Gamma\ (\var[2] : \ty) = \textit{(...)}$
    }
      & \makecell[l]{
          $O(1) + (\text{cost}$ of adding $\var[2]$ to \\
          the value for $\var$ in the array$)$
        } \\\hline
    \makecell[l]{
      $\tScope\Gamma\ty : \EnvM{(\Gamma, \var : \ty)}{\ty[2]} \ra \EnvM\Gamma{(\Pair{\ty[2]}{\ty})}$ \\
      $\tScope\Gamma\ty\ (m : \EnvM{(\Gamma, \var : \ty)}{\ty[2]}) = \textit{(...)}$
    }
      & $O(1) + (\text{cost of $m$})$ \\\hline
    \makecell[l]{
      $\tRun{}{} : \EnvM\Gamma\ty \ra \ET\Gamma \ra \Pair{\ty}{\ET\Gamma}$
      \\
      $\tRun{(m : \EnvM\Gamma\ty)}{(\mathit{env} : \ET\Gamma)} = \textit{(...)}$ \\
    }
      & $O(1) + \crungam \cdot |\Gamma| + (\text{cost of $m$})$
  \end{tabular}
\end{center}
\caption{\label{fig:envm-api}
  % \ifthenelse in a caption doesn't work, needs to be \protect'ed.
  The API of $\protect\EnvM\Gamma\ty$ with the complexities that we assume.
}
\end{table}

To understand the complexity of $\tRun{}{}$ that we require, note that its implementation has to allocate and deallocate an array, serialising the input and the output environments (of type $\ET\Gamma$) to and from that array; this is $O(|\Gamma|)$ work.
We assume here that we are allowed to return cotangents in our sparse format; if not, the $|\Gamma|$ term in its complexity would increase to $\sum_{\var:\ty\in\Gamma} \size\ty$ (where $\size\ty$ is proportional to the time required to (de)serialise a value of type $\ty$, and thus to convert back and forth to our sparse representation).
The reason why the cost of $\tRun{}{}$ is not written as the slightly weaker $O(|\Gamma|) + (\text{cost of $m$})$ is to allow some cancellation when computing with $\tRun{}{}$ in the proof: $O(|\Gamma|) - O(|\Gamma|)$ is not necessarily $0$, but $\crungam \cdot |\Gamma| - \crungam \cdot |\Gamma| = 0$.

\paragraph{Implementation}
Assuming an implementation in Haskell, and assuming that we want to support the parallelisation discussed in \cref{sec:parallelism} below, we need to resort to an implementation in terms of the \texttt{IO} monad.
For illustration, a possible definition in GHC Haskell is as follows:\footnote{For a full implementation of this monad in GHC Haskell, see \ifTomACMVersion{\cite[Appendix E]{efficient-chad-arxiv}}{\cref{app:evm-ghc-haskell}}.}
\begin{equation*}
  \EnvM\Gamma\ty \defeq \texttt{Int} \ra \texttt{IORef}\ (\texttt{IOVector}\ \texttt{Any}) \ra \texttt{IO}\ \ty
\end{equation*}
This is a reader monad, the reader context being the length of the environment in this subterm ($|\Gamma|$) together with a pointer (\texttt{IORef}) to a mutable vector (\texttt{IOVector}, from e.g.\ the \texttt{vector} package) of untyped values (\texttt{Any}); the values are untyped because we are implementing a mutable \emph{heterogeneous} vector.
Such a thing is not predefined in Haskell, so we must simulate it using a homogeneous vector of untyped values to and from which we \texttt{unsafeCoerce}.
We use a pointer to the vector because in the $\tScope{}{}$ method, if the vector is not large enough for the extended environment, we have to reallocate the array.\footnote{If a growing array resizes to twice its size each time the underlying buffer is exhausted, the total amount of reallocation~and copying work is linear in the final array length (as $\sum_{i=0}^{\lceil \log_2 n \rceil} 2^i = 2^{\lceil \log_2 n \rceil + 1} - 1 = O(n)$), and can thus be amortised away.}

In the non-parallel context, \texttt{ST}~\cite{fp-1994-st-monad} is sufficient as a replacement for \texttt{IO}.
Meanwhile, in an imperative language such as OCaml, the required (atomic, in the case of parallelism) mutability is already present everywhere, meaning that $\EnvM\Gamma\ty \defeq \texttt{Int} \ra \text{Array}\ \texttt{Any} \ra \ty$ suffices, with \texttt{Any} standing for the uninformative type, such as \texttt{Object} or \texttt{void*}.

Let us briefly look at some alternative implementations.
To get a more functional-style implementation (but unfortunately not parallel!), one can use resource-linear types (such as those implemented in Linear Haskell~\cite{fp-2018-linear-haskell}).
In this case, the monad looks as follows:
\begin{equation*}
  \EnvM\Gamma\ty \defeq \texttt{Int} \ra \EnvArr\Gamma \lra \Pair{\ur\tau}{\EnvArr\Gamma}
  % \label{eq:evm-def-linear}
\end{equation*}
where $\EnvArr\Gamma$ is a mutable array with methods analogous to those of $\EnvMap\Gamma$.
The changes are the usual ones for a resource-linear mutable array (giving back the input array in the result and changing some arrows to resource-linear ones); for an example, see the mutable arrays in the \texttt{linear-base} package.\footnote{\url{https://hackage.haskell.org/package/linear-base-0.4.0/docs/Data-Array-Mutable-Linear.html}}
$\ur\ty$ denotes an unrestricted type, e.g.\ \texttt{Ur}~$\ty$ in Linear Haskell.

If the reader is unfamiliar with resource-linear types, a simple intuitive stand-in is to use the original, purely functional State monad ($\EnvM\Gamma\ty \defeq \EnvMap\Gamma \ra \Pair{\tau}{\EnvMap\Gamma}$), but assume that $\EnvMap{}$ somehow has a magical implementation where $\mPopx{}$, $\mModify\var\ty\Gamma{f}{}$ and $\mGet\var\ty\Gamma{}$ from \cref{fig:env-map-interface} all run in $O(1)$ (apart from the cost of calling $f$ once in $\mModify{}{}{}{}{}$).

In the proof sketch in the following section, we will assume the complexities in \cref{fig:envm-api}, plus some properties about the semantics of these methods; these semantical properties are satisfied by all implementations discussed in this section, and will be left unstated here but are formulated precisely in the Agda formalisation in \texttt{spec/LACM.agda} (see \cref{sec:agda-formalisation}).

\section{Complexity Proof}\label{sec:complexity-proof}

Because the derivative program now lives in a monad, the (generalised) complexity criterion (\cref{eq:complexity-criterion-generalised}) has to be modified to include its handler, $\tRun{}{}$:
\begin{gather}
  \begin{array}{@{}l}
    \exists c, c'>0.\ 
    \forall (\var_1 \sqoftype \ty[2]_1, \sqldots, \var_n  \sqoftype \ty[2]_n \vdash \trm \sqoftype \ty[1]).\ 
    \forall \var_1 \sqoftype \ty[2]_1, \sqldots, \var_n \sqoftype \ty[2]_n.\ 
    \forall d \sqoftype \Dsyn{\ty[1]}_2.\ 
    \forall d_1 \sqoftype \Dsyn{\ty[2]_1}_2, \sqldots, d_n \sqoftype \Dsyn{\ty[2]_n}_2. \\
    \quad \cost{\tRun{(\tSnd{\Dsyn[\Gamma]{\trm}}\ d)}{\mathit{env}_0}}{\var_1=\var_1,\sqldots,\var_n=\var_n, d=d, \mathit{env}_0 = \tTuple{d_1, \sqldots, d_n}} \\
    \qquad \leq c' + c \cdot \cost{\trm}{\var_1=\var_1, \sqldots, \var_n=\var_n} + \crungam \cdot n
  \end{array}
  \label{eq:complexity-criterion-monad}
  \raisetag{17pt}
\end{gather}
% where the $\tZero$ is of type $\ET{\var_1 : \ty_1, \ldots, \var_n : \ty_n} = \Pair{\Pair{\ty_1}{\ldots}}{\ty_n}$.
% Here we also generalised the statement to work for terms that have more than a single variable in their environment.
Since $\tRun{}{}$ takes an initial environment cotangent (to be accumulated into) as an additional argument, we need to provide one; we generalise over which environment the monad is initialised with to enable a proof by induction.
The constant $\crungam$ is from \cref{fig:envm-api} (see below for a discussion).
% Note that we have added the $c'' \cdot n$ term because the (de)serialisation inside $\tRun{}{}$ and the construction of the tuple $\tTuple{\tZero[{\ty[2]_1}],\ldots,\tZero[{\ty[2]_n}]}$ introduce $O(n)$ work that was not there before.

% To make the statement amenable to a proof by induction, we slightly modify it. Specifically,~we:
% \begin{itemize}
% \item allow any initial environment derivative instead of only zeroes;
% % \item allow for the serialisation costs ($n$) of $\tRun{}{}$ and the cost ($c'$) of $\tSnd{}$, application, etc.\ at top-level in the scrutinised expression;
% \item account for amortisation using the potential function in the cost of the derivative expression;
% \item reduce $c''$ to the precise constant $\crungam$: the (de)serialisation inside $\tRun{}{}$ is now the only $O(n)$ work left.
% \end{itemize}
The next step is to account for amortisation, by subtracting incoming potential from the cost of the scrutinised expression (i.e.\ making incoming potential available as free computation steps), and adding outgoing potential to its cost (i.e.\ declaring outgoing potential as additional computation steps).
This yields the following criterion: (the same as \cref{eq:complexity-criterion-monad} except for the highlighted third line)
\begin{gather}
  \begin{array}{@{}l@{}}
    \exists c, c' > 0.\ 
    \forall (\var_1 \sqoftype \ty[2]_1, \sqldots, \var_n \sqoftype \ty[2]_n \vdash \trm \sqoftype \ty[1]).\ 
    \forall \var_1 \sqoftype \ty[2]_1, \sqldots, \var_n \sqoftype \ty[2]_n.\ 
    \forall d \sqoftype \Dsyn{\ty[1]}_2.\ 
    \forall d_1 \sqoftype \Dsyn{\ty[2]_1}_2, \sqldots, d_n \sqoftype \Dsyn{\ty[2]_n}_2. \\
    \quad \cost{\tRun{(\tSnd{\Dsyn[\Gamma]{\trm}}\ d)}{\mathit{env}_0}}{\var_1=\var_1,\sqldots,\var_n=\var_n, d=d, \mathit{env}_0 = \tTuple{d_1, \sqldots, d_n}}
    \\ \qquad\quad
    \textcolor{red}{{} - \poten{d} - \sum_{i=1}^n \poten{d_i} + \sum_{i=1}^n \poten{\mathit{res}_i}} \\
    \qquad \leq c' + c \cdot \cost{\trm}{\var_1=\var_1, \sqldots, \var_n=\var_n} + \crungam \cdot n
  \end{array}
  \label{eq:theorem-phi}
  \raisetag{26pt}
\end{gather}
where we have abbreviated $\tPair{\_}{\tTuple{\mathit{res}_1, \ldots, \mathit{res}_n}} = \tRun{(\tSnd{\Dsyn[\Gamma]{\trm}}\ d)}{\tTuple{d_1, \ldots, d_n}}$.
We prove this statement by induction on $\trm$; \cref{sec:proof-sketch} gives a proof sketch illustrating the main ideas.

From \cref{eq:theorem-phi} we derive a corollary that more directly states what a user can expect from our optimised version of CHAD.
We initialise $\mathit{env}_0$ with a tuple of zeros, because the gradient program computes $\text{gradient} + \mathit{env}_0$, and we want the gradient; furthermore, we bound $\poten{}$ using the fact that $\poten{\tZero[{\ty[2]_i}]} \le \poten{\mathit{res}_i}$ and recall that $\poten{d} = \cpoten \cdot \size{d}$.
This eliminates $\poten{}$ from the theorem statement:
% By specialising to a zero initial environment derivative again, and by bounding $\poten{}$,\footnote{In particular, we use that $\poten{\tZero[{\ty[2]_i}]} \le \poten{\mathit{res}_i}$ and that $\poten{d} \leq \size{\ty}$.} we get the following as a corollary from \cref{eq:theorem-phi}:
\begin{equation}
  \begin{array}{l}
    \exists c, c', c'' > 0.\ 
    \forall (\var_1 : \ty[2]_1, \ldots, \var_n : \ty[2]_n \vdash \trm : \ty[1]).\ 
    \forall \var_1 : \ty[2]_1, \ldots, \var_n : \ty[2]_n.\ 
    \forall d : \Dsyn{\ty[1]}_2. \\
    \quad \cost{\tRun{(\tSnd{\Dsyn[\Gamma]{\trm}}\ d)}{\tTuple{\tZero[{\ty[2]_1}],\ldots,\tZero[{\ty[2]_n}]}}}{\var_1=\var_1,\ldots,\var_n=\var_n, d=d} \\
    \qquad \leq
      c'
      + c \cdot \cost{\trm}{\var_1=\var_1, \ldots, \var_n=\var_n}
      + c'' \cdot n
      + \cpoten \cdot \size{d}
  \end{array}
  \label{eq:theorem-cor}
\end{equation}
Note that $c''$ captures both $\crungam$ and the $O(n)$ creation of the tuple~$\tTuple{\tZero[{\ty[2]_1}],\ldots,\tZero[{\ty[2]_n}]}$.

This is our final complexity theorem, and our Agda formalisation proves \cref{eq:theorem-cor}.\footnote{The constants used there are $c = 34$, $c' = 5$ and $c'' = 4$; it assumes for simplicity that $\cpoten = 1$. We stress that this constant 34 is not meaningful in practice because our cost model does not distinguish between various constant-time operations.}
Note that \cref{eq:theorem-cor} is the same as \cref{eq:complexity-criterion-generalised} apart the additional $c'' \cdot n + \size{d}$ term on the right-hand side (as well as inserting the call to $\tRun{}{}$).
However, this term is usually small, because even if a program has many inputs, these are typically organised in data structures rather than being passed as a large set of $n$ separate inputs; furthermore, for most applications, we feed reverse-mode differentiated code very sparse cotangents $d$ (e.g. basis vectors) for which `$\size{d}$' is small.
And it is actually unsurprising: realistically, any reverse AD algorithm will need some setup work per argument, if only allocating an array for them, and the incoming cotangent $d$ is typically going to need some processing.
Our mentioning it explicitly here is just to be faithful to our formalised proof.

\paragraph{Cost Model}
So far, we have left the cost model of our complexity proof implicit, but to write such a proof one of course has to have a fixed cost model.
\ifTomACMVersion{The extended version~\cite[Appendix C]{efficient-chad-arxiv}}{\cref{app:cost-model}} has a full definition, as does of course the Agda formalisation (see \cref{sec:agda-formalisation}), but it can be briefly summarised as follows:
\begin{itemize}
\item 
  The model assumes standard call-by-value semantics.
\item
  The model is very conservative: all computations that could cost time are given non-zero cost.
  For example, $\letin{\var}{\trm[2]}{\trm}$ is given cost $1 + (\text{cost of $\trm[2]$}) + (\text{cost of $\trm$})$: we consider variable binding to be a potentially costly operation.
  Furthermore, the expression $\funabs{\var}{\trm}$ costs $1$ plus the number of free variables of $t$: we count the allocation of the closure as potentially costly.
\item
  Simultaneously, we do not care about the relative cost of various constant-time operations.
  Scalar multiplication has cost 1, as does allocating a fixed amount of memory.
  These operations are not at all comparable in their practical runtime, but we consider them both constant-time.
  This is because the proof is only about asymptotic complexity, not about absolute runtime.
\end{itemize}

\subsection{Proof Sketch: Induction on $\trm$}\label{sec:proof-sketch}

The statement being proved here is \cref{eq:theorem-phi}.
We consider three representative cases of terms $\trm$ in the induction: (1) the very simplest case of $\trm=\tUnit$ (where we do not require an induction hypothesis), to illustrate some of the basic book-keeping about potential flowing in and out of computations; (2) a slightly more complex case of $\trm=\tPair{\trm_1}{\trm_2}$ to show how we invoke the induction hypothesis; and (3) the case of variable references $\trm=\var$, which is the only case in the whole proof where real work happens as this is where the potential is actually used, so we cannot merely cancel out the incoming and outgoing potentials.

The proof sketch uses ``$O(1)$'' as notation to indicate some unspecified bounded value whose bounds are independent of any of the other variables in the proof.
We use this to abbreviate the cost of some constant-time work whose exact cost is immaterial to the argument it appears in.

\paragraph{Simple Case}
Let us first consider the simplest case: $\trm = \tUnit$ in the environment $\var_1 : \ty_1, \ldots, \var_n : \ty_n$; its transformation rule can be found in \cref{fig:monadic-chad}.
The expression $\tRun{(\tSnd{\tPair{\tUnit}{\funabs{d}{\tReturn{\tUnit}}}}\ d)}{\mathit{env}_0}$, which \cref{eq:theorem-phi} scrutinises for this $\trm$, evaluates in $O(1) + \crungam \cdot n$ steps ($\crungam \cdot n$ are necessary to serialise and deserialise $\mathit{env}_0$) to $\tPair{\tUnit}{\mathit{env}_0}$.
Further, $\mathit{res}_i = (\mathit{env}_0)_i = d_i$ and thus $-\sum_{i=1}^n \poten{d_i} + \sum_{i=1}^n \poten{\mathit{res}_i} = 0$.
Noting that $\poten{d} = 1$ for all $d:\LUnit$, \cref{eq:theorem-phi} simplifies to the following clearly true inequality:
\begin{equation*}
  \begin{array}{@{\hspace{-1em}}l@{\hspace{-1.5em}}}
    \exists c, c' > 0.\ O(1) + \crungam \cdot n- 1 \leq c' + c + \crungam \cdot n.
  \end{array}
\end{equation*}
Note what happened with the potential: the first source of incoming potential (in $d : \Dsyn{\Unit}_2$) was ignored, resulting in an extra $-1$ term on the left-hand side of the inequality.
This only made proving the theorem ``easier'': we got some unused free computation steps.
The second source of incoming potential (in $d_1, \ldots, d_n$) cancelled exactly against the outgoing potential (in $\mathit{res}_1, \ldots, \mathit{res}_n$) because we did not change the accumulated environment cotangent.

\paragraph{Subterms and Sparsity}
Now let us consider the term $\trm = \tPair{\trm_1}{\trm_2}$.
When we evaluate the scrutinised expression, $\tRun{(\tSnd{\Dsyn[\Gamma]{\trm}}\ d)}{\mathit{env}_0}$, in addition to some constant-cost work we are going to do the following things (see \cref{fig:monadic-chad}):
(1) Evaluate $\Dsyn[\Gamma]{\trm_1}$;
(2) Evaluate $\Dsyn[\Gamma]{\trm_2}$;
(3) Call $\tSnd{\Dsyn[\Gamma]{\trm_1}}$ on the argument $\tlFst{d}$;
(4) Call $\tSnd{\Dsyn[\Gamma]{\trm_2}}$ on the argument $\tlSnd{d}$;
(5) Sequence the results of (3) and (4) in the monad and $\tRun{}{}$ the result.

In short, this is equivalent (apart from some constant-cost work) to evaluating the expression:
\[
  \tRun{(\tSnd{\Dsyn[\Gamma]{\trm_1}}\ (\tlFst{d}) \tSeq \tSnd{\Dsyn[\Gamma]{\trm_2}}\ (\tlSnd{d}))}{\mathit{env}_0}
\]
Because of the implementation as a state monad, we have, after defining $\tPair{v}{\mathit{env}'} \defeq \tRun{a}{\mathit{env}}$:
\begin{equation}
  \begin{array}{l}
    \cost{\tRun{(a \tBind \funabs{\var}{b})}{e}}{e=\mathit{env}, \Gamma} \\
    \quad = O(1)
      + \cost{\tRun{a}{e}}{e=\mathit{env}, \Gamma}
      + \cost{\tRun{b}{e}}{e=\mathit{env}', \var=v, \Gamma}
      - \crungam \cdot |\Gamma|
  \end{array}
  \label{eq:bind-cost-lemma}
\end{equation}
where the term $-\crungam \cdot |\Gamma|$ arises because in the left-hand side, we (de)serialise $\mathit{env}$ only once whereas in the right-hand side we do so twice: we have to subtract one of the two to make the left and right-hand sides equal.
Now define for convenience in the below:
\[
  \tPair{\_}{\mathit{env}'} \defeq \tRun{(\tSnd{\Dsyn[\Gamma]{\trm_1}}\ (\tlFst{d}))}{\tTuple{d_1, \ldots, d_n}}
  \hspace{0.7cm}
  \tPair{\_}{\mathit{env}''} \defeq \tRun{(\tSnd{\Dsyn[\Gamma]{\trm_2}}\ (\tlSnd{d}))}{\mathit{env}'}
\]
Using the above lemma about the cost of bind (\cref{eq:bind-cost-lemma}) in simplifying \cref{eq:theorem-phi}, we get the following:
\begin{equation}
  \begin{array}{l}
    \exists c, c' > 0.\ 
    \forall (\var_1 : \ty[2]_1, \ldots, \var_n : \ty[2]_n \vdash \trm : \ty[1]).\ 
    \forall \var_1 : \ty[2]_1, \ldots, \var_n : \ty[2]_n.\ 
    \forall d : \Dsyn{\ty[1]}_2. \\
    \forall d_1 : \Dsyn{\ty[2]_1}_2, \ldots, d_n : \Dsyn{\ty[2]_n}_2. \\
    \quad O(1)
            + \cost{\tRun{(\Dsyn[\Gamma]{\trm_1}\ d)}{\mathit{env}_0}}{\var_1=\var_1,\ldots,\var_n=\var_n, d=\tlFst{d}, \mathit{env}_0 = \tTuple{d_1, \ldots, d_n}} \\
    \qquad\quad {} + \cost{\tRun{(\Dsyn[\Gamma]{\trm_2}\ d)}{\mathit{env}}}{\var_1=\var_1,\ldots,\var_n=\var_n, d=\tlSnd{d}, \mathit{env} = \mathit{env}'} \\
    \qquad\quad {} - \crungam \cdot n - \poten{d} - \sum_{i=1}^n \poten{d_i} + \sum_{i=1}^n \poten{\mathit{res}_i} \\
    \qquad \leq c \cdot \cost{\tPair{\trm_1}{\trm_2}}{\var_1=\var_1, \ldots, \var_n=\var_n} + c' + \crungam \cdot n
  \end{array}
  \label{eq:proof-pair-simpl}
\end{equation}
Now the two big `$\cost{}{}$' calls match the ones in the induction hypotheses for $\trm_1$ and $\trm_2$ (\cref{eq:theorem-phi}); adding the two induction hypotheses together we get the following proposition:
\begin{equation}
  \begin{array}{l}
    \cost{\tRun{(\tSnd{\Dsyn[\Gamma]{\trm_1}}\ d)}{\mathit{env}_0}}{\var_1=\var_1,\ldots,\var_n=\var_n, d=\tlFst{d}, \mathit{env}_0 = \tTuple{d_1, \ldots, d_n}} \\
    \qquad {} - \poten{(\tlFst{d})} - \sum_{i=1}^n \poten{d_i} + \sum_{i=1}^n \poten{\mathit{env}'_i} \\
    \qquad {} + \cost{\tRun{(\tSnd{\Dsyn[\Gamma]{\trm_2}}\ d)}{\mathit{env}}}{\var_1=\var_1,\ldots,\var_n=\var_n, d=\tlSnd{d}, \mathit{env} = \mathit{env}'} \\
    \qquad {} - \poten{(\tlSnd{d})} - \sum_{i=1}^n \poten{\mathit{env}'_i} + \sum_{i=1}^n \poten{\mathit{env}''_i} \\
    \quad \leq c \cdot \cost{\trm_1}{\var_1=\var_1, \ldots, \var_n=\var_n} + c \cdot \cost{\trm_2}{\var_1=\var_1, \ldots, \var_n=\var_n} + 2c' + 2\crungam \cdot n
  \end{array}
  \label{eq:proof-pair-prop}
\end{equation}
Subtract \cref{eq:proof-pair-prop} from \cref{eq:proof-pair-simpl}:
\begin{equation*}
  \begin{array}{l}
    O(1) - \crungam \cdot n - \poten{d} + \poten{(\tlFst{d})} + \poten{(\tlSnd{d})} - \sum_{i=1}^n \poten{\mathit{env}''_i} + \sum_{i=1}^n \poten{\mathit{res}_i}
    \leq c \cdot 1 - c' - \crungam \cdot n
  \end{array}
\end{equation*}
where we simplified the right-hand side using our cost model, which gives that $\cost{\tPair{\trm_1}{\trm_2}}{\Gamma} = 1 + \cost{\trm_1}{\Gamma} + \cost{\trm_2}{\Gamma}$.
We can further simplify by observing that $\mathit{env}'' = \mathit{res}$ and by cancelling the two occurrences of $\crungam \cdot n$.

Then, to handle the $\poten{d}$ terms, we need to consider sparsity: we need to analyse the cases where $d = \tInl{\tUnit}$ and where $d = \tInr{\tPair{d'_1}{d'_2}}$.
In the first case, $\tlFst{d} = \tZero$ and $\tlSnd{d} = \tZero$, hence $\poten{d} = \poten{(\tlFst{d})} = \poten{(\tlSnd{d})} = 1$, thus $-\poten{d} + \poten{(\tlFst{d})} + \poten{(\tlSnd{d})} = 1 = O(1)$.
In the second case, $\poten{d} = 1 + \poten{(\tlFst{d})} + \poten{(\tlFst{d})}$, so the same expression evaluates to $-1$, which is also $O(1)$.
Hence we can merge the three $\poten{}$-terms into the $O(1)$ that is already there, yielding:
\begin{equation*}
  \begin{array}{l}
    O(1) \leq c - c'
  \end{array}
\end{equation*}
which is clearly true for sufficiently large $c$.

Again, note what happened to the potential.
The potential in the incoming cotangent, $d$, was mostly immaterial: it contributed $+1$ or $-1$ depending on how sparsely it was represented, but did not do anything significant.
This is expected, since in $\Dsyn[\Gamma]{\tPair{\trm_1}{\trm_2}}$ we do not build or consume any non-trivial fragments of cotangent values.
As for the potential in the environment cotangent accumulator: the outgoing potential of $\tSnd{\Dsyn[\Gamma]{\trm_1}}$, equal to $\sum_{i=1}^n \poten{\mathit{env}'_i}$, cancels precisely against the incoming potential (via the environment cotangent) of $\tSnd{\Dsyn[\Gamma]{\trm_2}}$, which is again expected because the environment cotangent itself is passed as-is from $\trm_1$ to $\trm_2$.

In general, this is what always happens in the proof: as long as we do not do anything material to cotangents, they at most consume a bounded number of evaluation steps in stored potential, and as long as we do not modify the environment cotangent ourselves, all the corresponding $\poten{}$ terms cancel.
The only case where we do something material to all of these, and where the $\poten{}$ terms do not immediately cancel, is for variable references.

\paragraph{Variable References: Amortisation}
Taking $\trm = \var_i : \ty[2]_i$ in the environment $\var_1 : \ty[2]_1, \ldots, \var_n : \ty[2]_n$ and inlining into \cref{eq:theorem-phi}, we get:
\begin{equation}
  \begin{array}{l}
    \exists c, c' > 0.\ 
    \forall v : \ty[2]_i.\ 
    \forall d : \Dsyn{\ty[2]_i}_2.\ 
    \forall d_1 : \Dsyn{\ty[2]_1}_2, \ldots, d_n : \Dsyn{\ty[2]_n}_2. \\
    \quad O(1) +
          \cost{\tRun{(\tOne{\var_i}{\Dsyn{\ty[2]_i}_2}{\Dsyn{\Gamma}_2}{d})}{\mathit{env}_0}}{d=d, \mathit{env}_0 = \tTuple{d_1, \ldots, d_n}} \\
    \qquad\quad {} - \poten{d} - \sum_{j=1}^n \poten{d_j} + \sum_{j=1}^n \poten{\mathit{res}_j} \\
    \qquad \leq c \cdot \cost{\var_i}{\var_i=v} + c' + \crungam \cdot n
  \end{array}
  \label{eq:prf-var-1}
\end{equation}
Here we use $v$ for the value of the variable $\var_i$ in the current evaluation environment.
In our cost model, $\cost{\var_i}{\Gamma} = 1$ for a variable $x_i$, and furthermore we know that that for all $j \not= i$, we have $\mathit{res}_j = d_j$.
For $\mathit{res}_i$, we know from the semantics of $\tOne{}{}{}$ that $\mathit{res}_i = d + d_i$.
The complexity property of $\tOne{}{}{}$ specialises to the following:
\begin{equation*}
  \begin{array}{l}
    \cost{\tRun{(\tOne{\var_i}{\Dsyn{\ty[2]_i}_2}{\Dsyn{\Gamma}_2}\ d)}{\mathit{env}_0}}{d = d, \mathit{env}_0 = \tTuple{d_1, \ldots, d_n}} \\
    \quad = O(1) + \cost{d + d_i}{d = d, d_i = d_i}
  \end{array}
\end{equation*}
Thus \cref{eq:prf-var-1} simplifies to:
\begin{equation}
  \begin{array}{l}
    \exists c, c' > 0.\ 
    \forall d : \Dsyn{\ty[2]_i}_2.\ 
    \forall d_i : \Dsyn{\ty[2]_i}_2. \\
    \quad O(1) + \cost{d + d_i}{d = d, d_i = d_i}
      - \poten{d} - \poten{d_i} + \poten{(d + d_i)}
      \leq c + c' + \crungam \cdot n
  \end{array}
  \label{eq:prf-var-2}
\end{equation}
Now we use the central amortisation property of $(+)$ (\cref{eq:plus-amortisation-law} in \cref{sec:informal-amortisation}), which implies that:
\begin{equation}
  \cost{d + d_i}{d = d, d_i = d_i} - \poten{d} - \poten{d_i} + \poten{(d + d_i)} \leq 0
  \label{eq:prf-var-plus}
\end{equation}
Subtracting \cref{eq:prf-var-plus} from \cref{eq:prf-var-2} gives us that it is enough to show that $O(1) \leq c + c' + \crungam \cdot n$, which is immediate for sufficiently large $c$.

Unlike before, we have actually used potential here: we received potential for $d$ and $d_i$ and needed to return potential for their sum.
Because the sum will contain less potential than the inputs to $(+)$, we can use the excess potential to pay for the execution of $(+)$ itself, without needing to count more than a bounded number of evaluation steps here for the transform of a variable reference.

\paragraph{The Other Cases}
The other cases in the proof are mostly analogous to the cases discussed above.
For $\letin{\var}{\trm_1}{\trm_2}$, we end up needing the lemma that the CHAD primal of a term is equal to the result of the original term (i.e.\ that $\tFst{\Dsyn[\Gamma]{\trm}}$ returns the same result as $\trm$ when run in the same environment); this is required because we need to relate the cost of evaluating $\trm_2$ to that of evaluating $\tRun{(\tSnd{\Dsyn[\Gamma]{\trm_2}}\ d)}{\mathit{env}_0}$, and these two evaluations happen in the same environment only if $\tFst{\Dsyn[\Gamma]{\trm_1}}$ and $\trm_1$ return the same result.

\subsection{Agda Formalisation}\label{sec:agda-formalisation}
% The phantom is to put _something_ here so that microtype doesn't shift the options into the margin.
We have formalised the above complexity proof in Agda ($\geq$2.6.3, \texttt{\vphantom{a}-{}-safe \vphantom{a}-{}-without-K}).
Agda~\cite{agda-2007-norell} is a dependently-typed functional language and proof assistant, and one of the standard proof assistants in the domain~of programming languages.
While not typically used for proofs with integer reasoning, it admits~a very natural encoding of the problem statement.
Our full development can be found
\ifACMAnonymous
  {in the supplementary material to this anonymous submission.}
  {online\footnote{\agdagithublink} and archived at~\cite{efficient-chad-artifact}%
   \ifTomACMVersion
     {.}
     {; the \emph{statements} of the theorems, and the definitions required to write those statements, are included in \cref{app:agda-spec}.}}

In the development, the source and target language are encoded as a fully well-typed well-scoped (De Bruijn) inductive data type in the standard fashion; the cost model is encoded in the evaluator (\texttt{eval}), which evaluates an expression of type $\ty$ to a (meta-language, i.e.\ Agda) value of type $\sem{\ty} \times \Z$.
The integer contains the number of `steps' taken in evaluating the expression: our cost model.

In a way, the Agda proof is somewhat more generic than the sketch above, because it defines the methods of $\EnvM{}{}$\footnote{The concrete implementation there called \texttt{LACM} for ``Linear ACcumulation Monad''.} together with properties about their semantics and complexity in an \texttt{abstract} block.
This means that the rest of the proof cannot use the \emph{implementation} of the methods and the monad type itself, but only their \emph{types} (and the properties of those methods that we provide in the block).
Because all three of the monad implementations that we outlined in \cref{sec:log-factors} satisfy those properties, we know that our Agda proof works for all three, regardless of exactly which concrete monad implementation we choose for the Agda formalisation.\footnote{The actual Agda implementation is a state monad with cons-list state, but with costs as if it was a constant-time version.}

\section{Arrays}\label{sec:efficient-chad-on-array-types}
Adding arrays to our language, which is so far simply first-order, is not difficult, if somewhat tedious.
We will show how the elements of a complexity proof for array operations are analogous to the cases already discussed in \cref{sec:complexity-proof}, but we leave a full proof to future work.

An array is a product type, hence we should take inspiration from the handling of binary products ($\Pair{\ty}{\ty[2]}$), where we introduced sparsity in order to efficiently represent zeros $\tZero[\Dsyn{\Pair{\ty}{\ty[2]}}_2]$ and one-hot cotangents $\tPair{\var}{\tZero[{\ty[2]}]}$ and $\tPair{\tZero[\ty]}{\var}$.
Thus our choice of $\Dsyn{\Array\ty}_2$ should allow efficient zeros and one-hots as well.

A one-hot array cotangent is a pair of an index (of type $\Z$)\footnote{With arrays we also need $\Z$ in our type system; being a discrete type, $\Dsyn{\Z}_1\defeq \Z$ and $\Dsyn{\Z}_2 \defeq \LUnit$ suffices.} and a cotangent for that cell (of type $\Dsyn{\ty}_2$), hence a sufficient choice seems to be to let $\Dsyn{\Array\ty}_2$ be some collection $\Bag{(\Pair{\Z}{\Dsyn{\ty}_2})}$ of pairs that we will convert to an array of pairs using $\taCollect{} : \Bag\ty \ra \Array\ty$ once we are done constructing it. For efficient differentiation of discarding, indexing and sharing, this bag should furthermore support constant-time creation of empty and singleton collections, as well as constant-time combination of two collections.
It turns out to be sufficient to simply defunctionalise the operations that we want to be efficient and use the following type definition (i.e.\ make them constant-time by construction):\footnote{The fact that $\Bag{}$ is a free monoid is unsurprising (because among the operations we defunctionalised are zero and plus), but also not very fundamental: in \cref{sec:constant-factors} we will add more constructors to $\Bag{}$ to avoid pessimising other operations.}
% data Bag a = BagEmpty | BagSingle a | BagPlus (Bag a) (Bag a)
\[
  \textbf{data}\ \Bag{\ty} = \tBagEmpty \mid \tBagOne{\ty} \mid \tBagPlus{(\Bag\ty)}{(\Bag\ty)}
\]
Observe that `$\taCollect{}$'ing such a $\Bag\ty$ costs at most as much time as was spent constructing it.
Hence, if we use the $\Bag{\ty}$ affinely, which we do, the cost of `$\taCollect{}$'ing it can be amortised against its creation. 
So in terms of asymptotic complexity, we cannot do better than this, absent parallelism.

% D₁[Ar τ] = Ar D₁[τ]
% D₂[Ar τ] = Bag (Int, D₂[τ])
Thus define $\Dsyn{\Array\ty}_1\defeq \Array{\Dsyn{\ty}_1}$ and $\Dsyn{\Array\ty}_2 \defeq \Bag{(\Pair{\Z}{\Dsyn{\ty}_2})}$.
%We still have have $\Dsyn{\ty}_1 = \ty$, for all types.

\paragraph{Operations}
We discuss three array operations here: elementwise construction, indexing and associative reduction.
Their typing rules are as follows:
\[
% GENERATE:
%
% Γ |- s:Int   Γ,i:Int |- t : τ
% -----------------------------
%  Γ |- build s (i. t) : Ar τ
  \frac{\Gamma \vdash \trm[2] : \Z \quad \Gamma, i : \Z \vdash \trm : \ty}
       {\Gamma \vdash \taBuild{\trm[2]}{i}{\trm} : \Array\ty}
  \qquad
% INDEXING:
%
% Γ |- s : Ar τ   Γ |- t : Int
% ----------------------------
%       Γ |- s ! t : τ
  \frac{\Gamma \vdash \trm : \Array\ty \quad \Gamma \vdash \trm[2] : \Z}
       {\Gamma \vdash \taIndex{\trm}{\trm[2]} : \ty}
  \qquad
% FOLD:
%
% Γ, x:(τ,τ) |- t : τ    Γ |- s : Ar τ
% ------------------------------------
%      Γ |- fold (x. t) s : τ
  \frac{\Gamma, \var : \Pair{\ty}{\ty} \vdash \trm : \ty \quad \Gamma \vdash \trm[2] : \Array\ty}
       {\Gamma \vdash \taFold{\var}{\trm}{\trm[2]} : \ty}
\]
The informal semantics are as follows: $\taBuild{n}{i}{\trm} = [\subst{\trm}{\sfor{i}{0}}, \ldots, \subst{\trm}{\sfor{i}{n-1}}]$, $\taIndex{[\var_1, \ldots, \var_n]}{i} = \var_i$, and $\taFold{\var}{\tFst{\var}\star\tSnd{\var}}{[\var_1, \ldots, \var_n]} = \var_1 \star \cdots \star \var_n$.
`$\taFold{}{}{}$' requires its array argument to be of non-zero length, and performs a reduction in unspecified order, assuming that $\trm$ is associative.\footnote{This is a typical operation in parallel array languages, such as \texttt{fold} in Accelerate and \texttt{reduce} in Futhark. The requirement that the array be non-empty is typically lifted by adding an additional initial value for the reduction, but that would make this section more verbose without adding interesting new problems.}

Parallel array languages typically have other operations as well, including special cases of `$\taBuild{}{}{}$' such as gather, transpose, stencils/convolutions and more (which can be implemented more efficiently than the general case), but also independent operations such as various scans as well as scatter (forward array permutation).
In practice one will need a specialised derivative for each of these, the former for efficiency and the latter for expressivity, but here we restrict ourselves to the given three, which are together already powerful enough to express most machine learning models.

\paragraph{Derivatives}
\begin{figure}
\begin{equation*}
  \begin{array}{l}
% D{Γ}[build s (i. t)] =
%   let (n, _) = D{Γ}[s]
%   let a = build n (i. D{Γ,i:Int}[t])
%   let (a1, a2) = unzip a
%   in (a1
%      ,λd. let pairs = collect d
%           let d2 = scatter (build n (i. 0{D₂τ})) pairs
%           let d3 = zipWith (f d'. scope{Γ,Int} (f d') >> return ()) a2 d2
%           in sequence d3)
  \Dsyn[\Gamma]{\taBuild{\trm[2]}{i}{\trm}}=
  \begin{array}[t]{@{}l@{}}
    \letin{\tPair{n}{\_}}{\Dsyn[\Gamma]{\trm[2]}}{}\\
    \letin{a}{\taBuild{n}{i}{\Dsyn[\Gamma,i:\Z]{\trm}}}{}\\
    \letin{\tPair{a_1}{a_2}}{\taUnzip{a}}{}\\
    \tPairOpen{a_1}\tPairSep{}\funabs{d}{}
    \begin{array}[t]{@{}l@{}}
      \letin{\mathit{pairs}}{\taCollect{d}}{}\\
      \letin{d_2}{\taScatter{(\taBuild{n}{i}{\tZero[\Dsyn{\ty}_2]})}{\mathit{pairs}}}{}\\
      \letin{d_3}{\taZipWith{f}{d'}{\tScope{\Dsyn{\Gamma}_2}{\LUnit}\ (f\ d')\tSeq \tReturn{\tUnit}}{a_2}{d_2}}{}\\
      \taSequence{d_3} \tSeq \tReturn{\tUnit}
      \tPairClose
    \end{array}
  \end{array}\\
% D{Γ}[s ! t : τ] =
%   let (s1, s2) = D{Γ}[s]
%   let (i, _) = D{Γ}[t]
%   in (s1 ! i, λd. s2 (BagSingle (i, d)))
  \Dsyn[\Gamma]{\taIndex{\trm}{\trm[2]}} =
  \begin{array}[t]{@{}l@{}}
    \letin{\tPair{\var_1}{\var_2}}{\Dsyn[\Gamma]{\trm}}{} \\
    \letin{\tPair{i}{\_}}{\Dsyn[\Gamma]{\trm[2]}}{} \\
    \tPair{\taIndex{\var_1}{i}}{\funabs{d}{\var_2\ (\tBagOne{\tPair{i}{d}})}}
  \end{array}\\
% 
% D{Γ}[fold (p. t) s] =
%   let (s1, s2) = D{Γ}[s]
%   let tree = fold (x'. let x = (getA (fst x'), getA (snd x'))
%                        let (y, f) = D{Γ,x:(τ,τ)}[t]
%                        in Node (fst x') y f (snd x'))
%                   (map (x. Leaf x) s1)
%   in (getA tree
%      ,λd. do lf <- unTree (λd'. λf. do ((), (d1, d2)) <- scope (f d')
%                                        return (d1, d2))
%                           (λd'. d')
%                           d
%                           tree
%              s2 (fromList (lf [])))
%
%% cf with old derivative of foldr in terms of scanr/scanl
\Dsyn[\Gamma]{\taFold{p}{\trm}{\trm[2]}} =
\begin{array}[t]{@{}l@{}}
  \letin{\tPair{\trm[2]_1}{\trm[2]_2}}{\Dsyn[\Gamma]{\trm[2]}}{}\\
  \mathbf{let}\ \mathit{tree} =
    \taFold{}{}{}
    \ \begin{array}[t]{@{}l@{}}
        (p'.\ 
           \begin{array}[t]{@{}l@{}}
             \letin{p}{\tPair{\tGetA {(\tFst{p'})}}{\tGetA {(\tSnd{p'})}}}\\
             \letin{\tPair{\var[2]}{f}}{\Dsyn[\Gamma,p:\Pair{\ty}{\ty}]{\trm}}{}\\
             \tNode {(\tFst p')} {\var[2]} {f} {(\tSnd p')})
           \end{array} \\
        (\taMap {\var} {\tLeaf x} {\trm[2]_1})\ \mathbf{in}
      \end{array} \\
  \begin{array}[t]{@{}l@{}}
     \tPairOpen\tGetA{\mathit{tree}}\tPairSep\\
     \funabs{d}{\tDo{
       \mathit{lf} \leftarrow
         \tUnTree{}{}{}{}\ 
           \begin{array}[t]{@{}l@{}}
             (\funabs{d'}{\funabs{f}{\tDo{
               \tPair{\tUnit}{\tPair{d_1}{d_2}} \leftarrow \tScope{\Dsyn{\Gamma}_2}{\Dsyn{\Pair{\ty}{\ty}}_2}\ (f\ d') \\
               \tReturn{\tPair{d_1}{d_2}})
             }}} \\
             d\ \mathit{tree}
           \end{array} \\
       \trm[2]_2\ (\taFromList{(\mathit{lf}\ [])}) \tPairClose
     }}
  \end{array}
\end{array}
\end{array}
\end{equation*}
\caption{\label{fig:array-op-derivatives}
  The derivative of the build, array indexing and fold operators.
}
\end{figure}

\begin{figure}
\begin{equation*}
% data Tree a f = Node (Tree a f) a f (Tree a f) | Leaf a
% 
% getA :: Tree a f -> a
% getA (Node _ x _ _) = x
% getA (Leaf x) = x
% 
% unTree :: Monad m => (d -> f -> m (d, d)) -> (d -> r) -> d -> Tree a f -> m ([r] -> [r])
% unTree f1 f2 d (Node t1 _ f t2) = do
%   (d1, d2) <- f1 d f
%   rs1 <- unTree f1 f2 d1 t1
%   rs2 <- unTree f1 f2 d2 t2
%   return (rs1 ∘ rs2)
% unTree f1 f2 d (Leaf _) = λl. f2 d : l
\begin{array}{@{}l@{}}
  \textbf{data}\ \Tree{a}{f}=\tNode{(\Tree{a}{f})}{a}{f}{(\Tree{a}{f})}\mid \tLeaf{a}\\[0.2em]
  \tGetA{} : \Function{\Tree{a}{f}}{a}\\
  \tGetA{(\tNode \_ x \_ \_)} = x \\
  \tGetA{(\tLeaf x)} = x \\[0.2em]
  \tUnTree{}{}{}{} : \text{Monad}\ m \Rightarrow \Function{(\Function{d}{\Function{f}{m\ (\Pair{d}{d})}})}{\Function{d}{\Function{\Tree{a}{f}}{m\ (\Function{\List{d}}{\List{d}})}}}\\
  \tUnTree{g}{d}{(\tNode{t_1}{\_}{f}{t_2})}=\tDo{
    \tPair{d_1}{d_2}\leftarrow g\ d\ f\\
    rs_1\leftarrow \tUnTree{g}{d_1}{t_1}\\
    rs_2\leftarrow \tUnTree{g}{d_2}{t_2}\\
    \tReturn{(\funabs{l}{rs_1\ (rs_2\ l))}}
  }\\
  \tUnTree{g}{d}{(\tLeaf \_)}=\tReturn{(\funabs{l}{d :: l})} \qquad \textcolor{separate}{\textit{--- $(::)$ is list cons}}
\end{array}
\end{equation*}
\caption{\label{fig:array-op-der-untree}
  The definitions of $\protect\Tree{}{}$ and $\protect\tUnTree{}{}{}{}$ used in $\Dsyn[\Gamma]{\protect\taFold{}{}{}}$.
  $\protect\List{r}$ are cons-lists of $r$; $(::)$ is their cons operator.
}
\end{figure}

We show the derivatives of the three operations in \cref{fig:array-op-derivatives}.
The simplest of the three, $\Dsyn[\Gamma]{\taIndex{\trm}{\trm[2]}}$, should not be surprising given the choice of $\Dsyn{\Array\ty}_2$:
it behaves similarly to the derivative of a tuple projection ($\tFst{}$ and $\tSnd{}$), and its complexity is clearly sound for the same reasons.
For $\taBuild{}{}{}$, we use three additional array operations with the following types:
\begin{gather*}
  \begin{array}{l}
    \taUnzip{} : \Array{(\Pair{\ty}{\ty[2]})} \ra \Pair{(\Array\ty)}{(\Array{\ty[2]})} \\
    \taScatter{}{} : \text{Monoid}\ \ty \Rightarrow \Array\ty \ra \Array{(\Pair{\Z}{\ty})} \ra \Array\ty
  \end{array} \\
  \frac{\Gamma, \var : \ty[2], \var[2] : \ty \vdash \trm[3] : \ty[3] \qquad \trm[2] : \Array{\ty[2]} \qquad \trm : \Array\ty}
       {\Gamma \vdash \taZipWith{\var}{\var[2]}{\trm[3]}{\trm[2]}{\trm}: \Array{\ty[3]}}
\end{gather*}
where `$\taScatter{}{}$' ``adds'' the values in its second argument to the indicated positions in the first argument using the ``add'' operation from its monoid structure.
(The `$\text{Monoid}\ \ty \Rightarrow$' notation indicates a constraint on $\ty$, using Haskell syntax.)
`$\taUnzip{}$' and `$\taZipWith{}{}{}{}{}$' can be defined in terms of `$\taBuild{}{}{}$' and indexing; `$\taScatter{}{}$' is a new primitive running in time linear in the size of its inputs in the sequential case.
Finally we also use
%the eliminator of $\Bag{}$ 
$\taCollect{} : \Bag\ty \ra \Array\ty$ and the monadic sequence operation ($\taSequence{} : \Array{(\EnvM{\Gamma}{\ty})} \ra \EnvM{\Gamma}{(\Array\ty)}$).\footnote{This sequencing is the thing that seems to prevent a parallel implementation here, but see \cref{sec:parallelism}.}

In the primal of $\Dsyn[\Gamma]{\taBuild{\trm[2]}{i}{\trm}}$, we simply build an array of the primal results of $\trm$ for each $i$, return the array of first components as the primal result, and retain the array of backpropagators for use in the backpropagator of `$\taBuild{}{}{}$'.\footnote{Note that $\tSnd{\Dsyn[\Gamma]{\trm[2]}}$ is thrown away, because being a linear function with $\LUnit$ as domain, it cannot compute anything useful. Said differently, $\trm[2]$ being of discrete type ($\Z$), it cannot continuously depend on anything.}
Then, when the backpropagator is called, we receive a sparse array cotangent in the form of a $\Bag{}$ of pairs.
After eliminating the $\Bag{}$ structure, we construct the full cotangent using `$\taScatter{}{}$' (in $O(n)$), and then we run each of the element backpropagators on the corresponding cotangent from $d_2$.
Finally, `$\taSequence{}$' runs all the resulting monad actions.
The appearance of `scatter' (forward array permutation) is unsurprising, because `build' is the quintessential gathering (backward array permutation) operation, and reverse differentiation dualises data flow.

As for the complexity for `$\taBuild{}{}{}$': all array operations in $\Dsyn[\Gamma]{\taBuild{\trm[2]}{i}{\trm}}$ operate on arrays of the same length and can run in linear time in the sequential setting --- in addition to the expected invocations of the backpropagators $f$.
The derivative functions resulting from the execution of $\Dsyn[\Gamma]{\trm[2]}$ and $\Dsyn[\Gamma,i:\Z]{\trm}$ are executed at most once, and cotangent values are treated affinely as required.
Hence, the complexity proof should extend analogously to the cases shown in \cref{sec:proof-sketch}.

The derivative of `$\taFold{}{}{}$' in \cref{fig:array-op-derivatives} is a bit more involved.\footnote{An alternative is given in~\cite{ad-2021-diff-scan}; see \cref{sec:related-work}.}
The approach taken here is to record (in a `$\Tree{}{}$' --- see \cref{fig:array-op-der-untree}) the reduction tree taken in the primal pass by the `$\taFold{}{}{}$' combinator, and to unfold over that same reduction tree, but now from the root instead of from the leaves, in the reverse pass (with `$\tUnTree{}{}{}{}$')\footnote{The `$\List d \ra \List d$' type is a Cayley-transformed list / ``difference list'' for constant-time concatenation~\cite{fp-1986-difference-lists}.}.
In practice, one would implement `$\tUnTree{}{}{}{}$' as a primitive operation together with the `$\Tree{}{}$' data type, and hide this complexity from users.
(See also \cref{sec:parallelism}.)
Finally we also use a new array primitive: $\taFromList{} : \List\ty \ra \Array\ty$, clearly also linear-time in the length of the list.
The complexity for $\taFold{}{}{}$ is sound for the same reasons as for `$\taBuild{}{}{}$' above: its direct work is within bounds, and it calls backpropagators of its subterms at most (here precisely)~once.

% generate, indexing, fold
% collect :: Bag a -> Ar a
% scatter :: Ar a -> Ar (Int, a) -> Ar a
% length  :: Ar a -> Int
% explain and give type of following, but don't define: scatter, length, unzip, zipWith, collect, sequence
%
% \[
% \begin{array}{l}
% \taCollect{}:\Function{\Bag\ty}{\Array\ty}\\
% \taScatter{}{}:\Function{\Array\ty}{\Function{\Array{(\Pair{\Z}{\ty})}}{\Array{\ty}}}\\
%\taUnzip{}:\Function{\Array{(\Pair{\ty}{\ty[2]})}}{\Pair{(\Array{\ty})}{(\Array{\ty[2]})}}\\
% \taLength{}:\Function{\Array\ty}{\Z}
% \taZipWith{}{}{}{}{} as sugar
% \taSequence as part of monad API
% \end{array}  
% \]
% \[
% \begin{array}{l}
% % \taUnzip{a}=\letin{a'}{a}\tPair{\taBuild{(\taLength{a'})}{i}{\tFst{(\taIndex{a'}{i})}}}{\taBuild{(\taLength{a'})}{i}{\tSnd{(\taIndex{a'}{i})}}}
% \end{array}
% \]

\section{Parallelism and Practical Efficiency}\label{sec:parallelism}

Automatic differentiation is typically applied to programs with inherent parallelism, so it would be a shame if the derivative program was forced to be sequential.
So far, the prime inhibitor to parallelisation of the derivative program seems to be the monad $\EnvM{}{}$.
Of course, we cannot run the left-hand and right-hand side of a bind operation ($\tBind$) in parallel, but in our code transformation (see \cref{fig:monadic-chad,fig:array-op-derivatives}) such binding is only used when there was an actual dependency in the source program already.
For independent source expressions, the corresponding monad actions are independently sequenced using ($\tSeq$) and `$\taSequence{}$'.\footnote{Note that `$\tDo{\trm[2]; \trm}$' is syntactic sugar for $\trm[2] \tSeq \trm$, and $\tDo{\var \leftarrow \trm[2]; \trm}$ is sugar for $\trm[2] \tBind \funabs\var{\trm}$.}
Can we run those actions in parallel?

Because our monad is a (local) accumulation monad, all updates to the individual cells \emph{add} a new contribution to the value already in that cell.
Addition is commutative and associative, hence it does not matter in which order we add these contributions: the inevitable reordering resulting from concurrent updates is fine.
We just need to ensure that the individual contributions do not get corrupted by concurrent access to the same mutable cells; for this locks or atomic updates suffice.\footnote{Locks are tricky here: they must not be too fine-grained (e.g.\ around individual cotangent scalars) nor too coarse-grained (e.g.\ around derivatives of large context variables). Atomic updates to individual scalars/pointers are more straightforward.}

Reimplementing the API of $\EnvM{}{}$ in this way, it becomes safe to execute the monadic actions sequenced with ($\tSeq$) and `$\taSequence{}$' in parallel, resulting in parallelism corresponding to independent expressions in the source program.
In an actual implementation it would be prudent to have both sequential and parallel versions of ($\tSeq$) and `$\taSequence{}$', because for some uses, parallelisation will cost more in overhead than it gains in useful parallelism.

This covers all operations expressed with the mentioned parallelisable combinators; what is left is the $\tUnTree{}{}{}{}$ function from \cref{fig:array-op-der-untree}, which we need to execute in such a way that the parallelism of the original `$\taFold{}{}{}$' reduction is reflected in its derivative.
Fortunately, it suffices to execute the two recursive calls to $\tUnTree{}{}{}{}$ in parallel: this is possible because they are independent (which one could formalise by putting them in a 2-element array to `$\taSequence{}$'
or by using applicative functor operators%
).
The resulting task-parallelism mirrors the reduction tree structure, and thus the parallelism structure, of the original `$\taFold{}{}{}$'.

\paragraph{Complexity}
Unlike the sequential case, it is unclear what a proper complexity criterion should be for the parallel case.
It is not hard to see that the total amount of \emph{work} performed even in the parallel derivative is proportional to that of the input program, but requiring a constant factor slowdown over the source program in overall runtime (the \emph{span} of the program) is impossible: parallel replication ($\taBuild{n}{i}{\var}$) seemingly has, in a naive cost model, constant runtime given enough parallel execution units (i.e.\ constant span), whereas its derivative, which must perform a parallel reduction (summing all entries in the incoming cotangent value), surely has span at least logarithmic in $n$.
Note that this $n$ need not be a visible, or even easily computable, property of the source program if it is a computed value, making it hard to even formulate the optimal complexity criterion for reverse AD on parallel array programs.
For this reason, we leave a formal complexity analysis of the parallel case to future work.

\subsection{Constant Factors and Execution on Wide-Vector Machines}\label{sec:constant-factors}

The approach described above will work acceptably on multicore CPU
%(multiple-instruction multiple-data, MIMD)
platforms with relatively low core counts, once some care has been taken to avoid excessive parallelism overhead by switching to sequential execution for subexpressions that are already executed in a sufficiently parallel manner.
However, to work on GPU/TPU platforms or similar wide-vector machines%
%(single-instruction multiple-data, SIMD)
, as well as to gain more performance on MIMD architectures, it will be necessary to:
\begin{enumerate}
\item find alternate implementations of the tree-like structures: $\Bag{}$ (\cref{sec:efficient-chad-on-array-types}) and $\Tree{}{}$ (\cref{fig:array-op-der-untree});
\item analyse and optimise the output derivative program, recognising places where our base transformation was too general and a special-case approach would be more efficient.
\end{enumerate}

\paragraph{`$\protect\Tree{}{}$' on Vector Machines}
In $\Dsyn[\Gamma]{\taFold{}{}{}}$, `$\Tree{}{}$' is used to record the reduction tree in the primal pass so that we can replay it in reverse order in the reverse pass.
In practice on wide vector machines such as GPUs, the reduction tree structure of a $\taFold{}{}{}$ is statically determined to a certain extent.
For example, for the (still competitive) approach described in~\cite{array-2016-nvidia-decoupled-lookback-scan} (see their Fig.\ 5), despite the fact that the block aggregates are combined in some nondeterministic order, the tree of intermediate values corresponding to the more classical chained-scan approach (their Fig.\ 4) is still computed and stored.
Keeping these stored intermediate values around until the reverse pass allows assuming the chained-scan reduction order in the reverse pass, meaning that we only need to store the block size (an integer), the block aggregates and the block-local sequential aggregates, where our \cref{fig:array-op-derivatives} stored the full `$\Tree{}{}$'.
In effect, we thus specialise `$\Tree{}{}$' to the practical (strongly reduced) space of possible reduction orders in the actual implementation, and choose a more compact --- and in this case non-recursive! --- representation for `$\Tree{}{}$' that describes just this smaller space.
Doing so allows us to convert the task-parallelism in $\tUnTree{}{}{}{}$ to data-parallelism (for suitable combination functions), mirroring the data-parallel reduction in the primal pass.

\paragraph{`$\protect\Bag{}$' on Vector Machines}
First note that we may only \emph{add} constructors to $\Bag{}$, not subtract, as we certainly need the current ones (zero, plus, and singleton) for general source programs that use array indexing.
But sometimes we can, and need to, do better.
For example, when differentiating the following program, which first computes some array $a$ and then multiplies $a$ by 2 pointwise:\footnote{Assuming the addition of $\taLength{} : \Array\ty \ra \Z$ in the source language; because its return type is discrete, its derivative is trivial: $\Dsyn[\Gamma]{\taLength{\trm}} = \tPair{\taLength{(\tFst{\Dsyn[\Gamma]{\trm}})}}{\funabs{\_}{\tReturn\tUnit}}$.}
\begin{equation}
  \ldots
  \vdash
    \letin{a}{\taBuild{(\ldots)}{i}{\ldots}}
    {\taBuild{(\taLength{a})}{i}{(\taIndex{a}{i}) \cdot 2}}
  : \Array\ty
  \label{eq:arrays-sample-prog-gather}
\end{equation}
the derivative program will create (in $\tSnd{\Dsyn[\Gamma]{\taIndex{a}{i}}}$, which is called `$\taLength{a}$' times) many $\tBagOne{}$ values, add those together into a large tree (in `$\taSequence{}$' in the $\Dsyn[\Gamma]{\taBuild{}{}{}}$ for the result, thus --- in a parallel context --- of nondeterministic associativity), serialise that tree to an array (with `$\taCollect{}$' in the $\Dsyn[\Gamma]{\taBuild{}{}{}}$ of $a$), and finally perform a `$\taScatter{}{}$' to construct the gradient of $a$.
However, clearly a more efficient derivative is to simply multiply the result cotangent by $2$ pointwise, and while what we generate is indeed ``only'' a constant factor off in a sequential setting, this constant factor is in fact very large, and furthermore it parallelises poorly.

This program exhibits a pattern known as a \emph{gather} operation,\footnote{The derivative of `gather' is `$\taScatter{}{}$', already used above. Using a single $\taScatter{}{}$, even if no more efficient form is found based on the particular index mapping used in the program, will be much faster than creating large numbers of $\tBagOne{}$ values and having to combine those in a log-depth tree with many uses of $\tBagPlus{}{}$.} which is the program shape `$\taBuild{n}{i}{\taIndex{\trm}{\trm[2]}}$' where $i$ does not occur freely in $\trm$.
In this case, aside from the operations that we already made constant-time by construction by making them constructors of $\Bag{}$ directly (zero, plus, and singleton), we also want to be able to insert a full array of cotangents.\footnote{This constructor will be constant-time, but its cost will be inflated to the size of the array, so that there is sufficient potential in the $\Bag{}$ for `$\taCollect{}$' to amortise against later. This linear cost is acceptable because $\tBagArray{}$ will replace other linear-time operations.}
Thus we can improve the situation with a principled change to our data structure, adding a constructor ($\tBagArray{(\Array\ty)}$) to the $\Bag{}$ data type.
To then productively use this constructor in differentiating the sample program in \cref{eq:arrays-sample-prog-gather}, the implementation should either recognise the gather shape of the source program pre-differentiation and rewrite it to use some gather-style primitive, or should recognise the (inefficient) pattern resulting from naively differentiating a gather-like build and optimise that to the special-purpose form using $\tBagArray{}$.

% Fortunately, we can fix this problem with a small, principled change to $\Bag{}$ together with standard array fusion optimisations.
% The constructors of $\Bag{}$ were chosen to make the operations they represent constant-time by construction.
% We would like the following operation to be constant-time as well:
% \begin{equation*}
%   \text{arrayBag} : \Array\ty \ra \Bag{\ty}
% \end{equation*}
% and we can do so by adding a suitable constructor ($\tBagArray{(\Array\ty)}$) to the $\Bag{}$ data type.
% Now the result cotangent can be presented in the concise form offered by $\tBagArray{}$ (and can be constructed in that form in the middle of the program with an analysis that detects trivial indexing patterns like these), and after making derivative of `$\taBuild{}{}{}$' aware of this possibility, the derivative simplifies to something much closer to the optimal pointwise-$\tfrac12$-multiplication.

As with all compiler optimisations, however, such tricks cannot cover all possible programs, but the common cases can be dealt with in this manner.
We leave a more thorough investigation of the problems and solutions to future work, in particular how to efficiently handle on GPU hardware the $\Bag{}$ operations that do \emph{not} fall into the nice, vectorisable case like described above.

\section{Higher-Order CHAD}\label{sec:efficient-chad-on-function-types}

\paragraph{Naive CHAD of Higher-Order Functions}
We give here the (semantically verified) recipe of \cite{vakar-2021-higher-order-reverse-ad,vakar-2022-chad} for differentiating function types using CHAD, where we omit any use of abstract or linear types and work directly with their default implementation.
We write $[]$ for an empty list and $\lplus$ for concatenation, and $\mathbf{fold}\ {v}\ \mathbf{with}\ {z},{acc}\to {\trm}\ \mathbf{from}\ {acc_0}$ for the usual sequential (right) fold elimination of the list $v$, starting the accumulator $acc$ from an initial value $acc_0$ and folding in values $z$ using the operation $\trm$.
The rules are as follows:
\[
\begin{array}{l@{\quad}l}
\Dsyn{\Function{\ty}{\ty[2]}}_1 \defeq \Function{\Dsyn{\ty}_1}{(\Pair{\Dsyn{\ty[2]}_1}{(\Function{\Dsyn{\ty[2]}_2}{\Dsyn{\ty}_2})})}
&
\Dsyn{\Function{\ty}{\ty[2]}}_2 \defeq \List{(\Pair{\Dsyn{\ty}_1}{\Dsyn{\ty[2]}_2})}
\\
\Dsyn[\Gamma]{\funabs{\var:\ty}{\trm}} \defeq
\begin{array}[t]{@{}l@{}}
  \tPairOpen
  \begin{array}[t]{@{}l@{}}
    \funabs{\var:\Dsyn{\ty}_1}{
      \begin{array}[t]{@{}l@{}}
        \letin{\tPair{\var[2]}{\var[2]'}}{\Dsyn[\Gamma,\var:\ty]{\trm}}{} \\
        \tPair{\var[2]}{\funabs{d}{\tSnd(\tSplit{}{}\ (\var[2]'\,d))}} \tPairSep
      \end{array}
    } \\
  \funabs{d}{}
  \begin{array}[t]{@{}l@{}}
    % Sorry for writing out the \tFold macro, it was just too cumbersome
    \mathbf{fold}\ d\ \mathbf{with}\ \var[3], \mathit{acc} \ra \\
    \quad \begin{array}[t]{@{}l@{}}
      \letin{\tPair{\var}{\var[2]}}{\var[3]}{} \\
      \mathit{acc} + \mathrlap{\tFst{(\tSplit{}{}\ (\tSnd \Dsyn[\Gamma,\var:\ty]{\trm}\,\var[2]))}}
    \end{array} \\
    \mathbf{from}\ \tZero[\Dsyn{\Gamma}_2] \tPairClose
  \end{array}
\end{array}
\end{array}
&
\Dsyn[\Gamma]{\trm\,\trm[2]}\defeq
\begin{array}[t]{@{}l@{}}
  \letin{\tPair{\var}{\var'}}{\Dsyn[\Gamma]{\trm}}{} \\
  \letin{\tPair{\var[2]}{\var[2]'}}{\Dsyn[\Gamma]{\trm[2]}}{} \\
  \letin{\tPair{\var[3]}{\var[3]'}}{\var\,\var[2]}{} \\
  \tPair{\var[3]}{\funabs{d}{\var[2]'(\var[3]'\,d)} + \var'\,[\tPair{\var[2]}{d}]}
\end{array}
\\
\tZero[\List{(\Pair{\Dsyn{\ty}_1}{\Dsyn{\ty[2]}_2})}]\defeq []
&
\tPlus[\List{(\Pair{\Dsyn{\ty}_1}{\Dsyn{\ty[2]}_2})}]{\trm}{\trm[2]}\defeq \trm\lplus\trm[2]\\
\end{array}
\]
The key idea is that in a function application, the incoming cotangent must be propagated backwards through the $\lambda$-abstraction being called, \emph{both to the function argument and to the captured context variables of the closure}.
CHAD separates these two parts of the derivative and handles the former with $\Dsyn{\Function{\ty}{\ty[2]}}_1$ 
and the latter with $\Dsyn{\Function{\ty}{\ty[2]}}_2 \ra \EnvM{\Dsyn{\Gamma}_2}{\Unit}$, which is the type of $\tSnd{\Dsyn[\Gamma]{\trm}}$ if $\trm$ is of type $\Function{\ty}{\ty[2]}$.
The list $\Dsyn{\Function{\ty}{\ty[2]}}_2$ is a log of all invocations of the function, containing for each invocation its input primal and its output cotangent.

\paragraph{Identifying the Complexity Issues}
It is precisely this separation of the derivative that leads to real\footnote{Technically the linear-time cost of $\lplus$ is also a problem, but that can be resolved by Cayley-transforming or using $\Bag{}$.} complexity problems.
In particular, because the derivative of a function value is split in two parts, it is impossible (from $\Dsyn[\Gamma]{\funabs{\var:\ty}\trm}$) to get the full gradient of a function, i.e.\ with respect to both its argument and its context, in one pass through $\Dsyn[\Gamma,\var:\ty]{\trm}$.
And for function application, which is the only eliminator of functions in the source language, $\tSnd{\Dsyn[\Gamma]{\trm\ \trm[2]}}$ does indeed need the full gradient of the function that was called (i.e.\ $\trm$).
It must therefore resort to using both halves of the function's derivative separately, meaning that we end up differentiating through a function \emph{twice} each time it is called.
If that function contains other function applications inside its body, the derivatives of the functions called there are evaluated four times, etc.

This behaviour can be exploited to violate our complexity criterion \cref{eq:complexity-criterion}.
Indeed, the programs $\trm_n$ for each $n$ (which are nested identity applications, and thus semantically just the identity):\footnote{It is unnecessary for the function being applied to be a \emph{literal} lambda expression --- defining the lambda somewhere and calling it elsewhere calls the same backpropagators in the end.
The example is just written this way for conciseness.}
\begin{align*}
% &\var_1 : \R \vdash \trm_1 \defeq \var_1 : \R \\
% &\var_2 : \R \vdash t_2 \defeq (\funabs{\var_1}{\var_1})\ \var_2 : \R \\
&\var_n : \R \vdash \trm_n \defeq (\funabs{\var_{n-1}}{(\cdots \funabs{\var_2}{(\funabs{\var_1}{\var_1})\ \var_2}\cdots)\ \var_{n-1}})\ \var_n : \R
\end{align*}
execute in $O(n)$ time.
Their transposed derivatives $\tSnd{\Dsyn[\var_n:\R]{\trm_n}}$ using the CHAD formulas above, however, take $O(2^n)$ time
to execute: because they contain $n$ nested pairs of an application of an abstraction, the backpropagator of each will execute the backpropagator of its body twice.
% \begin{align*}
%   x : \R &\vdash x * x \\
%   x : \R &\vdash (\fun x. \letin{y}{x + x}{\letin{z}{x * x}{z * z}}) x \\
%   x : \R &\vdash (\fun x. \begin{array}[t]{@{}l@{}}\letin{y}{x + x}{}\\\letin{z}{x * x}{}\\
%     (\fun x. \letin{y}{x + x}{\letin{z}{x * x}{z * z}}) (z * z)) x\end{array} \\
%     \\ 
%     & \text{core[var]} = \var * \var\\
%     & \text{nest}[f][\var] = (\fun x. \letin{y}{x + x}{\letin{z}{x * x}{f[z]}}) \var,\\
%     \\
%     &\qquad \text{core}[x], \text{nest}[\text{core}][x], \text{nest}[\text{nest}[\text{core}]][x], \text{nest}[\text{nest}[\text{nest}[\text{core}]]][x], \text{etc.}
% \end{align*}

% Then, we obtain the following combined action.
% \[
% \begin{array}{ll}
% \Dsyn{\Function{\ty}{\ty[2]}}_1\defeq   \BigSum{\tvar:\Type}(\Function{(\Pair{\Dsyn{\tvar}_1}{\Dsyn{\ty}_1})}{\Pair{\Dsyn{\ty[2]}_1}{(\Function{\Dsyn{\ty[2]}_2}{(\Pair{\Dsyn{\tvar}_2}{\Dsyn{\ty}_2})})}})\\
% \Dsyn{\Function{\ty}{\ty[2]}}_2\defeq \BigSum{\tvar:\Type}\Unit.
% \end{array}
% \]

\paragraph{Solving Complexity Issues Through Defunctionalisation}
Clearly, defunctionalisation \cite{fp-1998-defunctionalisation} translates away function types into a language that we can already differentiate efficiently,
% by implementing function types $\Function{\ty}{\ty[2]}$ as a sum type of tuples $\BigSum{lam\in \Lambda(\ty,\ty[2])}{\Pair{\Pair{\ty[3]_1^{lam}}{\cdots}}{\ty[3]_{n(lam)}^{lam}}}$, where $\Lambda(\ty,\ty[2])$ is the (finite) set of syntactic $\lambda$-abstractions $\var^{\funabs{\var[2]}{\trm}}_1:\ty[3]_1^{\funabs{\var[2]}{\trm}},\cdots,\var^{\funabs{\var[2]}{\trm}}_{n(\funabs{\var[2]}{\trm})}:\ty[3]_{n(\funabs{\var[2]}{\trm})}^{\funabs{\var[2]}{\trm}}\vdash \funabs{\var[2]}{\trm}:\Function{\ty}{\ty[2]}$ in the (global) program with captured context variables $\var^{\funabs{\var[2]}{\trm}}_1:\ty[3]_1^{\funabs{\var[2]}{\trm}},\cdots,\var^{\trm}_{n(\funabs{\var[2]}{\trm})}:\ty[3]_{n(\funabs{\var[2]}{\trm})}^{\funabs{\var[2]}{\trm}}$.
by implementing function types $\Function{\ty}{\ty[2]}$ as a sum type of tuples $\Pair{\Pair{\ty[3]_1^\ell}{\cdots}}{\ty[3]_{n_\ell}^\ell}$ for each syntactic lambda-abstraction $\ell$  of type $\Function{\ty}{\ty[2]}$ in the program, writing $\ty[3]^\ell_1,\ldots,\ty[3]^\ell_{n_\ell}$ for the list of types of $\ell$'s captured context variables.
(This list is a subset of the types in $\ell$'s environment.)
% $\lambda$-abstractions then get replaced by their corresponding sum constructor and function applications get replaced by an evaluator that pattern matches on the form of a function and continues to evaluate the corresponding body.
As such, we can simply defunctionalise (a well-known strategy for  compiling code with function types) before applying CHAD and then call it a day.
\emph{Why} exactly does this solve the problem of inefficient function types though?
The key observation is to decompose defunctionalisation into the composition of:
\begin{itemize}
  \item the local program transformation of (typed) closure conversion \cite{DBLP:conf/popl/MinamideMH96}: we convert every function into a closure, which is a pair of (a subset of) its environment and a function that does not capture any context variables (a ``closed'' function);
  \item the global program transformation of ``deexistentialisation'' that replaces an existential type with the finite sum type of all of its instantiations found in the whole program.
  A global program analysis is needed here to be able to use a finite rather than an infinite sum type; we need to analyse precisely which instances of the existential are actually used.
\end{itemize}
\ifTomACMVersion{It}{As we show in \cref{app:closure-conversion}, it} is the first part of the transformation that solves the efficiency~problems of CHAD on function types\ifTomACMVersion{ (see \cite[Appendix B]{efficient-chad-arxiv})}{}:
by replacing all functions with closed functions.
As a consequence, we avoid the need to propagate back any cotangents to captured context variables, removing the need for one half of a function's CHAD derivative: with only one half left, the duplication is gone, eliminating the complexity problem.
In particular, it will now suffice to take $\Dsyn{\Function{\ty}{\ty[2]}}_2 \defeq \LUnit$, which simplifies~the term-level derivatives accordingly.
The resulting CHAD transformation remains local.

This idea of using closure conversion to speed up AD of higher-order functions is first used by \cite{ad-2008-reverse-functional-ad} (later distilled to its essence by  \cite{ad-2021-alvarez-picallo}).
More recently, the short paper \cite{ad-2019-vytiniotisdifferentiable} suggested its use in the context of CHAD-like AD transformations. This section can be seen as an elaboration of the suggested idea of the latter paper.

% \section{Future Work}

% \begin{itemize}
% \item parallelism proof
% \item constant factors
% \item more array primitives: scan, scatter
% \item optimised special case derivatives of array primitives that can be expressed as generate: map, zipwith, stencils, backpermute/gather
% \item application to, or compile to, Accelerate or Futhark?
% \end{itemize}

\section{Related Work}\label{sec:related-work}

This paper shows how the basic reverse AD algorithm of CHAD can be made efficient.
The basic CHAD technique for a first-order language with tuples in combinator form was originally introduced by \citeN{adfp-2018-categories-ad}. 
\cite{vakar-2021-higher-order-reverse-ad,vakar-2022-chad,nunes-2022-chad-expressive} show how it applies to a $\lambda$-calculus with various type formers, giving a correctness proof, but no complexity proof.
\citeN{ad-2022-kerjean:hal-03123968} point out that the resulting code transformation closely resembles the Diller-Nahm variant of the Dialectica interpretation.

Similar optimisation techniques to the ones we use to make CHAD efficient (notably, sparse vectors and functional mutability) were previously used by  \citeN{ad-2021-krawiec-kmett-ad} and \citeN{ad-dualrev-th} to make \emph{dual-numbers} reverse AD efficient.
Their approach has $\Dsyn{\R} = \Pair{\R}{(\LR \lra \underline{\Gamma})}$ and $\Dsyn{\Pair{\ty}{\ty[2]}} = \Pair{\Dsyn{\ty}}{\Dsyn{\ty[2]}}$ (``the pair in the leaves''), very different from our split between $\Dsyn{-}_1$ and $\Dsyn{-}_2$ (``the pair at the root''); it turns out that they essentially do classical tracing.
The required ID-generation in their approach makes it less clear how it might apply to parallel programs.

Mutability in functional AD tends to be used for accumulation: this occurs in~\cite{ad-2021-krawiec-kmett-ad} as well as Dex~\cite{dex-2021-ad} and Futhark~\cite{ad-2022-futhark-partial-recompute}.
(In~\cite{ad-2023-yolo}, which describes the basic structure of Dex' AD algorithm (linearise-then-transpose), mutation is not yet necessary due to the simplicity of their input language.)
Dex extends the method to a richer source language and needs to use mutability with an algebraic effect for (parallel) accumulation, similar to the solution in this paper; Futhark uses uniqueness types to implement the same idea.
We instead use a monad to implement this effect.

Previous work in computer-formalised proofs about AD are, to the best of our knowledge, limited to 
\cite{thesis-2020-curtis-fwd-ad-gradient-compiler-opts}, which formalises the correctness proofs for the dual numbers forward-mode AD transformation of \citeN{ad-2019-fwd-ad-gradient-compiler-opts,ad-2020-sam-mathieu-matthijs,DBLP:conf/esop/BartheCLG20,DBLP:journals/corr/abs-2007-05282,vakar-staton-huot-2021,nunes-2022-dual-numbers-long} in Coq,
and \cite{ad-2021-verifying}, which gives a Coq proof of the correctness of an effect handler-based variant of the 
reverse-mode AD techniques of 
\cite{ad-2018-cps-by-callstack,ad-2019-delimited-continuations,ad-2018-rev-delimited-continuations} (which rely on non-functional control flow).
Both papers focus on the correctness of AD, rather than its complexity.

In currently used industrial systems (such as TensorFlow \cite{ad-2016-tensorflow}, PyTorch \cite{ad-2017-pytorch} or JAX \cite{ad-2018-jax}), AD is typically performed on first-order (data-parallel) functional array processing languages.
AD of second order functional array languages as a source transformation has been considered recently by 
\citeN{ad-2022-futhark-partial-recompute}.
They allow some recompution and a resulting suboptimal complexity to achieve a simpler and more practically efficient algorithm, in the hope that such recomputation is rare in practice.
By contrast, here we study how to avoid all recomputation.
\citeN{ad-2021-diff-scan} present a derivative for scans just in terms of standard second-order array combinators, but this version has the downside of being not quite complexity-efficient --- it has a complexity blowup in the case of nesting fold in the combination function of fold.
We avoid this blow-up with a custom primitive for the derivative (of fold, in our case, but we expect scans to work similarly).

The idea of using closure conversion to make AD of higher order functions efficient first appears buried in the details of
VLAD/Stalin$\nabla$ \cite{ad-2008-reverse-functional-ad,ad-2016-scheme-higher-order-ad}.
The idea is again present in \cite{ad-2019-vytiniotisdifferentiable} (in the context of CHAD) and \cite{ad-2021-alvarez-picallo} (for an AD algorithm using string diagram rewrites), without precisely demontrating its importance.
We have made an effort to spell out and motivate the idea in the present paper, making clear how it arises as a natural solution to the complexity problems of higher-order CHAD.

\citeN{ad-2022-elsman-combinatorial} show that treating multi-linear operations as special cases can result in very nice derivatives using a generalised product-rule where the general approach produces unwieldy code. In a future publication we will show how to make multi-linear operations more first-class in CHAD as well in order to benefit from this, especially since their setting is similar (although expressed on a smaller language, lacking tuples, array indexing and first-class control flow).

Recently, \citeN{van2024forward} made clever use of type classes to present various AD algorithms in a uniform way.
Their considerations are orthogonal to our concerns in this work.

\citeN{shaikhha2023nabla} discuss how to  differentiate \emph{source code} that uses sparse array operations efficiently.
By contrast, we use a sparse array representation in the \emph{generated derivative code} to achieve efficiency.

\newpage

% Unfortunately we can't use \ifACMAnonymous here because passing an environment as an argument to that doesn't fly.
\makeatletter
\if@ACM@anonymous
\makeatother
% No acknowledgements in anonymous mode
\else
\makeatother
\begin{acks}
We would like to thank Gershom Bazerman for suggesting Okasaki's banker's method, which inspired the amortisation argument in this paper.

This project has received funding via NWO Veni grant number VI.Veni.202.124.
\end{acks}
\fi

\bibliography{bibliography}

\ifTomACMVersion{}{
\clearpage 

\appendix
\section*{Appendix}

\section{Why \emph{union} Is Not Efficient}\label{app:union-not-efficient}

In \cref{sec:monadic-lifting} we changed the transformation so that the transformed code passes around a growing environment cotangent in state-passing style, instead of simply returning the local environment contributions upwards from each branch of the program.
Not only does this provide the right program structure to later swap out the (log-time) functional persistent tree map for a mutable array in \cref{sec:log-factors}, but it is also necessary from a complexity perspective: simply keeping $\mUnion{}{}$ in is inefficient, even if one has a magical tree map that has linear instead of linearithmic complexity.

\newcommand\tmagic{t_{\text{magic}}}

The actual map union\footnote{As defined in the Haskell \texttt{containers} library, and proved optimal in~\cite{algorithms-1979-merging-tarjan}.} has runtime $O(m \log\mleft(\frac{n}{m}+1\mright))$, where $m$ is the size of the smaller argument to $\mUnion{}{}$ and $n$ the size of the larger.
For small $m > 0$ this simplifies to $O(\log n)$, and for $m \leq n$ in $O(n)$ it simplifies to $O(n)$.

Suppose that we had access to a magical union running in time $O(m)$ where $m$ is the size of the smaller argument to $\mUnion{}{}$.
(Note that this is always strictly better than the complexity reported above.)
This is still too expensive, as witnessed by $x_1 : \R, \ldots, x_n : \R \vdash \tmagic : \R$ defined as follows:
\begin{center}
\begin{tikzpicture}[xscale=0.5, yscale=0.6]
\node (a) at (0, -0.2) {$\star$};
\node (b1) at (-2, -1) {$\star$};
\node (b2) at (2, -1) {$\star$};
\node (c1) at (-3, -2) {$\star$};
\node (c2) at (-1, -2) {$\star$};
\node (c3) at (1, -2) {$\star$};
\node (c4) at (3, -2) {$\star$};
\node (v1) at (-3, -2.5) {$\vdots$};
\node (v2) at (-1, -2.5) {$\vdots$};
\node (v3) at (1, -2.5) {$\vdots$};
\node (v4) at (3, -2.5) {$\vdots$};
\node (d1) at (-3, -3.25) {$\star$};
\node (d2) at (-1, -3.25) {$\star$};
\node (d3) at (1, -3.25) {$\star$};
\node (d4) at (3, -3.25) {$\star$};
\node (e1) at (-3.5, -4.4) {$x_1$};
\node (e2) at (-2.5, -4.4) {$x_2$};
\node (e3) at (-1.5, -4.4) {$...$};
\node (e4) at (-0.5, -4.4) {$...$};
\node (e5) at (0.5, -4.4) {$...$};
\node (e6) at (1.5, -4.4) {$...$};
\node (e7) at (2.5, -4.4) {$...$};
\node (e8) at (3.5, -4.4) {$x_{2^r}$};
\draw (a) -- (b1);
\draw (a) -- (b2);
\draw (b1) -- (c1);
\draw (b1) -- (c2);
\draw (b2) -- (c3);
\draw (b2) -- (c4);
\draw (d1) -- (e1);
\draw (d1) -- (e2);
\draw (d2) -- (e3);
\draw (d2) -- (e4);
\draw (d3) -- (e5);
\draw (d3) -- (e6);
\draw (d4) -- (e7);
\draw (d4) -- (e8);
\node[anchor=west] (l0) at (5, -0.2) {layer 0};
\node[anchor=west] (l1) at (5, -1.1) {layer 1};
\node[anchor=west] (l2) at (5, -2) {layer 2};
\node[anchor=west] (lv) at (5, -2.5) {\quad\ \ $\vdots$};
\node[anchor=west] (l3) at (5, -3.3) {layer $r-1$};
% \node[anchor=west] (l4) at (5, -4.4) {layer $r$};
\end{tikzpicture}
\end{center}
i.e.\ a complete binary tree adding $x_1, \ldots, x_{2^r}$ where $r \defeq \lfloor \log_2(n) \rfloor$, so that $2^r \leq n < 2^{r+1}$.
All variables in the leaves are distinct.
For $0 \leq i \leq r-1$, layer $i$ has $2^i$ occurrences of $\star$, and hence in $\Dsyn[\Gamma]{\tmagic}$ we will get $2^i$ applications of $\mUnion{}{}$ on maps of size $2^{r-1-i}$ on layer $i$.
The total computational cost of all these unions is (under the magical union assumption):
\[
  \sum_{i = 0}^{r - 1} 2^i \cdot O(2^{r-1-i})
  = O(r \cdot 2^r)
  = O(\log(n) \cdot n)
\]
which is asymptotically larger than $O(n)$, the runtime cost of $\tmagic$.
(Note that the total number of $\star$ operations in $\tmagic$ itself is equal to $\sum_{i=0}^{r-1} 2^i = 2^r - 1 = O(n)$.)
Hence even with this magical union, CHAD does not yet attain the correct complexity for reverse AD, and the state passing modification in \cref{sec:monadic-lifting} is necessary.

\section{Solving Complexity Issues Through Closure Conversion}
\label{app:closure-conversion}
The idea behind typed closure conversion \cite{DBLP:conf/popl/MinamideMH96} is as follows.
We transform types, using a fresh type variable $\tvar:\Type$, 
\begin{align*}
  &\TCC{\R}\defeq \R\quad \TCC{\Unit}\defeq \Unit\quad \TCC{[\Pair{\ty}{\ty[2]}]}\defeq \Pair{\TCC{\ty}}{\TCC{\ty[2]}}\quad 
  \TCC{[\Sum{\ty}{\ty[2]}]}\defeq \Sum{\TCC{\ty}}{\TCC{\ty[2]}}\\ 
  &\TCC{[\Function{\ty}{\ty[2]}]}\defeq \textcolor{red}{\BigSum{\tvar:\Type}\Pair{\tvar}{(\CFunction{(\Pair{\tvar}{\TCC\ty})}{\TCC{\ty[2]}})}}
\end{align*}
and, writing $\FV{\trm}$ for the free variables occurring in the term $\trm$, transform the programs
\begin{align*}
\TCC{\var} &\defeq \var\\
\TCC{[\letin{\var}{\trm}{\trm[2]}]}&\defeq \letin{\var}{\TCC\trm}{\TCC{\trm[2]}}\\
\TCC{\tUnit} & \defeq \tUnit\\
\TCC{[\tPair{\trm}{\trm[2]}]} & \defeq \tPair{\TCC\trm}{\TCC{\trm[2]}}\\
\TCC{[\tFst\trm]}&\defeq \tFst\TCC\trm\\
\TCC{[\tSnd\trm]}&\defeq \tSnd\TCC\trm\\
\TCC{[\tInl\trm]}&\defeq \tInl\TCC\trm\\
\TCC{[\tInr\trm]}&\defeq \tInr\TCC\trm\\
\TCC{\left[\tSAMatchCpt{\trm}{\var}{\trm[2]}{\var[2]}{\trm[3]}\right]}&\defeq  \tSAMatchCpt{\TCC\trm}{\var}{\TCC{\trm[2]}}{\var[2]}{\TCC{\trm[3]}}\\
\TCC{[\funabs{\var}{\trm}]}&\defeq \textcolor{red}{\tPack[\TCC{\TypeOf(\tTuple{\FV{\trm}\setminus \{\var\}})}]{\tPair{\tTuple{\FV{\trm}\setminus \{\var\}}}{\funabs{\tPair{\tTuple{\FV{\trm}\setminus \{\var\}}}{\var}}{\TCC\trm}}}}  \\
\TCC{[\trm\,{\trm[2]}]} &\defeq  \textcolor{red}{\tPackMatchTyped{\TCC\trm}{\tvar}{\var[3]}{\letin{\tPair{cvars}{f}}{\var[3]}{f\,\tPair{cvars}{\TCC{\trm[2]}}}}} \\
\TCC{r}&\defeq r\\
\TCC{[\tSign\trm]} & \defeq \tSign\TCC\trm\\
\TCC{[\tOpApp{\trm_1,\ldots,\trm_n}]}&\defeq \tOpApp{\TCC{\trm_1},\ldots,\TCC{\trm_n}}
\end{align*}
Then, $\Gamma\vdash \trm:\ty$ implies $\TCC{\Gamma}\vdash \TCC{\trm}:\TCC{\ty}$, where 
$\TCC{[\var_1:\ty_1,\ldots,\var_n:\ty_n]}\defeq \var_1:\TCC{\ty_1},\ldots,\var_n:\TCC{\ty_n}$.
In the above, $\CFunction{\ty}{\ty[2]}$ is a type of \emph{closed} functions with the typing rules
$$
\inferrule{\textcolor{red}{\var:\ty}\vdash \trm:\ty[2] }{
\Gamma\vdash \funabs{\var}{\trm}:\CFunction{\ty}{\ty[2]}}\qquad 
\qquad 
\inferrule{\Gamma\vdash \trm:\CFunction{\ty}{\ty[2]}\quad \Gamma\vdash \trm[2]:\ty}{\Gamma\vdash \trm\,\trm[2]:\ty[2]}
$$
The idea is that $\CFunction{\ty}{\ty[2]}$ only holds \emph{closed} functions from $\ty$ to $\ty[2]$, i.e. ones that do not capture any context variables.
Further, the type $\BigSum{\tvar:\Type}\ty$ is a sum type indexed by the kind of types\footnote{We formulate closure conversion using a tagged sum type, rather than an untyped existential type as is sometimes done. The motivation is that we need to use equality checks on type tags at runtime for the casts and addition in the CHAD transformation.}:
$$
\inferrule{\Gamma\vdash \trm:\subst{\ty}{\sfor{\tvar}{\ty[2]}}}
{\Gamma\vdash \tPack[{\ty[2]}]{\trm}:\BigSum{\tvar:\Type}{\ty}}
\qquad 
\inferrule{
\Gamma\vdash \trm:\BigSum{\tvar:\Type}\ty\qquad
\tvar:\Type,\Gamma,\var:\ty\vdash \trm[2]:\ty[3]
}{
\Gamma\vdash\tPackMatchTyped{\trm}{\tvar}{\var}{\trm[2]}:\ty[3],
}
$$
where $\tvar$ can occur freely in $\ty$ and $\trm[2]$ in the elimination rule.
That is, we assume that we have an (impredicative) type universe $\Type$ with decidable equality in our type system.
Such a universe can be implemented for our type system, for example, in Haskell by using GADTs.

Crucially, $\TCC{\trm}$ computes the same function\footnote{Indeed, for any program $\trm$ between first order types (types on which $\TCC{[-]}$ acts as the identity), $\trm$ is $\beta\eta$-equal to 
$\TCC{\trm}$, so in particular is observationally equivalent.} as $\trm$ and does so in the same computational complexity.
In fact, most functional languages compile function types via closure conversion.

After applying closure conversion, we can apply the CHAD as follows to types without free type variables (monomorphic types) as well as their programs: 
% $\tvar:\Type\vdash \Dsyn{\tvar}_1:\Type$ and  $\tvar:\Type\vdash \Dsyn{\tvar}_2:\Type$:
\[
\begin{array}{ll}
\Dsyn{\CFunction{\ty}{\ty[2]}}_1\defeq \CFunction{\Dsyn{\ty}_1}{(\Pair{\Dsyn{\ty[2]}_1}{(\Function{\Dsyn{\ty[2]}_2}{\Dsyn{\ty}_2})})}\qquad& 
\Dsyn{\CFunction{\ty}{\ty[2]}}_2\defeq \Unit \\
\Dsyn{\BigSum{\tvar:\Type}\ty}_1\defeq \BigSum{\tvar:\Type}\Dsyn{\ty}_1\qquad& \Dsyn{\BigSum{\tvar:\Type}\ty}_2\defeq\LBigSum{\tvar:\Type}\Dsyn{\ty}_2
\\
%\Dsyn{\tvar}_1\defeq \Dsyn{\tvar}_1\qquad& \Dsyn{\tvar}_2\defeq \Dsyn{\tvar}_2\\
\Dsyn[\Gamma]{\funabs{\var:\ty}{\trm}}\defeq \tPair{\funabs{\var:\Dsyn{\ty}_1}\Dsyn[\var:\ty]{\trm}}{\funabs{\_:\Unit}\tZero}
&
\Dsyn[\Gamma]{\trm\,\trm[2]}\defeq
\begin{array}[t]{@{}l@{}}
\letin{\tPair{\var}{\_}}{\Dsyn[\Gamma]{\trm}}{}\\
\letin{\tPair{\var[2]}{\var[2]'}}{\Dsyn[\Gamma]{\trm[2]}}{}\\
\letin{\tPair{\var[3]}{\var[3]'}}{\var\,\var[2]}{}\\
\tPair{\var[3]}{\funabs{v}\var[2]'(\var[3]'\,v)}
\end{array}\\
\Dsyn[\Gamma]{\tPack[{\ty[3]}]{\trm}}\defeq 
\begin{array}[t]{@{}l@{}}
\letin{\tPair{\var}{\var'}}{\Dsyn[\Gamma]{\trm}}{}\\
\tPair{\tPack[{\ty[3]}]{\var}}{\funabs{v}{\var'(\tCast[{\ty[3]}]{v})}}
\end{array}
&
\begin{array}[t]{@{}l@{}}
  \Dsyn[\Gamma]{\tPackMatchTyped{\trm}{\tvar}{\var}{\trm[2]}}\defeq \\
  \quad\letin{\tPair{\var[3]}{\var[3]'}}{\Dsyn[\Gamma]{\trm}}{}\\
  \quad\tPackMatchTyped{\var[3]}{{\tvar}}{\var}{}\\
  \qquad\letin{\tPair{\var[2]}{\var[2]'}}{\Dsyn[\Gamma]{\trm[2]}}{}\\
  \qquad\tPairOpen{\var[2]}\tPairSep{\funabs{v}{\aletin{\tPair{w_1}{w_2}}{\tSplit{}{}(\var[2]'\,v)}{w_1+\var[3]'\,(\tPack[{\ty[3]}]{w_2})\tPairClose}}}
  \end{array}
\end{array}  
\]
Here, $\LBigSum{\tvar:\Type}{\ty}$ has the following API:
\[
\begin{array}{ll}
\tPack[{\ty[2]}]:\subst{\ty}{\sfor{\tvar}{\ty[2]}}\to \LBigSum{\tvar:\Type}{\ty}
\qquad\;
  &
\tCast[{\ty[2]}]:(\LBigSum{\tvar:\Type}{\ty})\to \subst{\ty}{\sfor{\tvar}{\ty[2]}}\\
\tZero[{\LBigSum{\tvar:\Type}{\ty}}]:\LBigSum{\tvar:\Type}{\ty}& (\tPlus[\LBigSum{\tvar:\Type}{\ty}]{}{}):\Pair{(\LBigSum{\tvar:\Type}{\ty})}{(\LBigSum{\tvar:\Type}{\ty})}\to \LBigSum{\tvar:\Type}{\ty},
\end{array}
\]
which we can implement, completely analogously to the case of binary sum types, by representing $\LBigSum{\tvar:\Type}{\ty}$ as $\Sum{\Unit}{(\BigSum{\tvar:\Type}\ty)}$.
To implement this API, it is crucial that we work with $\Sigma$-types that hold type tags with decidable equality rather than untagged existentials.
Observe that similarly to our treatment of sum types (with $\tlCast{}$ and $\trCast{}$), the differentiation of existential types requires some runtime casting\footnote{This throws an error if it fails -- such an error will never be hit by the code generated by CHAD after closure conversion.}, in the absence of a type system with full dependent types.
We can prove, however, that all required casts are type-safe in a stronger dependently-typed type system.

If we are willing to do a global program analysis, we can identify at compile-time the finite subset of components of the sum types that are actually used in practice. 
This allows us to simply replace the infinite sum type with a finite sum type (a transformation that we have referred to as ``deexistentialisation'' in \cref{sec:efficient-chad-on-function-types}).
We have now effectively arrived at our combination of closure conversion and defunctionalisation that we discussed in \cref{sec:efficient-chad-on-function-types}.

\iffalse
environment cotangent / environment derivative / environment vector:
\begin{itemize}
\item An \emph{environment vector} is a product of types from an environment.
\item An \emph{environment cotangent} is the cotangent of the context of some term.
\end{itemize}
Thus, in the first half of the paper, we represent \emph{environment cotangents} as \emph{environment vectors} of $\Dsyn{\Gamma}_2$, which is a \emph{cotangent environment} (due to being an environment of cotangents).

derivative program / target program:
\begin{itemize}
\item The input to the code transformation is the \emph{source program}.
\item The output of the code transformation is the \emph{derivative program}.
\end{itemize}
\fi

\section{Cost Model}\label{app:cost-model}

\newcommand\eval[2]{\mathrm{eval}(#1; #2)}
\newcommand\FreeVars[1]{\mathsf{FV}(#1)}

In the table below, we describe the cost model (using call-by-value evaluation) used in the formalised proof in natural language.
In the Agda formalisation (\cref{app:agda-spec}), this is embedded in the \texttt{eval} function, namely in its second component, as well as (for the $\EnvM{}{}$ methods) in their implementation in the \texttt{spec.LACM} module.
We separately describe the model here to aid in understanding what is encoded in the formal specification.

As in the paper, we use $\cost{\trm}{\Gamma}$ to denote the cost of evaluating $\trm$ in the evaluation environment $\Gamma$.
We furthermore use $\eval{\trm}{\Gamma}$ to denote the \emph{result} of evaluating $\trm$ in that evaluation environment.
$\FreeVars{\trm}$ denotes the set of free variables of the term $\trm$ (only used for lambda abstraction to measure the size of the closure to allocate).

The term language that we analyse is \texttt{Term}, in \texttt{spec.agda}.

\begin{tabular}{l|l}
\textbf{Term $\trm$} & \textbf{Cost:} $\cost{\trm}{\Gamma}$ \\\hline
$\var$ \textit{(variable)} & 1 \\
$\letin{\var}{\trm[2]}{\trm}$ & $1 + \cost{\trm[2]}{\Gamma} + \cost{\trm}{\var = \eval{\trm[2]}{\Gamma}, \Gamma}$ \\
$\funabs{\var}{\trm}$ & $1 + |\FreeVars{\funabs{\var}{\trm}}|$ \\
$\trm\ \trm[2]$ &
  \makecell[l]{
    $1 + \cost{\trm}{\Gamma} + \cost{\trm[2]}{\Gamma} + \cost{f\ x}{f = \eval{\trm}{\Gamma}, x = \eval{\trm[2]}{\Gamma}, \Gamma} - 2$ \\
    \quad \textit{(The $- 2$ compensates for the two superfluous variable references $f$ and $x$.)}
  } \\
$\op(\trm)$ &
  \makecell[l]{
    $1 + \cost{\trm}{\Gamma}$ \textit{(for simplicity we assume unary operators only;} \\
    \quad \textit{$n$-ary operators take tuples, e.g.\ $(+) : \Pair\R\R \ra \R$)}
  } \\
$\tUnit$ & $1$ \\
$\tPair{\trm}{\trm[2]}$ & $1 + \cost{\trm}{\Gamma} + \cost{\trm[2]}{\Gamma}$ \\
$\tFst{\trm}$ & $1 + \cost{\trm}{\Gamma}$ \\
$\tSnd{\trm}$ & $1 + \cost{\trm}{\Gamma}$ \\
$\tInl{\trm}$ & $1 + \cost{\trm}{\Gamma}$ \\
$\tInr{\trm}$ & $1 + \cost{\trm}{\Gamma}$ \\
$\tSAMatch{\trm[3]}{\var[1]}{\trm[2]}{\var[2]}{\trm[1]}$ & $\tSAMatch{\eval{\trm[3]}{\Gamma}}{\var[1]'}{1 + \cost{\trm[3]}{\Gamma} + \cost{\trm[2]}{\var[1]=\var[1]', \Gamma}}{\var[2]'}{1 + \cost{\trm[3]}{\Gamma} + \cost{\trm[1]}{\var[2]=\var[2]', \Gamma}}$ \\
% $ $ & $ $ \\
% $\tReturn{\trm}$ & $1 + \cost{\trm}{\Gamma}$ \\
% $\tRun{(\trm \tBind \funabs{\var}{\trm[2]})}{\mathit{env}}$ & $1 + \cost{\mathit{env}}{\Gamma} + \cost{\tRun{\trm}{e}}{e=\mathit{env}, \Gamma} + \cost{\trm[2]}{\var = \eval{\tRun{\trm}{e}}{\Gamma}, e=\mathit{env}, \Gamma}$ \\
% $\tRun{(\tOne{\var}{\ty}{\Delta}\ d)}{\mathit{env}}$ & $1 + \cost{d}{\Gamma} + \cost{\tRun{\trm}{e}}{e=\mathit{env}, \Gamma} + \cost{\trm[2]}{\var = \eval{\tRun{\trm}{e}}{\Gamma}, e=\mathit{env}, \Gamma}$ \\
\end{tabular}

Because the implementations of the cotangent monoids are kept abstract in the term language used in the formalisation (the definitions can be found in the \texttt{spec.linear-types} module), \texttt{Term} has separate constructors for them and thus they get a separate treatment in the cost model.
To understand the costs here, refer to the semantics of these operations given in \cref{sec:sparsity} ($\LPair{\ty}{\ty[2]}$) and \cref{fig:naive-api} ($\LSum{\ty}{\ty[2]}$, which was unchanged in \cref{sec:efficiency-problems}).
If you are reading along with \texttt{eval} in the formalisation, note note that \texttt{snd (zerov $\tau$)} is always 1 currently; this `1' is inlined in the costs in the table below.

\begin{tabular}{l|l}
\textbf{Term $\trm$} & \textbf{Cost:} $\cost{\trm}{\Gamma}$ \\\hline
$\tlUnit$ & $1$ \\
$\tlPair{\trm}{\trm[2]}$ & $1 + \cost{\trm}{\Gamma} + \cost{\trm[2]}{\Gamma}$ \\
$\tlFst{\trm}$ & $\tSAMatchMaybe{\eval{\trm}{\Gamma}}{1 + \cost{\trm}{\Gamma} + 1 \qquad\textit{(additional `1' to compute the $\tZero$)}}{\_}{1 + \cost{\trm}{\Gamma}}$ \\
$\tlSnd{\trm}$ & $\tSAMatchMaybe{\eval{\trm}{\Gamma}}{1 + \cost{\trm}{\Gamma} + 1 \qquad\textit{(idem)}}{\_}{1 + \cost{\trm}{\Gamma}}$ \\
$\tlInl{\trm}$ & $1 + \cost{\trm}{\Gamma}$ \\
$\tlInr{\trm}$ & $1 + \cost{\trm}{\Gamma}$ \\
$\tlCast{\trm}$ & $\tSAMatchMaybe{\eval{\trm}{\Gamma}}{1 + \cost{\trm}{\Gamma} + 1 \qquad\textit{(idem)}}{\var}{\tSMatch{\var}{\_}{1 + \cost{\trm}{\Gamma}}{\_}{\tError}}$ \\
$\trCast{\trm}$ & $\tSAMatchMaybe{\eval{\trm}{\Gamma}}{1 + \cost{\trm}{\Gamma} + 1 \qquad\textit{(idem)}}{\var}{\tSMatch{\var}{\_}{\tError}{\_}{1 + \cost{\trm}{\Gamma}}}$ \\
$\smash{\tZero[\LPair{\ty}{\ty[2]}]}$ & $1$ \\
$\smash{\tZero[\LSum{\ty}{\ty[2]}]}$ & $1$
\end{tabular}

As for $\tZero[\LR]$: the formalisation defines a primitive operation $\texttt{LZERO} : \LUnit \ra \LR$, hence computing $\tZero[\LR]$ takes $\cost{\texttt{LZERO}(\tlUnit)}{\Gamma} = 1 + 1 = 2$ steps.
The choice for this design was fairly arbitrary.

Finally, for the local accumulation monad ($\EnvM{}{}$; the specific implementation in Agda is called \texttt{LACM}), the situation is slightly more complex because the cost of a particular monadic computation depends on the incoming derivative vector, which is not known at the point where the methods are invoked.
Thus the Agda code splits the cost model in two parts:
\begin{enumerate}
\item \label{item:cost-model-monad-impl}
  The implementation of \texttt{LACM} tracks the number of steps taken within the monad methods, as well as in the continuation of $(\tBind)$.
  This is possible because we modified the type of $(\tBind)$ (\texttt{bind}) as follows, abridged from the Agda code:
  \[ \text{bind} : \texttt{LACM}\ \Gamma\ \ty[2] \ra (\ty[2] \ra \texttt{LACM}\ \Gamma\ \ty \times \mathbb Z) \ra \texttt{LACM}\ \Gamma\ \ty \]
  That is to say, the continuation additionally returns the number of steps taken therein.
  This accumulated total number of steps is finally returned as the cost of calling $\tRun{}{}$ on the whole computation.
  The intended semantics is that the monadic computation does not actually run until, well, $\tRun{}{}$-ning it, at which point the computation runs to completion, collecting the number of steps taken, which then gets returned.
\item \label{item:cost-model-monad-eval}
  The evaluator does not know anything about the internals of the monad, and simply accounts constant cost for adding the operation in question to the pending computation in memory --- except for $\tRun{}{}$, where of course the returned cost is added to the total.
\end{enumerate}
The costs in point (\ref{item:cost-model-monad-impl}) are available to the proof through a list of \emph{properties} about the monad; the actual monad implementation is hidden through the use of an \texttt{abstract} block.
The types of the monad methods as the proof sees them are as follows:
\begin{align*}
  \texttt{pure} &: \ty \ra \texttt{LACM}\ \Gamma\ \ty \\
  \texttt{bind} &: \texttt{LACM}\ \Gamma\ \ty[2] \ra (\ty[2] \ra \Pair{\texttt{LACM}\ \Gamma\ \ty}{\Z}) \ra \texttt{LACM}\ \Gamma\ \ty \\
  \texttt{run} &: \texttt{LACM}\ \Gamma\ \ty \ra \smash{\ET{\Gamma}} \ra \Pair{\ty}{(\Pair{\smash{\ET{\Gamma}}}{\Z})} \\
  \texttt{add} &: \texttt{Idx}\ \Gamma\ \ty \ra \ty \ra \texttt{LACM}\ \Gamma\ \Unit \\
  \texttt{scope} &: \ty \ra \texttt{LACM}\ (\ty \mathbin{::} \Gamma)\ \ty[2] \ra \texttt{LACM}\ \Gamma\ (\Pair{\ty}{\ty[2]})
\end{align*}
The mentioned properties can be found in the \texttt{spec.LACM} module in \cref{app:agda-spec}.

% Should we typeset them here? Lots of work, and redundant.

% \begin{tabular}{l|l}
% \textbf{Name} & \textbf{Property} \\\hline
% \texttt{run-pure} &
%   $\begin{array}{@{}l@{}}
%     \forall \Gamma\, \ty\, (\var : \ty)\, (\mathit{env} : \ET{\Gamma}). \\
%     \quad \fst\ (\snd\ (\texttt{run}\ (\texttt{pure}\ \var)\ \mathit{env})) = \mathit{env} \\
%     \quad \land\ \snd\ (\snd\ (\texttt{run}\ (\texttt{pure}\ \var)\ \mathit{env})) = 1 + |\Gamma| + 1 \\
%   \end{array}$ \\\hline
% \texttt{run-bind} &
%   $\begin{array}{@{}l@{}}
%     \forall \Gamma\, \ty[2]\, \ty\, (m_1 : \texttt{LACM}\ \Gamma\ \ty[2])\, (k : \ty[2] \ra \Pair{\texttt{LACM}\ \Gamma\ \ty}{\Z})\, (\mathit{env} : \ET{\Gamma}). \\
%     \quad \textbf{let}\ (r_1, (\mathit{env}', c)) = \texttt{run}\ m_1\ \mathit{env} \\
%     \quad \hphantom{\textbf{let}\ } m_2 = 
%     \quad \fst\ (\snd\ (\texttt{run}\ (\texttt{bind}\ m_1\ k)\ \mathit{env})) = ? \\
%     \quad \land\ c = c_1 + c_{\text{call}} + c_2 - |\Gamma| \\
%   \end{array}$ \\\hline
% \end{tabular}

The costs accounted by \texttt{eval} are as follows, unsurprising as usual:

\begin{tabular}{l|l}
\textbf{Term $\trm$} & \textbf{Cost:} $\cost{\trm}{\Gamma}$ \\\hline
$\tReturn{\trm}$ & $1 + \cost{\trm}{\Gamma}$ \\
$\trm[2] \tBind \trm$ & $1 + \cost{\trm[2]}{\Gamma} + \cost{\trm}{\Gamma}$ \\
$\tRun{\trm[2]}{\trm}$ & $1 + \cost{\trm[2]}{\Gamma} + \cost{\trm}{\Gamma} + \snd\ (\snd\ (\texttt{run}\ (\eval{\trm[2]}{\Gamma})\ (\eval{\trm}{\Gamma})))$ \\
$\tOne{}{}{}\ \trm$ & $1 + \cost{\trm}{\Gamma}$ \\
$\tScope{}{}\ \trm[2]\ \trm$ & $1 + \cost{\trm[2]}{\Gamma} + \cost{\trm}{\Gamma}$
\end{tabular}

\section{Agda Formalisation Specification}\label{app:agda-spec}

The specification of the Agda proof, which provides only those definitions necessary to state the main complexity theorems (but not prove them), follows on subsequent pages.
This specification consists of three modules; each module starts on a new page.

\ifACMAnonymous
  {The full formalisation is also included in the supplementary material.}
  {The full formalisation can be found at \agdagithublink.}

% Print the agda-generated HTML using Firefox, with the following settings:
% - Paper size: A4
% - Scale 100%
% - Custom margins: top 1.03, bottom 0.80, left 0.71, right 0.71 inches
% - Print NO headers and footers


\section*{Supplementary material}

```agda
module spec.linear-types where

open import Agda.Builtin.Float using (Float; primFloatPlus)
open import Agda.Builtin.Maybe using (Maybe; nothing; just)
open import Agda.Builtin.Sigma using (_,_; fst; snd)
open import Agda.Builtin.Unit using (⊤; tt)

open import Data.List using (List; []; _::_)
open import Data.Integer using (ℤ; _+_; + _)
open import Data.Product using (_×_)
open import Data.Sum using (_⊎_; inj₁; inj₂)


-- The linear (i.e. monoidal) types. These types have a monoid structure, and
-- have a potential function (φ) defined on them.
data LTyp : Set where
  LUn LR : LTyp
  _:*!_ : LTyp -> LTyp -> LTyp
  _:+!_ : LTyp -> LTyp -> LTyp

-- A linear typing environment is a list of linear types.
LEnv : Set
LEnv = List LTyp

-- The representation (semantics) of the linear types; the representation of
-- normal types follows in `spec.agda`.
LinRep : LTyp -> Set
LinRep LUn = ⊤
LinRep LR = Float
LinRep (σ :*! τ) = Maybe (LinRep σ × LinRep τ)
LinRep (σ :+! τ) = Maybe (LinRep σ ⊎ LinRep τ)

-- Linear environment tuple: a tuple of all the types in a linear environment.
-- This is used to pass a linear environment as a _value_ into, and out of,
-- the monadic computation in the target program.
LEtup : LEnv -> Set
LEtup [] = ⊤
LEtup (τ :: Γ) = LinRep τ × LEtup Γ

-- An index into a typing environment
data Idx {n} {typ : Set n} : List typ -> typ -> Set n where
  Z : {e : List typ} {τ : typ} -> Idx (τ :: e) τ
  S : {e : List typ} {τ τ' : typ} -> Idx e τ -> Idx (τ' :: e) τ

one : ℤ
one = + 1

-- The zero part of the monoid structure of the linear types. Aside from
-- returning the value, this also returns an integer recording the number of
-- evaluation steps taken during the operation. This integer is used for
-- complexity analysis.
-- Because zerov and plusv are not implemented in terms of evaluation of terms,
-- we simply use an approximation here that is proportional to the actual
-- number of steps. In practice this means that we can take c_φ = 1.
zerov : (τ : LTyp) -> LinRep τ × ℤ
zerov LUn = tt , one
zerov LR = 0.0 , one
zerov (σ :*! τ) = nothing , one
zerov (σ :+! τ) = nothing , one


-- The addition part of the monoid structure of the linear types. Similarly,
-- the number of evaluation steps is returned.
--
```

```agda
-- For sum types, we return zero on adding incompatible values instead of
-- throwing an error. This prevents D2τ (σ :+ τ) from being a monoid, but of
-- course, if a proper implementation that would throw errors does not in fact
-- error on a given input program, the implementation here would not introduce
-- values that violate the monoid laws either.
--
-- In particular, because the dependently-typed variant of CHAD is correct (see
-- Nunes, Vákár. "CHAD for expressive total languages." MSCS 2023 (to appear)
-- (also arXiv:2110.00446)), there is an external proof that those error cases
-- would be impossible, and thus that the cases that violate the monoid laws
-- here are also impossible.
--
-- We put up with this infelicity because it allows us to avoid having to model
-- partiality in our language, which is no fundamental issue but introduces a
-- large amount of administration everywhere that makes the proof harder to
-- read and to write.
plusv : (τ : LTyp) -> LinRep τ -> LinRep τ -> LinRep τ × ℤ
plusv LUn tt tt = tt , one
plusv LR x y = primFloatPlus x y , one
plusv (σ :*! τ) nothing y = y , one
plusv (σ :*! τ) x nothing = x , one
plusv (σ :*! τ) (just (x , y)) (just (x' , y')) =
  let xr , cx = plusv σ x x'
      yr , cy = plusv τ y y'
  in just (xr , yr) , one + cx + cy
plusv (σ :+! τ) x nothing = x , one
plusv (σ :+! τ) nothing y = y , one
plusv (σ :+! τ) (just (inj₁ x)) (just (inj₁ y)) =
  let z , cz = plusv σ x y
  in just (inj₁ z) , one + cz
plusv (σ :+! τ) (just (inj₂ x)) (just (inj₂ y)) =
  let z , cz = plusv τ x y
  in just (inj₂ z) , one + cz
plusv (σ :+! τ) _ _ = nothing , one   -- NOTE: a proper implementation would error here.


-- Add the value 'val' into the position 'idx' in the environment tuple.
addLEτ : {Γ : LEnv} {τ : LTyp} -> (idx : Idx Γ τ) -> (val : LinRep τ) -> LEtup Γ -> LEtup
addLEτ Z val (x , env) = fst (plusv _ val x) , env
addLEτ (S i) val (x , env) = x , addLEτ i val env


-- Project a value out of an environment tuple.
_Eτ!!_ : {Γ : LEnv} {τ : LTyp} -> LEtup Γ -> Idx Γ τ -> LinRep τ
(x , env) Eτ!! Z = x
(x , env) Eτ!! (S i) = env Eτ!! i
```

```agda
module spec.LACM where

open import Agda.Builtin.Sigma using (_,_; fst; snd)
open import Agda.Builtin.Unit using (⊤; tt)

open import Data.Integer using (ℤ; _+_; +_; _-_)
open import Data.List using (_::_; length)
open import Data.Product using (_×_)
open import Relation.Binary.PropositionalEquality using (_≡_; refl; cong; trans)

open import Data.Integer.Solver using (module +-*-Solver)
open +-*-Solver using (solve; _:+_; _:-_; :-_; con; _:=_)

open import spec.linear-types


-- In Agda, an 'abstract' block prevents the contained definitions from
-- _reducing_ outside of the block. In effect, this means that outside of the
-- 'abstract' block, only the type signatures of its definitions are visible,
-- not the bodies. We use this to ensure that the complexity proof depends only
-- on the semantics and the complexities of the LACM interpretation, not its
-- actual implementation.
abstract
  -- Local accumulation monad.
  LACM : LEnv -> Set -> Set
  LACM Γ a = LEtup Γ -> a × LEtup Γ × ℤ

  -- The methods of the monad, including pure and bind.

  pure : ∀ {Γ : LEnv} {a : Set} -> a -> LACM Γ a
  pure x e = x , e , one

  -- Note that the continuation should also return the cost of evaluating the
  -- continuation; this cost will be included in the cost returned by 'run'
  -- when handling the top-level monadic computation.
  bind : ∀ {Γ : LEnv} {a b : Set} -> LACM Γ a -> (a -> LACM Γ b × ℤ) -> LACM Γ b
  bind f g e =
    let x , e1 , c1 = f e
        m2 , ccall = g x
        y , e2 , c2 = m2 e1
    in y , e2 , one + c1 + ccall + c2

  -- Returns computation result, the output environment, and the cost of
  -- evaluating the monadic computation.
  run : ∀ {Γ : LEnv} {a : Set} -> LACM Γ a -> LEtup Γ -> a × LEtup Γ × ℤ
  run {Γ} f e =
    let r , e' , c = f e
    in r , e' , one + + length Γ + c

  -- Add the given value to the value at the given index in the state.
  add : ∀ {Γ : LEnv} {τ : LTyp} -> Idx Γ τ -> LinRep τ -> LACM Γ ⊤
  add {τ = τ} Z x (y , e) =
    let z , cz = plusv τ x y
    in tt , (z , e) , one + cz
  add (S i) x (y , e) =  -- supposed to be O(1) access, so we don't cost this traversal
    let r , e' , c = add i x e
    in r , (y , e') , c

  -- Temporarily add a new cell to the state. The final value of this cell is
  -- returned inside the monad when 'scope' finishes.
  scope : ∀ {Γ : LEnv} {τ : LTyp} {a : Set}
        -> LinRep τ -> LACM (τ :: Γ) a -> LACM Γ (LinRep τ × a)
  scope x f e =
```

```
      let r , (x' , e') , c = f (x , e)
      in (x' , r) , e' , one + c

-- Properties about the monad methods that we need when reasoning about LACM
-- in the complexity proof. These four lemmas "push" 'run' down inside
-- 'pure', 'bind', 'add' and 'scope', and thereby define properties both
-- about the _semantics_ of the monad, as well as its _complexity_.
--
-- The cost counts are used abstractly here: a cost of 1 indicates some O(1)
-- work, not necessarily one single step in the underlying machine. Or,
-- alternatively: it is assumed that the underlying machine has support for
-- these (side-effectful) operations and can execute them in the indicated
-- number of "steps". Ultimately, this matters little, because this proof is
-- only concerned with complexity, not absolute performance (which would need
-- to be benchmarked on an actual machine anyway).

run-pure : ∀ {Γ : LEnv} {a : Set} -> (x : a)
        -> (env : LEtup Γ)
        -> let _ , env' , c = run {Γ} (pure x) env
           in (env' ≡ env) × (c ≡ one + + length Γ + one)

run-bind : ∀ {Γ : LEnv} {a b : Set} -> (m1 : LACM Γ a) -> (k : a -> LACM Γ b × ℤ)
        -> (env : LEtup Γ)
        -> let _ , env' , c = run (bind m1 k) env
               r1 , env1 , c1 = run m1 env
               m2 , ccall = k r1
               r2 , env2 , c2 = run m2 env1
           in (env' ≡ env2) × (c ≡ c1 + ccall + c2 - + length Γ)

run-add : ∀ {Γ : LEnv} {τ : LTyp}
        -> (idx : Idx Γ τ) -> (val : LinRep τ)
        -> (env : LEtup Γ)
        -> let tt , env' , c = run (add idx val) env
           in (env' ≡ addLEτ idx val env)
              × (c ≡ + 2 + snd (plusv τ val (env Eτ!! idx)) + + length Γ)

run-scope : ∀ {Γ : LEnv} {a : Set} {τ : LTyp}
         -> (m : LACM (τ ∷ Γ) a) -> (inval : LinRep τ)
         -> (env : LEtup Γ)
         -> let (outval1 , x1) , env1 , c1 = run (scope inval m) env
                x2 , (outval2 , env2) , c2 = run m (inval , env)
            in (x1 ≡ x2) × (outval1 ≡ outval2) × (env1 ≡ env2) × (c1 ≡ c2)

-- Proofs for the above properties -- these are elided in the paper appendix.
```

```agda
module spec where

open import Agda.Builtin.Bool using (true; false)
open import Agda.Builtin.Float
  using (Float; primFloatPlus; primFloatTimes; primFloatNegate; primFloatLess)
open import Agda.Builtin.Maybe using (nothing; just)
open import Agda.Builtin.Sigma using (_,_; fst; snd)
open import Agda.Builtin.Unit using (⊤; tt)

open import Data.Nat using (ℕ) renaming (_+_ to _+ℕ_)
open import Data.Fin using (Fin; zero; suc)
open import Data.List using (List; []; _∷_; length; map)
open import Data.Integer using (ℤ; _+_; _-_; _*_; -_; +_; _≤_)
open import Data.Product using (_×_)
open import Data.Sum using (_⊎_; inj₁; inj₂)
open import Function.Base using (id; _$_; _∘_; case_of_)
open import Relation.Binary.PropositionalEquality using (_≡_; refl; sym; subst; cong)

open import spec.linear-types public
import spec.LACM as LACM
open LACM using (LACM)


-------------------- UTILITY FUNCTIONS -----------------------------------------

-- Project from a list with a bounded index into that list. Not sure why this
-- is not in the standard library for lists.
_!!_ : ∀ {n} {a : Set n} -> (l : List a) -> Fin (length l) -> a
(x ∷ xs) !! zero = x
(x ∷ xs) !! suc i = xs !! i


-------------------- TYPES -----------------------------------------------------
-- Linear types were already defined in spec.linear-types; here are the
-- regular, non-linear types.

-- Types are indexed by whether they are primal types (i.e. types that occur in
-- the source program) or dual types (i.e. types that occur in the target
-- program).
data PDTag : Set where
  Pr Du : PDTag

-- The types of the object language. The source language is typed by
-- 'Typ Pr', which allows only a few types. The target language is typed by
-- 'Typ Du', which allows not only the source language types but also a couple
-- of other types, including functions and the linear types.
data Typ : PDTag -> Set where
  Un Inte R : ∀ {tag} -> Typ tag
  _:*_ : ∀ {tag} -> Typ tag -> Typ tag -> Typ tag
  _:+_ : ∀ {tag} -> Typ tag -> Typ tag -> Typ tag

  _:->_ : Typ Du -> Typ Du -> Typ Du
  -- Environment vector monad. This is the same EVM as in the paper; the
  -- implementation is LACM.
  EVM : LEnv -> Typ Du -> Typ Du

  -- The linear types (with embedded potential)
  Lin : LTyp -> Typ Du

-- A normal typing environment is a list of types.
Env : PDTag -> Set
Env tag = List (Typ tag)
```

```
-- The representation / semantics of our types. LinRep from spec.linear-types
-- does the same for the linear types LTyp.
Rep : ∀ {tag} -> Typ tag -> Set
Rep Un = ⊤
Rep Inte = ℤ
Rep R = Float
Rep (σ :* τ) = Rep σ × Rep τ
Rep (σ :+ τ) = Rep σ ⊎ Rep τ
Rep (σ :-> τ) = Rep σ -> Rep τ × ℤ
Rep (EVM Γ τ) = LACM Γ (Rep τ)
Rep (Lin τ) = LinRep τ


-- Convert a type from a source-language type to a target-language type. This
-- function is operationally the identity.
dut : Typ Pr -> Typ Du
dut Un = Un
dut Inte = Inte
dut R = R
dut (σ :* τ) = dut σ :* dut τ
dut (σ :+ τ) = dut σ :+ dut τ


-- The embedded counterpart of LEtup: a tuple of all the types in a linear
-- environment. This is used to pass a linear environment as a _value_ into,
-- and out of, the monadic computation in the target program.
LEτ : LEnv -> Typ Du
LEτ [] = Un
LEτ (τ :: Γ) = Lin τ :* LEτ Γ


-- LEtup and ⟦LEτ⟧ are the same thing.
LEtup-eq-LEτ : (Γ : LEnv) -> Rep (LEτ Γ) ≡ LEtup Γ
LEtup-eq-LEτ [] = refl
LEtup-eq-LEτ (τ :: Γ) rewrite LEtup-eq-LEτ Γ = refl


LEtup-to-LEτ : (Γ : LEnv) -> Rep (LEτ Γ) -> LEtup Γ
LEtup-to-LEτ [] x = x
LEtup-to-LEτ (τ :: Γ) (x , env) = x , LEtup-to-LEτ Γ env


LEτ-to-LEtup : (Γ : LEnv) -> LEtup Γ -> Rep (LEτ Γ)
LEτ-to-LEtup [] x = x
LEτ-to-LEtup (τ :: Γ) (x , env) = x , LEτ-to-LEtup Γ env



------------------- PRIMITIVE OPERATIONS -------------------------------

-- The primitive operations in our object language. Again, some operations are
-- available both in the source and in the target language, whereas others (the
-- 'Du'-indexed ones) are available only in the target language.
data Primop : (tag : PDTag) -> (σ τ : Typ) -> Set where
  ADD  : ∀ {tag} -> Primop tag (R :* R) R
  MUL  : ∀ {tag} -> Primop tag (R :* R) R
  NEG  : ∀ {tag} -> Primop tag R R
  LIT  : ∀ {tag} -> Float -> Primop tag Un R
  IADD : ∀ {tag} -> Primop tag (Inte :* Inte) Inte
  IMUL : ∀ {tag} -> Primop tag (Inte :* Inte) Inte
  INEG : ∀ {tag} -> Primop tag Inte Inte
  -- sign: (negative or positive) or zero/NaN
  SIGN : ∀ {tag} -> Primop tag R ((Un :+ Un) :+ Un)

  LZERO  : Primop Du (Lin LUn) (Lin LR)
  LADD   : Primop Du (Lin LR :* Lin LR) (Lin LR)
  LSCALE : Primop Du (Lin LR :* R) (Lin LR)
  LNEG   : Primop Du (Lin LR) (Lin LR)

-- Semantics of the primitive operations.
```

```
evalprim : ∀ {tag} {σ τ} -> Primop tag σ τ -> Rep σ -> Rep τ
evalprim ADD (x , y) = primFloatPlus x y
evalprim MUL (x , y) = primFloatTimes x y
evalprim NEG x = primFloatNegate x
evalprim (LIT x) tt = x
evalprim IADD (x , y) = x + y
evalprim IMUL (x , y) = x * y
evalprim INEG x = - x
evalprim SIGN x =
  case primFloatLess x 0.0 of
    λ where true  -> inj₁ (inj₁ tt)
            false -> case primFloatLess 0.0 x of
                       λ where true  -> inj₁ (inj₂ tt)
                               false -> inj₂ tt
evalprim LZERO tt = 0.0
evalprim LADD (x , y) = primFloatPlus x y
evalprim LSCALE (x , y) = primFloatTimes x y
evalprim LNEG x = primFloatNegate x


-------------------- OBJECT LANGUAGE -------------------------------------

-- The object language. The source language and the target language are both
-- expressed using the same Term data type, just with a different index: a
-- source term is of type 'Term Pr Γ τ', whereas a target term is of type
-- 'Term Du Γ τ'.
data Term : (tag : PDTag) -> (Γ : Env tag) -> (τ : Typ tag) -> Set where
  var   : ∀ {tag} {Γ : Env tag} {τ : Typ tag}
        -> Idx Γ τ -> Term tag Γ τ
  let'  : ∀ {tag} {Γ : Env tag} {σ τ : Typ tag}
        -> Term tag Γ σ -> Term tag (σ :: Γ) τ -> Term tag Γ τ

  prim  : ∀ {tag} {Γ : Env tag} {σ τ : Typ tag}
        -> Primop tag σ τ -> Term tag Γ σ -> Term tag Γ τ

  unit  : ∀ {tag} {Γ : Env tag}
        -> Term tag Γ Un

  pair  : ∀ {tag} {Γ : Env tag} {σ τ : Typ tag}
        -> Term tag Γ σ -> Term tag Γ τ -> Term tag Γ (σ :* τ)
  fst'  : ∀ {tag} {Γ : Env tag} {σ τ : Typ tag}
        -> Term tag Γ (σ :* τ) -> Term tag Γ σ
  snd'  : ∀ {tag} {Γ : Env tag} {σ τ : Typ tag}
        -> Term tag Γ (σ :* τ) -> Term tag Γ τ

  inl   : ∀ {tag} {Γ : Env tag} {σ τ : Typ tag}
        -> Term tag Γ σ -> Term tag Γ (σ :+ τ)
  inr   : ∀ {tag} {Γ : Env tag} {σ τ : Typ tag}
        -> Term tag Γ τ -> Term tag Γ (σ :+ τ)
  case' : ∀ {tag} {Γ : Env tag} {σ τ ρ : Typ tag}
        -> Term tag Γ (σ :+ τ)
        -> Term tag (σ :: Γ) ρ -> Term tag (τ :: Γ) ρ
        -> Term tag Γ ρ

  -- The Γ' is the closure of the lambda. We model this explicitly because the
  -- cost of evaluating 'lam' is linear in the size of its closure, so it is
  -- worth keeping it small.
  lam   : {Γ : Env Du} {τ : Typ Du}
        -> {σ : Typ Du}
        -> (Γ' : Env Du) -> ({ρ : Typ Du} -> Idx Γ' ρ -> Idx Γ ρ)  -- Γ' is a subset of Γ
        -> Term Du (σ :: Γ') τ -> Term Du Γ (σ :-> τ)
  app   : {Γ : Env Du} {σ τ : Typ Du}
        -> Term Du Γ (σ :-> τ) -> Term Du Γ σ -> Term Du Γ τ
```

```
pureevm : {Γ : Env Du} {Γ' : LEnv} {τ : Typ Du}
       -> Term Du Γ τ -> Term Du Γ (EVM Γ' τ)
bindevm : {Γ : Env Du} {Γ' : LEnv} {σ τ : Typ Du}
       -> Term Du Γ (EVM Γ' σ) -> Term Du Γ (σ :-> EVM Γ' τ) -> Term Du Γ (EVM Γ' τ)
runevm : {Γ : Env Du} {Γ' : LEnv} {τ : Typ Du}
       -> Term Du Γ (EVM Γ' τ) -> Term Du Γ (LEτ Γ') -> Term Du Γ (τ :* LEτ Γ')
addevm : {Γ : Env Du} {Γ' : LEnv} {τ : LTyp}
       -> Idx Γ' τ -> Term Du Γ (Lin τ) -> Term Du Γ (EVM Γ' Un)
scopeevm : {Γ : Env Du} {Γ' : LEnv} {τ : LTyp} {σ : Typ Du}
       -> Term Du Γ (Lin τ) -> Term Du Γ (EVM (τ :: Γ') σ)
       -> Term Du Γ (EVM Γ' (Lin τ :* σ))

lunit : {Γ : Env Du}
      -> Term Du Γ (Lin LUn)

lpair : {Γ : Env Du} {σ τ : LTyp}
      -> Term Du Γ (Lin σ) -> Term Du Γ (Lin τ) -> Term Du Γ (Lin (σ :*! τ))
lfst' : {Γ : Env Du} {σ τ : LTyp}
      -> Term Du Γ (Lin (σ :*! τ)) -> Term Du Γ (Lin σ)
lsnd' : {Γ : Env Du} {σ τ : LTyp}
      -> Term Du Γ (Lin (σ :*! τ)) -> Term Du Γ (Lin τ)
lpairzero : {Γ : Env Du} {σ τ : LTyp}
         -> Term Du Γ (Lin (σ :*! τ))

linl : {Γ : Env Du} {σ τ : LTyp}
     -> Term Du Γ (Lin σ) -> Term Du Γ (Lin (σ :+! τ))
linr : {Γ : Env Du} {σ τ : LTyp}
     -> Term Du Γ (Lin τ) -> Term Du Γ (Lin (σ :+! τ))
lcastl : {Γ : Env Du} {σ τ : LTyp}
       -> Term Du Γ (Lin (σ :+! τ))
       -> Term Du Γ (Lin σ)
lcastr : {Γ : Env Du} {σ τ : LTyp}
       -> Term Du Γ (Lin (σ :+! τ))
       -> Term Du Γ (Lin τ)
lsumzero : {Γ : Env Du} {σ τ : LTyp}
         -> Term Du Γ (Lin (σ :+! τ))


-------------------- OBJECT LANGUAGE UTILITIES --------------------------------
-- Utilities for working with the object language: weakening and some
-- alternative forms of constructors.

-- A data type representing weakenings.
--
-- The reason we have this explicit representation of reindexing mappings, as
-- opposed to a general sink function with the following type:
--   sink : {Γ Γ' : Env tag} {τ : Typ tag}
--       -> ({σ : Typ tag} -> Idx Γ σ -> Idx Γ' σ)
--       -> Term tag Γ τ -> Term tag Γ' τ
-- is that with the above representation we'd need (a very weak form of?)
-- functional extensionality to use certain lemmas in the complexity proof. The
-- reason for that is that we'd like to use multiple lemmas about the same
-- things together, and all of those lemmas return facts about terms that
-- normalise to the same thing but contain uses of 'sink' applied to unknown
-- indices inside them. If 'sink' took a function argument, then proving
-- equality here would involve proving equality of functions given equal
-- syntactic representation, which Agda does not do, despite being much weaker
-- than full functional extensionality.
--
-- This 'Weakening' type does not model all such Idx->Idx functions, but since
-- we need only a very limited set of them and this data type is sufficient to
-- describe those, we can make our lives easy and work with this simple
-- representation.
data Weakening {tag} : (Γ Γ' : Env tag) -> Set where
```

```
  WEnd  : {Γ : Env tag} -> Weakening Γ Γ
  WCut  : {Γ' : Env tag} -> Weakening [] Γ'
  WCopy : {Γ Γ' : Env tag} {τ : Typ tag} -> Weakening Γ Γ' -> Weakening (τ :: Γ) (τ :: Γ')
  WSkip : {Γ Γ' : Env tag} {τ : Typ tag} -> Weakening Γ Γ' -> Weakening Γ (τ :: Γ')


-- Apply a weakening to a single index.
weaken-var
  : ∀ {tag} {Γ Γ' : Env tag}
  -> (w : Weakening Γ Γ')
  -> {τ : Typ tag}
  -> Idx Γ τ
  -> Idx Γ' τ
weaken-var WEnd i = i
weaken-var (WCopy w) Z = Z
weaken-var (WCopy w) (S i) = S (weaken-var w i)
weaken-var (WSkip w) i = S (weaken-var w i)


-- Sink a term using a weakening (an index remapping). A typical special case
-- is in 'sink1' below.
sink : ∀ {tag} {Γ Γ' : Env tag}
     -> {τ : Typ tag}
     -> Weakening Γ Γ'
     -> Term tag Γ τ
     -> Term tag Γ' τ
sink w (var i) = var (weaken-var w i)
sink w (let' e1 e2) = let' (sink w e1) (sink (WCopy w) e2)
sink w (lam Γ' inj e) = lam Γ' (weaken-var w ∘ inj) e
sink w (app e1 e2) = app (sink w e1) (sink w e2)
sink w (prim op e) = prim op (sink w e)
sink w unit = unit
sink w (pair e1 e2) = pair (sink w e1) (sink w e2)
sink w (fst' e) = fst' (sink w e)
sink w (snd' e) = snd' (sink w e)
sink w (inl e) = inl (sink w e)
sink w (inr e) = inr (sink w e)
sink w (case' e1 e2 e3) = case' (sink w e1) (sink (WCopy w) e2) (sink (WCopy w) e3)
sink w (pureevm e) = pureevm (sink w e)
sink w (bindevm e1 e2) = bindevm (sink w e1) (sink w e2)
sink w (runevm e1 e2) = runevm (sink w e1) (sink w e2)
sink w (addevm i e) = addevm i (sink w e)
sink w (scopeevm e1 e2) = scopeevm (sink w e1) (sink w e2)
sink w lunit = lunit
sink w (lpair e1 e2) = lpair (sink w e1) (sink w e2)
sink w (lfst' e) = lfst' (sink w e)
sink w (lsnd' e) = lsnd' (sink w e)
sink w lpairzero = lpairzero
sink w (linl e) = linl (sink w e)
sink w (linr e) = linr (sink w e)
sink w (lcastl e) = lcastl (sink w e)
sink w (lcastr e) = lcastr (sink w e)
sink w lsumzero = lsumzero


-- Add one additional free variable to the bottom of the term's free variable
-- list (here of type σ). This, for example, allows one to put a term under one
-- additional let-binding (whose variable is unused in the term).
sink1 : ∀ {tag} {Γ : Env tag} {σ τ : Typ tag} -> Term tag Γ τ -> Term tag (σ :: Γ) τ
sink1 = sink (WSkip WEnd)


-- Build a closure. The 'lam' constructor in Term represents the inclusion of
-- the (smaller) closure environment into the larger containing environment
-- with an index remapping function, but writing those inline is cumbersome.
-- It's easier to simply give a list of indices into the containing environment
-- that you want to include in the closure. This 'lamwith' function allows you
-- to do that; said list is the list 'vars'. 'α' is the argument type of the
```

```
-- lambda.
lamwith : {α : Typ Du} {Γ : Env Du} {τ : Typ Du}
        -> (vars : List (Fin (length Γ)))
        -> Term Du (α :: map (\i -> Γ !! i) vars) τ
        -> Term Du Γ (α :-> τ)
lamwith {_} {Γ} vars body =
  lam (map (Γ !!_) vars)
      (buildinj vars)
      body
  where
    buildidx : {Γ : Env Du} -> (i : Fin (length Γ)) -> Idx Γ (Γ !! i)
    buildidx {[]} ()
    buildidx {_ :: _} zero = Z
    buildidx {_ :: _} (suc i) = S (buildidx i)

    buildinj : {Γ : Env Du} {ρ : Typ Du}
             -> (vars : List (Fin (length Γ)))
             -> Idx (map (\i -> Γ !! i) vars) ρ -> Idx Γ ρ
    buildinj (i :: vars) Z = buildidx i
    buildinj (i :: vars) (S idx) = buildinj vars idx

-- 'bindevm' from Term is '>>=' of the environment vector monad EVM; this is
-- '>>'. 'a >> b' is expanded to 'let x = b in a >>= \_ -> x'. Note the
-- creation of a closure using 'lamwith' containing one entry, namely x.
thenevm : {Γ : LEnv} {Γ' : Env Du}
        -> Term Du Γ' (EVM Γ Un) -> Term Du Γ' (EVM Γ Un) -> Term Du Γ' (EVM Γ Un)
thenevm a b =
  let' b $
    bindevm (sink1 a) (lamwith (zero :: []) (var (S Z)))

-- Generic index retyping utility. An index of type τ into an environment Γ can
-- be retyped as an index of modified type into a modified environment.
convIdx : ∀ {n} {typ typ' : Set n} {Γ : List typ} {τ : typ}
        -> (f : typ -> typ')
        -> Idx Γ τ -> Idx (map f Γ) (f τ)
convIdx f Z = Z
convIdx f (S i) = S (convIdx f i)


-------------------- DIFFERENTIATION ----------------------------------------
-- Derivative type mappings and derivatives of primitive operations.

-- The primal type mapping, written D[τ]₁ in the paper.
D1τ : Typ Pr -> Typ Du
D1τ Un = Un
D1τ Inte = Inte
D1τ R = R
D1τ (σ :* τ) = D1τ σ :* D1τ τ
D1τ (σ :+ τ) = D1τ σ :+ D1τ τ

-- The dual type mapping, written D[τ]₂ in the paper. Dual types are linear
-- (i.e. have a monoid structure).
D2τ' : Typ Pr -> LTyp
D2τ' Un = LUn
D2τ' Inte = LUn
D2τ' R = LR
D2τ' (σ :* τ) = D2τ' σ :*! D2τ' τ
D2τ' (σ :+ τ) = D2τ' σ :+! D2τ' τ

-- Dual type as a target language type.
D2τ : Typ Pr -> Typ Du
D2τ τ = Lin (D2τ' τ)

-- Primal environment mapping. This is D[Γ]₁ in the paper.
```

```
D1Γ : Env Pr -> Env Du
D1Γ = map D1τ

-- Dual environment mapping. Recall LEτ from above. This is \overline{D[Γ]₂} in
-- the paper.
D2Γtup : Env Pr -> Typ Du
D2Γtup Γ = LEτ (map D2τ' Γ)

-- The codomain of the backpropagator of a differentiated program. 'EVM' is the
-- environment vector monad, instantiated with the local accumulation monad
-- LACM. 'D2Γ' is used in the type of 'chad' below.
D2Γ : Env Pr -> Typ Du
D2Γ Γ = EVM (map D2τ' Γ) Un

-- Convert a _value_ of source-language type to a primal value in the
-- differentiated world. Because D1τ is the identity for non-function types,
-- this function is also the identity on values.
primal : (τ : Typ Pr) -> Rep τ -> Rep (D1τ τ)
primal Un tt = tt
primal Inte x = x
primal R x = x
primal (σ :* τ) (x , y) = primal σ x , primal τ y
primal (σ :+ τ) (inj₁ x) = inj₁ (primal σ x)
primal (σ :+ τ) (inj₂ y) = inj₂ (primal τ y)

-- Our primitive operations work on types of which the primal is the same as
-- the original type. This is of course true for _all_ our types in this Agda
-- development, but this ceases to be true once we add function types to the
-- source language. In that situation, we would thus require that primitive
-- operations do not take or return function values.
niceprim : {σ τ : Typ Pr} -> Primop Pr σ τ -> (D1τ σ ≡ dut σ) × (D1τ τ ≡ dut τ)
niceprim ADD = refl , refl
niceprim MUL = refl , refl
niceprim NEG = refl , refl
niceprim (LIT _) = refl , refl
niceprim SIGN = refl , refl
niceprim IADD = refl , refl
niceprim IMUL = refl , refl
niceprim INEG = refl , refl

-- The reverse derivative of a primitive operation. The returned term takes as
-- input (i.e. uses in its environment) the primal of its argument and the
-- cotangent of its output, and returns the cotangent of its argument. This is
-- wrapped in a more easily-used form below in 'dprim'.
dprim' : {σ τ : Typ Pr} -> Primop Pr σ τ -> Term Du (D2τ τ :: D1τ σ :: []) (D2τ σ)
dprim' ADD = lpair (var Z) (var Z)
dprim' MUL = lpair (prim LSCALE (pair (var Z) (snd' (var (S Z)))))
                   (prim LSCALE (pair (var Z) (fst' (var (S Z)))))
dprim' NEG = prim LNEG (var Z)
dprim' (LIT x) = lunit
dprim' SIGN = prim LZERO lunit
dprim' IADD = lpair lunit lunit
dprim' IMUL = lpair lunit lunit
dprim' INEG = lunit

-- More easy to use version of dprim' above, using let-bindings to take the two
-- input terms as separate arguments.
dprim : {Γ : Env Du} {σ τ : Typ Pr}
      -> Primop Pr σ τ -> Term Du Γ (D1τ σ) -> Term Du Γ (D2τ τ) -> Term Du Γ (D2τ σ)
dprim op p d =
  let' p $
  let' (sink1 d) $
    sink (WCopy (WCopy WCut)) (dprim' op)
```

```
-- Retype a source-language primitive operation as a target-language one.
duPrim : {σ τ : Typ Pr} -> Primop Pr σ τ -> Primop Du (dut σ) (dut τ)
duPrim ADD = ADD
duPrim MUL = MUL
duPrim NEG = NEG
duPrim (LIT x) = LIT x
duPrim SIGN = SIGN
duPrim IADD = IADD
duPrim IMUL = IMUL
duPrim INEG = INEG


-- Retype a source-language primitive operation as a target-language one
-- working on primal values. This is all the same because of 'niceprim'.
d1Prim : {σ τ : Typ Pr} -> Primop Pr σ τ -> Primop Du (D1τ σ) (D1τ τ)
d1Prim {σ} {τ} op =
  subst (\t -> Primop Du t (D1τ τ)) (sym (fst (niceprim op))) $
  subst (\t -> Primop Du (dut σ) t) (sym (snd (niceprim op))) $
    duPrim op



-------------------- EVALUATION --------------------------------------------

-- A valuation / value environment: one value for each type in the typing
-- environment.
data Val (tag : PDTag) : Env tag -> Set where
  empty : Val tag []
  push : {Γ : Env tag} {τ : Typ tag} -> Rep τ -> Val tag Γ -> Val tag (τ :: Γ)

-- Project a value from a valuation.
valprj : ∀ {tag} {Γ : Env tag} {τ : Typ tag} -> (env : Val tag Γ) -> Idx Γ τ -> Rep τ
valprj (push x env) Z = x
valprj (push x env) (S i) = valprj env i

-- Map 'primal' over a valuation, lifting a valuation from the
-- non-differentiated world into a valuation in source-language world.
primalVal : {Γ : Env Pr} -> Val Pr Γ -> Val Du (D1Γ Γ)
primalVal empty = empty
primalVal {τ :: _} (push x env) = push (primal τ x) (primalVal env)

-- Given an inclusion of Γ' in Γ, and a valuation of Γ, build a valuation of
-- Γ'. This is used for evaluation of closures in 'eval' below.
buildValFromInj : ∀ {tag} {Γ Γ' : Env tag}
               -> ({ρ : Typ tag} -> Idx Γ' ρ -> Idx Γ ρ) -> Val tag Γ -> Val tag Γ'
buildValFromInj {Γ' = []} inj env = empty
buildValFromInj {Γ' = τ :: Γ'} inj env =
  push (valprj env (inj Z))
       (buildValFromInj (inj ∘ S) env)

-- The semantics of the term language. Aside from returning the evaluation
-- result, this also returns an integer recording the number of evaluation
-- steps taken during evaluation. This integer is used for complexity analysis.
eval : ∀ {tag} {Γ : Env tag} {τ : Typ tag} -> Val tag Γ -> Term tag Γ τ -> Rep τ × ℤ
eval env (var i) = valprj env i , one
eval env (let' rhs e) =
  let rhs' , crhs = eval env rhs
      e' , ce = eval (push rhs' env) e
  in e' , one + crhs + ce
eval env (lam Γ' inj e) =
  (\x -> eval (push x (buildValFromInj inj env)) e) , one + + (length Γ')
eval env (app e1 e2) =
  let f , cf = eval env e1
      x , cx = eval env e2
      y , cy = f x
  in y , one + cf + cx + cy
```

```
eval env (prim op e) =
  let e' , ce = eval env e
  in evalprim op e' , one + ce
eval env unit = tt , one
eval env (pair e1 e2) =
  let e1' , ce1 = eval env e1
      e2' , ce2 = eval env e2
  in (e1' , e2') , one + ce1 + ce2
eval env (fst' e) =
  let e' , ce = eval env e
  in fst e' , one + ce
eval env (snd' e) =
  let e' , ce = eval env e
  in snd e' , one + ce
eval env (inl e) =
  let e , ce = eval env e
  in inj₁ e , one + ce
eval env (inr e) =
  let e , ce = eval env e
  in inj₂ e , one + ce
eval env (case' e1 e2 e3) =
  let v , cv = eval env e1
  in case v of
       λ where (inj₁ x) -> let z , cz = eval (push x env) e2
                           in z , one + cv + cz
               (inj₂ y) -> let z , cz = eval (push y env) e3
                           in z , one + cv + cz
eval env (pureevm {Γ' = Γ'} e) =
  let e' , ce = eval env e
  in LACM.pure e' , one + ce
eval env (bindevm {Γ' = Γ'} e1 e2) =
  let e1' , ce1 = eval env e1
      e2' , ce2 = eval env e2
  in LACM.bind e1' e2' , one + ce1 + ce2
eval env (runevm {Γ' = Γ'} e1 e2) =
  let mf , ce1 = eval env e1
      denv , cdenv = eval env e2
      x , envctg , capp = LACM.run mf (LEtup-to-LEτ Γ' denv)
  in (x , LEτ-to-LEtup Γ' envctg) , one + ce1 + cdenv + capp
eval env (addevm {Γ' = Γ'} idx e) =
  let e' , ce = eval env e
  in LACM.add idx e' , one + ce
eval env (scopeevm e1 e2) =
  let e1' , ce1 = eval env e1
      e2' , ce2 = eval env e2
  in LACM.scope e1' e2' , one + ce1 + ce2
eval env lunit = tt , one
eval env (lpair e1 e2) =
  let e1' , ce1 = eval env e1
      e2' , ce2 = eval env e2
  in (just (e1' , e2')) , one + ce1 + ce2
eval env (lfst' {σ = σ} e) =
  let e' , ce = eval env e
  in case e' of
       λ where nothing -> let z , cz = zerov σ
                          in z , one + ce + cz
               (just (x , y)) -> x , one + ce
eval env (lsnd' {τ = τ} e) =
  let e' , ce = eval env e
  in case e' of
       λ where nothing -> let z , cz = zerov τ
                          in z , one + ce + cz
               (just (x , y)) -> y , one + ce
eval env lpairzero = nothing , one
```

```
eval env (linl e) =
  let e' , ce = eval env e
  in just (inj₁ e') , one + ce
eval env (linr e) =
  let e' , ce = eval env e
  in just (inj₂ e') , one + ce
eval env (lcastl {σ = σ} e) =
  let e' , ce = eval env e
  in case e' of
       λ where nothing ->
                    let z , cz = zerov σ
                    in z , one + ce + cz
                 (just (inj₁ x)) -> x , one + ce
                 (just (inj₂ _)) ->  -- NOTE: a proper implementation would error here.
                    let z , cz = zerov σ
                    in z , one + ce + cz
eval env (lcastr {τ = τ} e) =
  let e' , ce = eval env e
  in case e' of
       λ where nothing ->
                    let z , cz = zerov τ
                    in z , one + ce + cz
                 (just (inj₁ _)) ->  -- NOTE: a proper implementation would error here.
                    let z , cz = zerov τ
                    in z , one + ce + cz
                 (just (inj₂ y)) -> y , one + ce
eval env lsumzero = nothing , one

-- Project out the number of evaluation steps from 'eval'.
cost : ∀ {tag} {Γ : Env tag} {τ : Typ tag} -> Val tag Γ -> Term tag Γ τ -> ℤ
cost env e = snd (eval env e)


-------------------- CHAD ----------------------------------------------------

-- A term that produces the zero value of the given type.
zerot : {Γ : Env Du} -> (τ : Typ Pr) -> Term Du Γ (D2τ τ)
zerot Un = lunit
zerot Inte = lunit
zerot R = prim LZERO lunit
zerot (σ :* τ) = lpairzero
zerot (σ :+ τ) = lsumzero

-- The CHAD code transformation.
chad : {Γ : Env Pr} {τ : Typ Pr}
    -> Term Pr Γ τ
    -> Term Du (D1Γ Γ) (D1τ τ :* (D2τ τ :-> D2Γ Γ))
chad (var idx) =
  pair (var (convIdx D1τ idx))
       (lamwith [] (addevm (convIdx D2τ' idx) (var Z)))
chad (let' {σ = σ} e1 e2) =
  let' (chad e1) $
  let' (fst' (var Z)) $
  let' (sink (WCopy (WSkip WEnd)) (chad e2)) $
    pair (fst' (var Z))
         (lamwith (zero :: suc (suc zero) :: []) $
             bindevm
               (scopeevm (zerot σ) (app (snd' (var (S Z))) (var Z)))
               (lamwith (suc (suc zero) :: []) $
                  app (snd' (var (S Z))) (fst' (var Z))))
chad (prim op e) =
  let' (chad e) $
    pair (prim (d1Prim op) (fst' (var Z)))
         (lamwith (zero :: []) $
```

```
              app (snd' (var (S Z)))
                  (dprim op (fst' (var (S Z))) (var Z))
chad unit = pair unit (lamwith [] (pureevm unit))
chad (pair e1 e2) =
  let' (pair (chad e1) (chad e2)) $
    pair (pair (fst' (fst' (var Z)))
               (fst' (snd' (var Z))))
         (lamwith (zero :: []) $
            thenevm (app (snd' (fst' (var (S Z)))) (lfst' (var Z)))
                    (app (snd' (snd' (var (S Z)))) (lsnd' (var Z))))
chad (fst' {τ = τ} e) =
  let' (chad e) $
    pair (fst' (fst' (var Z)))
         (lamwith (zero :: []) $
            app (snd' (var (S Z)))
                (lpair (var Z) (zerot τ)))
chad (snd' {σ = σ} e) =
  let' (chad e) $
    pair (snd' (fst' (var Z)))
         (lamwith (zero :: []) $
            app (snd' (var (S Z)))
                (lpair (zerot σ) (var Z)))
chad (inl e) =
  let' (chad e) $
    pair (inl (fst' (var Z)))
         (lamwith (zero :: []) $
            app (snd' (var (S Z)))
                (lcastl (var Z)))
chad (inr e) =
  let' (chad e) $
    pair (inr (fst' (var Z)))
         (lamwith (zero :: []) $
            app (snd' (var (S Z)))
                (lcastr (var Z)))
chad (case' {σ = σ} {τ = τ} e1 e2 e3) =
  let' (chad e1) $
    case' (fst' (var Z))
      (let' (sink (WCopy (WSkip WEnd)) (chad e2)) $
         pair (fst' (var Z))
              (lamwith (zero :: suc (suc zero) :: []) $
                 bindevm
                   (scopeevm (zerot σ) (app (snd' (var (S Z))) (var Z)))
                   (lamwith (suc (suc zero) :: []) $
                      app (snd' (var (S Z))) (linl (fst' (var Z))))))))
      (let' (sink (WCopy (WSkip WEnd)) (chad e3)) $
         pair (fst' (var Z))
              (lamwith (zero :: suc (suc zero) :: []) $
                 bindevm
                   (scopeevm (zerot τ) (app (snd' (var (S Z))) (var Z)))
                   (lamwith (suc (suc zero) :: []) $
                      app (snd' (var (S Z))) (linr (fst' (var Z))))))))


--------------------- THE COMPLEXITY THEOREMS ------------------------------------

-- The potential function, here using c_φ = 1 because this suffices due to our
-- costing of plusv.
φ : (τ : LTyp) -> LinRep τ -> ℤ
φ LUn tt = one
φ LR _  = one
φ (σ :*! τ) nothing = one
φ (σ :*! τ) (just (x , y)) = one + φ σ x + φ τ y
φ (σ :+! τ) nothing = one
φ (σ :+! τ) (just (inj₁ x)) = one + φ σ x
```

```
φ (σ :+! τ) (just (inj₂ y)) = one + φ τ y

-- The potential function mapped over a list of linear types.
φ' : (Γ : LEnv) -> LEtup Γ -> ℤ
φ' [] tt = + 0
φ' (τ :: Γ) (x , env) = φ τ x + φ' Γ env

-- The statement of the complexity theorem including potential. A value of this
-- type (i.e. a proof of the theorem) is given in `chad-cost.agda`.
TH1-STATEMENT : Set
TH1-STATEMENT =
  {Γ : Env Pr} {τ : Typ Pr}
  -> (env : Val Pr Γ)
  -> (ctg : Rep (D2τ τ))
  -> (denvin : LEtup (map D2τ' Γ))
  -> (t : Term Pr Γ τ)
  -> let (_primal , bp) , crun = eval (primalVal env) (chad t)
         envf , ccall = bp ctg
         tt , denvout , cmonad = LACM.run envf denvin
     in
     crun
     + ccall
     + cmonad
     - φ (D2τ' τ) ctg
     - φ' (map D2τ' Γ) denvin
     + φ' (map D2τ' Γ) denvout
     - + length Γ
     ≤ + 34 * cost env t

-- In th2 we bound φ by the size of the incoming cotangent. This measures the
-- size of a cotangent value.
size : (τ : LTyp) -> LinRep τ -> ℕ
size LUn .tt = 1
size LR _ = 1
size (σ :*! τ) nothing = 1
size (σ :*! τ) (just (x , y)) = 1 +ℕ size σ x +ℕ size τ y
size (σ :+! τ) nothing = 1
size (σ :+! τ) (just (inj₁ x)) = 1 +ℕ size σ x
size (σ :+! τ) (just (inj₂ y)) = 1 +ℕ size τ y

-- In th2 we initialise the environment derivative accumulator to zero, because
-- that is how CHAD will be used in practice. This term creates a zero
-- environment derivative.
zero-env-term : {Γ' : Env Du} -> (Γ : Env Pr) -> Term Du Γ' (D2Γtup Γ)
zero-env-term [] = unit
zero-env-term (τ :: Γ) = pair (zerot τ) (zero-env-term Γ)

-- The statement of the corollary that bounds φ to not mention potential any
-- more. A value of this type (i.e. a proof of the theorem) is given in
-- `chad-cost.agda`.
TH2-STATEMENT : Set
TH2-STATEMENT =
  {Γ : Env Pr} {τ : Typ Pr}
  -> (env : Val Pr Γ)
  -> (ctg : Rep (D2τ τ))
  -> (t : Term Pr Γ τ)
  -> cost (push ctg (primalVal env))
          (runevm (app (snd' (sink (WSkip WEnd) (chad t)))
                       (var Z))
                  (zero-env-term Γ))
     ≤ + 5
       + + 34 * cost env t
       + + size (D2τ' τ) ctg
       + + 4 * + length Γ
```

## E   ENVIRONMENT VECTOR MONAD IMPLEMENTATION IN GHC HASKELL

The implementation spans two modules: EVM_IO, the actual monad implementation, and `ParMonoid`, a type class (with instances) for types supporting parallel accumulation.

```haskell
{-# LANGUAGE DataKinds                   #-}
{-# LANGUAGE DerivingStrategies          #-}
{-# LANGUAGE GADTs                       #-}
{-# LANGUAGE GeneralizedNewtypeDeriving  #-}
{-# LANGUAGE ImportQualifiedPost         #-}
{-# LANGUAGE KindSignatures              #-}
{-# LANGUAGE RoleAnnotations             #-}
{-# LANGUAGE ScopedTypeVariables         #-}
{-# LANGUAGE TypeApplications            #-}
{-# LANGUAGE TypeOperators               #-}

{-|
Environment vector monad, built using 'IO'. This implementation supports parallelism.

Internally this code uses 'unsafeCoerce' quite heavily, but the exposed API is type-safe.
-}
module EVM_IO (
    -- * EVM
    EVM,
    run,
    one,
    scope,
    parsequence,

    -- * Supporting types
    HList(..),
    hlength,
    Idx(..),
) where

import Control.Monad              (forM_)
import Control.Monad.Trans.Class  (lift)
import Control.Monad.Trans.Reader (ReaderT, runReaderT, local, ask)
import Data.Kind                  (Type)
import Data.Vector                qualified as V
import Data.Vector.Mutable        qualified as MV
import Data.IORef                 (IORef, newIORef, readIORef, writeIORef)
import GHC.Exts                   (Any)
import Unsafe.Coerce              qualified as Unsafe (unsafeCoerce)

import ParMonoid

-- | The environment vector monad.
newtype EVM (ts :: [Type]) a =
    -- The 'Int' is the length of 'ts'.
    -- The IORef contains the data: head of type-level list = end of the array.
    --   It is wrapped in an IORef so that we can reallocate it non-lexically.
    --   The 'Any' is the corresponding 't' from 'ts'.
    EVM (ReaderT (Int, IORef (MV.IOVector Any)) IO a)
  deriving newtype (Functor, Applicative, Monad)

-- Raise the 'ts' argument from 'phantom' to 'representational' to prevent coercion to
-- unrelated types, which would invalidate the uses of 'unsafeCoerce' in this file.
```

```haskell
type role EVM representational _

-- | A heterogeneous list.
data HList (ts :: [Type]) where
  HNil :: HList '[]
  HCons :: t -> HList ts -> HList (t ': ts)

hlength :: HList ts -> Int
hlength HNil = 0
hlength (HCons _ l) = 1 + hlength l

-- | An index into a type-level list.
data Idx (ts :: [Type]) t where
  Z :: Idx (t ': ts) t
  S :: Idx ts t -> Idx (u ': ts) t

-- | Handler of the 'EVM' monad. The integer is the initial capacity of the
-- environment array, if it is larger than the input 'HList'.
run :: Int -> EVM ts a -> HList ts -> IO (a, HList ts)
run initcap m initvec = do
    -- allocate the array
    let initlen = hlength initvec
    arr <- MV.replicate (max initcap initlen) dummy
    -- fill in the initial values; need to reverse because the head of the list should
    -- be at the end of the array
    forM_ (zip (reverse (anonymiseHList initvec)) [0..]) $ \(x, i) -> MV.write arr i x
    -- run the computation
    ref <- newIORef arr
    result <- let EVM f = m in runReaderT f (initlen, ref)
    -- read the final values; again, reversed
    outvec <- deanonymiseHList initvec <$> traverse (MV.read arr) [initlen-1, initlen-2 .. 0]
    -- return
    return (result, outvec)
  where
    anonymiseHList :: HList ts -> [Any]
    anonymiseHList HNil = []
    anonymiseHList (HCons x xs) = unsafeToAny x : anonymiseHList xs

    deanonymiseHList :: HList ts -> [Any] -> HList ts
    deanonymiseHList HNil _ = HNil
    deanonymiseHList (HCons _ model) (x : xs) =
      HCons (unsafeFromAny x) (deanonymiseHList model xs)
    deanonymiseHList (HCons _ _) [] = error "Will not happen"

-- | Accumulate a value into the environment vector.
one :: forall t ts. ParMonoid t => t -> Idx ts t -> EVM ts ()
one x idx = EVM $ do
  (veclen, arrref) <- ask
  -- compute the index in the array
  let idx' = veclen - 1 - idx2int idx
  -- read the current value from the array
  arr <- lift $ readIORef arrref
  var <- unsafeFromAny @(VarRepr t) <$> MV.read arr idx'
  -- update it with the combined result
  lift $ pmappend var x
  where
    idx2int :: Idx ts' i -> Int
```

```
    idx2int Z = 0
    idx2int (S i) = 1 + idx2int i

-- | Enter a scope with an additional cell in the accumulator array. The given
-- value is the initial value of the new cell.
scope :: forall t ts a. ParMonoid t => t -> EVM (t ': ts) a -> EVM ts (a, t)
scope zero m = EVM $ do
  (veclen, arrref) <- ask
  -- if we're at capacity, reallocate and update 'arrref'
  arr <- lift $ readIORef arrref
  let capacity = MV.length arr
  arr2 <- if veclen == capacity
          then do arr2 <- lift $ resize (2 * capacity) dummy arr
                  lift $ writeIORef arrref arr2
                  return arr2
          else return arr
  -- now that there's space, write the zero value in the next slot
  var <- lift $ pmakevar zero
  MV.write arr2 veclen (unsafeToAny @(VarRepr t) var)
  -- run the computation (note that 'arrref' was already updated above)
  result <- let EVM f = m in local (\_ -> (veclen + 1, arrref)) f
  -- read the result value
  outval <- lift $ pfinalise var
  -- return
  return (result, outval)


-- | Evaluate a vector of computations in parallel.
--
-- The given function is a parallelism strategy; it should evaluate the vector of IO
-- operations in parallel in IO somehow.
--
-- Technically this function allows you to evaluate arbitrary IO actions within EVM
-- by exploiting the flexible typing of the parallelism strategy. This is, however,
-- not the intent because we're unsure whether doing so is semantically sound.
parsequence :: (V.Vector (IO a) -> IO (V.Vector a))
            -> V.Vector (EVM ts a) -> EVM ts (V.Vector a)
parsequence parallel vm = EVM $ do
  (veclen, arrref) <- ask  -- we're unlifting through ReaderT to IO here
  lift $ parallel (V.map (\(EVM f) -> runReaderT f (veclen, arrref)) vm)


-- | Resize an 'IOVector' to the given bounds, filling new cells with the given initial value.
resize :: Int -> e -> MV.IOVector e -> IO (MV.IOVector e)
resize newlen initval arr = do
  let len = MV.length arr
  arr2 <- MV.generateM newlen $ \i ->
          if i < len then MV.read arr i
                     else return initval
  return arr2


dummy :: Any
dummy = unsafeToAny ()


unsafeToAny :: a -> Any
unsafeToAny = Unsafe.unsafeCoerce


unsafeFromAny :: Any -> a
unsafeFromAny = Unsafe.unsafeCoerce
```

```haskell
{-# LANGUAGE LambdaCase   #-}
{-# LANGUAGE TypeFamilies #-}
module ParMonoid where

import Control.Concurrent.STM
import Control.Monad             (when)

class ParMonoid a where
  type VarRepr a
  -- | Allocate a mutable cell.
  pmakevar :: a -> IO (VarRepr a)
  -- | Read the final value out of a mutable cell; this is NOT thread-safe!
  pfinalise :: VarRepr a -> IO a
  -- | Add a value into a mutable cell. This is thread-safe.
  pmappend :: VarRepr a -> a -> IO ()

instance ParMonoid () where
  type VarRepr () = ()
  pmakevar _ = return ()
  pfinalise _ = return ()
  pmappend _ _ = return ()

instance ParMonoid Float where
  type VarRepr Float = TVar Float
  pmakevar x = newTVarIO x
  pfinalise p = atomically $ readTVar p
  pmappend p x = atomically $ modifyTVar' p (+ x)

instance (ParMonoid a, ParMonoid b) => ParMonoid (a, b) where
  type VarRepr (a, b) = (VarRepr a, VarRepr b)
  pmakevar (x, y) = (,) <$> pmakevar x <*> pmakevar y
  pfinalise (p1, p2) = (,) <$> pfinalise p1 <*> pfinalise p2
  pmappend (p1, p2) (x, y) = pmappend p1 x >> pmappend p2 y

pmappendSparse :: (a -> IO va) -> (va -> a -> IO ()) -> TMVar va -> Maybe a -> IO ()
pmappendSparse _ _ _ Nothing = return ()
pmappendSparse makevar append p (Just x) =
  atomically (tryReadTMVar p) >>= \case
    Nothing -> do
      var <- makevar x
      success <- atomically $ tryPutTMVar p var
      when (not success) $ do
        var' <- atomically $ readTMVar p
        append var' x
    Just var -> append var x

instance ParMonoid a => ParMonoid (Maybe a) where
  type VarRepr (Maybe a) = TMVar (VarRepr a)
  pmakevar Nothing = newEmptyTMVarIO
  pmakevar (Just x) = pmakevar x >>= newTMVarIO
  pfinalise p = atomically (tryReadTMVar p) >>= \case
                  Nothing -> return Nothing
                  Just var -> Just <$> pfinalise var
  pmappend = pmappendSparse pmakevar pmappend

newtype LEither a b = LEither (Maybe (Either a b))
  deriving (Show)
```

```haskell
instance (ParMonoid a, ParMonoid b) => ParMonoid (LEither a b) where
  type VarRepr (LEither a b) = TMVar (Either (VarRepr a) (VarRepr b))
  pmakevar (LEither Nothing) = newEmptyTMVarIO
  pmakevar (LEither (Just (Left x))) = pmakevar x >>= newTMVarIO . Left
  pmakevar (LEither (Just (Right x))) = pmakevar x >>= newTMVarIO . Right
  pfinalise p = atomically (tryReadTMVar p) >>= \case
                  Nothing -> return (LEither Nothing)
                  Just (Left var) -> LEither . Just . Left <$> pfinalise var
                  Just (Right var) -> LEither . Just . Right <$> pfinalise var
  pmappend p (LEither mx) = pmappendSparse makeContents addContents p mx
    where
      makeContents (Left x) = Left <$> pmakevar x
      makeContents (Right x) = Right <$> pmakevar x

      addContents (Left var) (Left y) = pmappend var y
      addContents (Right var) (Right y) = pmappend var y
      addContents (Left _) (Right _) = error "ParMonoid: Left + Right"
      addContents (Right _) (Left _) = error "ParMonoid: Right + Left"
```



\section{Environment Vector Monad Implementation in GHC Haskell}\label{app:evm-ghc-haskell}

The implementation spans two modules: \texttt{EVM\_IO}, the actual monad implementation, and \texttt{ParMonoid}, a type class (with instances) for types supporting parallel accumulation.

\inputminted[fontsize=\footnotesize]{haskell}{hs/EVM_IO.hs}
\newpage

\inputminted[fontsize=\footnotesize]{haskell}{hs/ParMonoid.hs}

}  % ACM version

\end{document}

% vim: set et sw=2 ts=8:

%% file: defs.tex
\newcommand\R{\mathbb R}
\newcommand\B{\mathbb B}
\newcommand\Z{\mathbb Z}

\newcommand\ra\rightarrow
\newcommand\lra\multimap

\newcommand\snd{\textrm{snd}}

\newcommand\Map{\textrm{Map}}
\newcommand\coloneqqq{\mathbin{\raisebox{0.5pt}{::}{=}}}
\newcommand\Op{\mathsf{Op}}
\newcommand\sem[1]{\llbracket #1 \rrbracket}

\newcommand\ur{{!}}

\newcommand\figheading[1]{\textbf{#1}\hfill}

\newcommand\defeq{\coloneqq}

\newcommand\maybespace[1]{\ifthenelse{\isempty{#1}}{}{\ #1}}

\definecolor{separate}{rgb}{0.2,0.2,0.8}

\newcommand\varidxexpr[3]{#1 \ifthenelse{\isempty{#2}}{}{: #2} \ifthenelse{\isempty{#3}}{}{\in #3}}

\newcommand\computationtype[1]{\underline{#1}}
\makeatletter
\newcommand\@TyAlph[1]{%
\ifcase #1\or \tau\or \sigma\or \rho\else \@ctrerr \fi%
}
\newcommand\ty[1][1]{{\@TyAlph{#1}}}

\newcommand\@CTyAlph[1]{%
\computationtype{\ifcase #1\or \tau\or \sigma\or \rho\else \@ctrerr \fi}%
}
\newcommand\cty[1][1]{{\@CTyAlph{#1}}}

\newcommand\tvar[1][1]{{\@TyVarAlph{#1}}}
\newcommand\@TyVarAlph[1]{%
\ifcase #1\or \alpha\or \beta\or \gamma\else \@ctrerr \fi%
}
\newcommand\var[1][1]{{\@VarAlph{#1}}}
\newcommand\@VarAlph[1]{%
\ifcase #1\or x\or y\or z\or u\or v\or w\else \@ctrerr \fi%
}

\newcommand\trm[1][1]{{\@TermAlph{#1}}}
\newcommand\@TermAlph[1]{%
\ifcase #1\or t\or s\or r\else \@ctrerr \fi%
}

\newcommand\val[1][1]{%
\ifcase #1\or v\or w\or u\else \@ctrerr \fi%
}

\newcommand\op[1][1]{%
\ifcase #1\or \mathsf{op}\or \mathsf{op}'\or \mathsf{op}''\else \@ctrerr \fi%
}
\makeatother

\newcommand\letin[3]{\mathbf{let}\ #1=#2\ \mathbf{in}\ #3}
\newcommand\aletin[3]{
  \begin{array}[t]{@{}l@{}}
    \mathbf{let}\ #1=#2 \ \mathbf{in}\\
     #3
  \end{array}}
\newcommand\tFst[1]{\mathbf{fst}\maybespace{#1}}
\newcommand\tSnd[1]{\mathbf{snd}\maybespace{#1}}
\newcommand\funabs[2]{\lambda #1.\ #2}
\newcommand\tUnit{\langle \rangle}
\newcommand\tPairOpen{\langle}
\newcommand\tPairClose{\rangle}
\newcommand\tPairSep{,}
\newcommand\tPair[2]{\tPairOpen{#1}\tPairSep{#2}\tPairClose}
\newcommand\tTuple[1]{\tPairOpen{#1}\tPairClose}
\newcommand\tInl[1]{\mathbf{inl}\maybespace{#1}}
\newcommand\tInr[1]{\mathbf{inr}\maybespace{#1}}
\newcommand\tSAMatch[5]{
\begin{array}[t]{@{}l@{}}
    \mathbf{case}\ #1\ \mathbf{of}\ \{\\
    \quad \phantom{\mid\;}\tInl#2\ra #3\\
    \quad \mid \tInr#4\ra #5\\
    \}
\end{array}}
\newcommand\tSAMatchCpt[5]{
\begin{array}[t]{@{}l@{}}\\[-18pt]
    \mathbf{case}\ #1\ \mathbf{of}\:\{\,\tInl#2\ra #3\\
    \phantom{\mathbf{case}\ #1\ \mathbf{of}}  \mid \tInr#4\ra #5
    \}
\end{array}}
\newcommand\tSMatch[5]{\mathbf{case}\ #1\ \mathbf{of}\ \{\tInl#2\ra #3\mid \tInr#4\ra #5\}}
\newcommand\tSAMatchMaybe[4]{
\begin{array}[t]{@{}l@{}}
    \mathbf{case}\ #1\ \mathbf{of}\ \{ \\
    \quad \phantom{\mid\;}\Nothing \ra #2 \\
    \quad \mid \Just #3 \ra #4 \\
    \}
\end{array}}
\newcommand\tSign[1]{\mathsf{sign}\maybespace{#1}}
\newcommand\tDo[1]{
  \mathbf{do}\ %
  \begin{array}[t]{@{}l@{}}
    #1
  \end{array}
}
\newcommand\tOpApp[1]{\op(#1)}
\newcommand\tDOpApp[2]{D\op^t(#1; #2)}

\newcommand\DsynCaseSumHang{
  \raisebox{-1.8em}{\( \mathcal{D}_\Gamma \mleft[
    \begin{array}{@{}l@{}}
        \mathbf{case}\ \trm \textcolor{gray}{{}: \Sum{\ty}{\ty[2]}}\ \mathbf{of}\ \{ \\
        \quad \phantom{\mid{}}\tInl\var\ra \trm[2] \\
        \quad \mid \tInr\var[2]\ra \trm[3] \\
        \}
    \end{array}
  \mright] \)}
}

\newcommand\tOne[3]{\mathbf{one}_{\varidxexpr{#1}{#2}{#3}}}
\newcommand\tOnep[3]{\mathbf{one}'_{\varidxexpr{#1}{#2}{#3}}}
\newcommand\tSplit[2]{\mathbf{split}_{\ifthenelse{\isempty{#1#2}}{}{#1,#2}}}
\newcommand\tScope[2]{\mathbf{scope}_{\ifthenelse{\isempty{#1#2}}{}{#1,#2}}}
\newcommand\tRun[2]{\mathbf{run}\maybespace{#1}\maybespace{#2}}

\newcommand\tZero[1][]{\underline{0}_{#1}}
\newcommand\tPlus[3][]{#2+_{#1}#3}

\newcommand\lplus{\mathbin{+\!\!+}}

\newcommand\tlUnit{\langle\hspace{-2pt}{\mid}\hspace{-1pt}{\mid}\hspace{-2pt}\rangle}
\newcommand\tlFst[1]{\mathbf{lfst}\maybespace{#1}}
\newcommand\tlSnd[1]{\mathbf{lsnd}\maybespace{#1}}
\newcommand\tlPair[2]{\langle\hspace{-2pt}{\mid} #1, #2{\mid}\hspace{-2pt}\rangle}

\newcommand\tlCast[1]{\mathbf{lcastl}\maybespace{#1}}
\newcommand\trCast[1]{\mathbf{lcastr}\maybespace{#1}}
\newcommand\tlInl[1]{\mathbf{linl}\maybespace{#1}}
\newcommand\tlInr[1]{\mathbf{linr}\maybespace{#1}}
\newcommand\tError{\mathbf{error}}

\newcommand\tCast[2][]{\mathbf{cast}_{#1}\maybespace{#2}}

\newcommand\mEmpty{\textit{empty}}
\newcommand\mGet[4]{\textit{get}_{\varidxexpr{#1}{#2}{#3}}\maybespace{#4}}
\newcommand\mPopx[1]{\textit{pop}'\maybespace{#1}}
\newcommand\mPop[1]{\textit{pop}\maybespace{#1}}
\newcommand\mPush[1]{\textit{push}_{\tZero}\maybespace{#1}}
\newcommand\mModify[5]{\textit{modify}_{\varidxexpr{#1}{#2}{#3}}\maybespace{#4}\maybespace{#5}}
\newcommand\mUnion[2]{\textit{union}\maybespace{#1}\maybespace{#2}}

\newcommand\tBagEmpty{\textrm{BEmpty}}
\newcommand\tBagOne[1]{\textrm{BOne}\maybespace{#1}}
\newcommand\tBagPlus[2]{\textrm{BPlus}\maybespace{#1}\maybespace{#2}}
\newcommand\tBagArray[1]{\textrm{BArray}\maybespace{#1}}

\newcommand\Tree[2]{\textrm{Tree}\maybespace{#1}\maybespace{#2}}
\newcommand\tNode[4]{\textrm{Node}\maybespace{#1}\maybespace{#2}\maybespace{#3}\maybespace{#4}}
\newcommand\tLeaf[1]{\textrm{Leaf}\maybespace{#1}}
\newcommand\tGetA[1]{\textrm{getA}\maybespace{#1}}

\newcommand\tUnTree[3]{\textrm{unTree}\maybespace{#1}\maybespace{#2}\maybespace{#3}}

\newcommand\taBuild[3]{\textrm{build}\maybespace{#1}\ifthenelse{\isempty{#2#3}}{}{\ (#2.\ #3)}}
\newcommand\taIndex[2]{#1 \mathbin{!} #2}
\newcommand\taMap[3]{\textrm{map}\ifthenelse{\isempty{#1#2}}{}{\ (#1.\ #2)}\maybespace{#3}}

\newcommand\taFold[3]{\textrm{fold}\ifthenelse{\isempty{#1#2}}{}{\ (#1.\ #2)}\maybespace{#3}}

\newcommand\taUnzip[1]{\textrm{unzip}\maybespace{#1}}
\newcommand\taCollect[1]{\textrm{collect}\maybespace{#1}}
\newcommand\taSequence[1]{\textrm{sequence}\maybespace{#1}}
\newcommand\taScatter[2]{\textrm{scatter}\maybespace{#1}\maybespace{#2}}
\newcommand\taZipWith[5]{\textrm{zipWith}\ifthenelse{\isempty{#1#2#3}}{}{\ (#1\ #2.\ #3)}\maybespace{#4}\maybespace{#5}}

\newcommand\taLength[1]{\mathrm{length}\maybespace{#1}}

\newcommand\taFromList[1]{\textrm{fromList}\maybespace{#1}}
\newcommand\tBind{\mathbin{>\!\!>\!\!=}}

\newcommand\tSeq{\mathbin{>\!\!>}}
\newcommand\tReturn[1]{\mathbf{return}\maybespace{#1}}

\newcommand\Unit{\mathbf{1}}
\newcommand\Pair[2]{#1\times #2}
\newcommand\Function[2]{#1\ra #2}
\newcommand\Sum[2]{#1\sqcup #2}
\newcommand\BigSum[2]{\Sigma_{#1}\ #2}

\newcommand\LR{\underline{\R}}
\newcommand\LUnit{\underline{\mathbf{1}}}
\newcommand\LPair[2]{#1 \mathbin{\underline{\times}} #2}
\newcommand\LSum[2]{#1 \mathbin{\underline{\sqcup}} #2}

\newcommand\LBigSum[2]{{\underline{\Sigma}}_{#1}\ #2}
\newcommand\Type{\mathsf{Type}}

\newcommand\Just[1]{\textrm{Just}\maybespace{#1}}
\newcommand\Nothing{\textrm{Nothing}}
\newcommand\EnvMap[1]{\textrm{EMap}\maybespace{#1}}
\newcommand\EnvArr[1]{\textrm{EArr}\maybespace{#1}}
\newcommand\Bag[1]{\textrm{Bag}\maybespace{#1}}

\newcommand\Array[1]{\textrm{Array}\maybespace{#1}}

\newcommand\List[1]{\textrm{List}\maybespace{#1}}

\newcommand\ET[1]{\overline{#1}}

\newcommand\Dsyn[2][]{\mathcal{D}_{#1}\ifthenelse{\isempty{#2}}{}{\hspace{-1pt}[#2]}}
\newcommand\Env[1]{\mathbf{EV}\maybespace{#1}}
\newcommand\EnvM[2]{\mathbf{EVM}\maybespace{#1}\maybespace{#2}}

\newcommand\poten[2][]{\varphi_{#1}\maybespace{#2}}

\newcommand\cpoten{c_\varphi}
\newcommand\crungam{c_{\text{run}}}
\newcommand\cost[2]{\textrm{cost}\ifthenelse{\isempty{#1#2}}{}{(#1; #2)}}
\newcommand\size[2][]{\textrm{size}_{#1}\maybespace{#2}}
\newcommand\enventry[3]{#1 = #2 : #3}  % value environment entry: {var}{value}{type}

\newcommand\TCC[1]{#1^{CC}}
\newcommand\FV[1]{FV[#1]}
\newcommand\CFunction[2]{#1\stackrel{\square}{\ra} #2}
\newcommand\tPack[2][]{\mathbf{pack}_{#1}\,#2}

\newcommand\tPackMatchTyped[4]{\mathbf{case}\,#1\,\mathbf{of}\,\tPack[{#2}]{#3}\,\ra #4}
\newcommand\TypeOf{\mathsf{TypeOf}}

\newcommand\subst[2]{#1{}[#2]}
\newcommand\sfor[2]{^{#2}\!/\!_{#1}}

\newcommand\sqldots{.\hspace{1pt}.\hspace{1pt}.\hspace{1pt}}
\newcommand\sqoftype{\hspace{-1pt}: \hspace{-1pt}}

%% file: efficient.bbl
%%% -*-BibTeX-*-
%%% Do NOT edit. File created by BibTeX with style
%%% ACM-Reference-Format-Journals [18-Jan-2012].

\begin{thebibliography}{56}

%%% ====================================================================
%%% NOTE TO THE USER: you can override these defaults by providing
%%% customized versions of any of these macros before the \bibliography
%%% command.  Each of them MUST provide its own final punctuation,
%%% except for \shownote{}, \showDOI{}, and \showURL{}.  The latter two
%%% do not use final punctuation, in order to avoid confusing it with
%%% the Web address.
%%%
%%% To suppress output of a particular field, define its macro to expand
%%% to an empty string, or better, \unskip, like this:
%%%
%%% \newcommand{\showDOI}[1]{\unskip}   % LaTeX syntax
%%%
%%% \def \showDOI #1{\unskip}           % plain TeX syntax
%%%
%%% ====================================================================

\ifx \showCODEN    \undefined \def \showCODEN     #1{\unskip}     \fi
\ifx \showDOI      \undefined \def \showDOI       #1{#1}\fi
\ifx \showISBNx    \undefined \def \showISBNx     #1{\unskip}     \fi
\ifx \showISBNxiii \undefined \def \showISBNxiii  #1{\unskip}     \fi
\ifx \showISSN     \undefined \def \showISSN      #1{\unskip}     \fi
\ifx \showLCCN     \undefined \def \showLCCN      #1{\unskip}     \fi
\ifx \shownote     \undefined \def \shownote      #1{#1}          \fi
\ifx \showarticletitle \undefined \def \showarticletitle #1{#1}   \fi
\ifx \showURL      \undefined \def \showURL       {\relax}        \fi
% The following commands are used for tagged output and should be
% invisible to TeX
\providecommand\bibfield[2]{#2}
\providecommand\bibinfo[2]{#2}
\providecommand\natexlab[1]{#1}
\providecommand\showeprint[2][]{arXiv:#2}

\bibitem[\protect\citeauthoryear{Abadi, Barham, Chen, Chen, Davis, Dean, Devin,
  Ghemawat, Irving, Isard, Kudlur, Levenberg, Monga, Moore, Murray, Steiner,
  Tucker, Vasudevan, Warden, Wicke, Yu, and Zheng}{Abadi et~al\mbox{.}}{2016}]%
        {ad-2016-tensorflow}
\bibfield{author}{\bibinfo{person}{Mart{\'{\i}}n Abadi}, \bibinfo{person}{Paul
  Barham}, \bibinfo{person}{Jianmin Chen}, \bibinfo{person}{Zhifeng Chen},
  \bibinfo{person}{Andy Davis}, \bibinfo{person}{Jeffrey Dean},
  \bibinfo{person}{Matthieu Devin}, \bibinfo{person}{Sanjay Ghemawat},
  \bibinfo{person}{Geoffrey Irving}, \bibinfo{person}{Michael Isard},
  \bibinfo{person}{Manjunath Kudlur}, \bibinfo{person}{Josh Levenberg},
  \bibinfo{person}{Rajat Monga}, \bibinfo{person}{Sherry Moore},
  \bibinfo{person}{Derek~Gordon Murray}, \bibinfo{person}{Benoit Steiner},
  \bibinfo{person}{Paul~A. Tucker}, \bibinfo{person}{Vijay Vasudevan},
  \bibinfo{person}{Pete Warden}, \bibinfo{person}{Martin Wicke},
  \bibinfo{person}{Yuan Yu}, {and} \bibinfo{person}{Xiaoqiang Zheng}.}
  \bibinfo{year}{2016}\natexlab{}.
\newblock \showarticletitle{TensorFlow: {A} System for Large-Scale Machine
  Learning}. In \bibinfo{booktitle}{\emph{12th {USENIX} Symposium on Operating
  Systems Design and Implementation, {OSDI} 2016, Savannah, GA, USA, November
  2-4, 2016}}, \bibfield{editor}{\bibinfo{person}{Kimberly Keeton} {and}
  \bibinfo{person}{Timothy Roscoe}} (Eds.). \bibinfo{publisher}{{USENIX}
  Association}, \bibinfo{pages}{265--283}.
\newblock
\urldef\tempurl%
\url{https://www.usenix.org/conference/osdi16/technical-sessions/presentation/abadi}
\showURL{%
\tempurl}


\bibitem[\protect\citeauthoryear{{Accelerate contributors}}{{Accelerate
  contributors}}{2020}]%
        {accelerate-docs}
\bibfield{author}{\bibinfo{person}{{Accelerate contributors}}.}
  \bibinfo{year}{2020}\natexlab{}.
\newblock \bibinfo{title}{\texttt{Data.Array.Accelerate} (accelerate-1.3.0.0)}.
\newblock
  \bibinfo{howpublished}{\url{https://hackage.haskell.org/package/accelerate-1.3.0.0/docs/Data-Array-Accelerate.html}}.
\newblock
\newblock
\shownote{Accessed: 2020-11-28.}


\bibitem[\protect\citeauthoryear{Alvarez{-}Picallo, Ghica, Sprunger, and
  Zanasi}{Alvarez{-}Picallo et~al\mbox{.}}{2021}]%
        {ad-2021-alvarez-picallo}
\bibfield{author}{\bibinfo{person}{Mario Alvarez{-}Picallo},
  \bibinfo{person}{Dan~R. Ghica}, \bibinfo{person}{David Sprunger}, {and}
  \bibinfo{person}{Fabio Zanasi}.} \bibinfo{year}{2021}\natexlab{}.
\newblock \showarticletitle{Functorial String Diagrams for Reverse-Mode
  Automatic Differentiation}.
\newblock \bibinfo{journal}{\emph{CoRR}}  \bibinfo{volume}{abs/2107.13433}
  (\bibinfo{year}{2021}).
\newblock
\showeprint[arXiv]{2107.13433}
\urldef\tempurl%
\url{https://arxiv.org/abs/2107.13433}
\showURL{%
\tempurl}


\bibitem[\protect\citeauthoryear{Barthe, Crubill{\'{e}}, Lago, and
  Gavazzo}{Barthe et~al\mbox{.}}{2020}]%
        {DBLP:conf/esop/BartheCLG20}
\bibfield{author}{\bibinfo{person}{Gilles Barthe},
  \bibinfo{person}{Rapha{\"{e}}lle Crubill{\'{e}}}, \bibinfo{person}{Ugo~Dal
  Lago}, {and} \bibinfo{person}{Francesco Gavazzo}.}
  \bibinfo{year}{2020}\natexlab{}.
\newblock \showarticletitle{On the Versatility of Open Logical Relations -
  Continuity, Automatic Differentiation, and a Containment Theorem},
  \bibfield{editor}{\bibinfo{person}{Peter M{\"{u}}ller}} (Ed.).
\newblock \bibinfo{journal}{\emph{Programming Languages and Systems - 29th
  European Symposium on Programming, {ESOP} 2020, Held as Part of the European
  Joint Conferences on Theory and Practice of Software, {ETAPS} 2020, Dublin,
  Ireland, April 25-30, 2020, Proceedings}}  \bibinfo{volume}{12075},
  \bibinfo{pages}{56--83}.
\newblock
\urldef\tempurl%
\url{https://doi.org/10.1007/978-3-030-44914-8\_3}
\showDOI{\tempurl}


\bibitem[\protect\citeauthoryear{Baydin, Pearlmutter, Radul, and
  Siskind}{Baydin et~al\mbox{.}}{2017}]%
        {ad-2018-survey-automatic-differentiation}
\bibfield{author}{\bibinfo{person}{Atilim~Gunes Baydin},
  \bibinfo{person}{Barak~A. Pearlmutter}, \bibinfo{person}{Alexey~Andreyevich
  Radul}, {and} \bibinfo{person}{Jeffrey~Mark Siskind}.}
  \bibinfo{year}{2017}\natexlab{}.
\newblock \showarticletitle{Automatic Differentiation in Machine Learning: a
  Survey}.
\newblock \bibinfo{journal}{\emph{J. Mach. Learn. Res.}}  \bibinfo{volume}{18}
  (\bibinfo{year}{2017}), \bibinfo{pages}{153:1--153:43}.
\newblock
\urldef\tempurl%
\url{http://jmlr.org/papers/v18/17-468.html}
\showURL{%
\tempurl}


\bibitem[\protect\citeauthoryear{Bernardy, Boespflug, Newton, Jones, and
  Spiwack}{Bernardy et~al\mbox{.}}{2018}]%
        {fp-2018-linear-haskell}
\bibfield{author}{\bibinfo{person}{Jean{-}Philippe Bernardy},
  \bibinfo{person}{Mathieu Boespflug}, \bibinfo{person}{Ryan~R. Newton},
  \bibinfo{person}{Simon~Peyton Jones}, {and} \bibinfo{person}{Arnaud
  Spiwack}.} \bibinfo{year}{2018}\natexlab{}.
\newblock \showarticletitle{Linear Haskell: practical linearity in a
  higher-order polymorphic language}.
\newblock \bibinfo{journal}{\emph{Proc. {ACM} Program. Lang.}}
  \bibinfo{volume}{2}, \bibinfo{number}{{POPL}} (\bibinfo{year}{2018}),
  \bibinfo{pages}{5:1--5:29}.
\newblock
\urldef\tempurl%
\url{https://doi.org/10.1145/3158093}
\showDOI{\tempurl}


\bibitem[\protect\citeauthoryear{Bradbury, Frostig, Hawkins, Johnson, Leary,
  Maclaurin, Necula, Paszke, Vander{P}las, Wanderman-{M}ilne, and
  Zhang}{Bradbury et~al\mbox{.}}{2018}]%
        {ad-2018-jax}
\bibfield{author}{\bibinfo{person}{James Bradbury}, \bibinfo{person}{Roy
  Frostig}, \bibinfo{person}{Peter Hawkins}, \bibinfo{person}{Matthew~James
  Johnson}, \bibinfo{person}{Chris Leary}, \bibinfo{person}{Dougal Maclaurin},
  \bibinfo{person}{George Necula}, \bibinfo{person}{Adam Paszke},
  \bibinfo{person}{Jake Vander{P}las}, \bibinfo{person}{Skye
  Wanderman-{M}ilne}, {and} \bibinfo{person}{Qiao Zhang}.}
  \bibinfo{year}{2018}\natexlab{}.
\newblock \bibinfo{booktitle}{\emph{{JAX}: composable transformations of
  {P}ython+{N}um{P}y programs}}.
\newblock
\urldef\tempurl%
\url{http://github.com/google/jax}
\showURL{%
\tempurl}


\bibitem[\protect\citeauthoryear{Brown and Tarjan}{Brown and Tarjan}{1979}]%
        {algorithms-1979-merging-tarjan}
\bibfield{author}{\bibinfo{person}{Mark~R. Brown} {and}
  \bibinfo{person}{Robert~Endre Tarjan}.} \bibinfo{year}{1979}\natexlab{}.
\newblock \showarticletitle{A Fast Merging Algorithm}.
\newblock \bibinfo{journal}{\emph{J. {ACM}}} \bibinfo{volume}{26},
  \bibinfo{number}{2} (\bibinfo{year}{1979}), \bibinfo{pages}{211--226}.
\newblock
\urldef\tempurl%
\url{https://doi.org/10.1145/322123.322127}
\showDOI{\tempurl}


\bibitem[\protect\citeauthoryear{Chakravarty, Keller, Lee, McDonell, and
  Grover}{Chakravarty et~al\mbox{.}}{2011}]%
        {acc-2011-cuda}
\bibfield{author}{\bibinfo{person}{Manuel M.~T. Chakravarty},
  \bibinfo{person}{Gabriele Keller}, \bibinfo{person}{Sean Lee},
  \bibinfo{person}{Trevor~L. McDonell}, {and} \bibinfo{person}{Vinod Grover}.}
  \bibinfo{year}{2011}\natexlab{}.
\newblock \showarticletitle{Accelerating {Haskell} array codes with multicore
  {GPUs}}. In \bibinfo{booktitle}{\emph{Proceedings of the {POPL} 2011 Workshop
  on Declarative Aspects of Multicore Programming, {DAMP} 2011, Austin, TX,
  USA, January 23, 2011}}, \bibfield{editor}{\bibinfo{person}{Manuel Carro}
  {and} \bibinfo{person}{John~H. Reppy}} (Eds.). \bibinfo{publisher}{{ACM}},
  \bibinfo{address}{New York, NY, USA}, \bibinfo{pages}{3--14}.
\newblock
\urldef\tempurl%
\url{https://doi.org/10.1145/1926354.1926358}
\showDOI{\tempurl}


\bibitem[\protect\citeauthoryear{Chin Jen~Sem}{Chin Jen~Sem}{2020}]%
        {thesis-2020-curtis-fwd-ad-gradient-compiler-opts}
\bibfield{author}{\bibinfo{person}{Curtis Chin Jen~Sem}.}
  \bibinfo{year}{2020}\natexlab{}.
\newblock \showarticletitle{Formalized Correctness Proofs of Automatic
  Differentiation in {C}oq}.
\newblock \bibinfo{journal}{\emph{Master's Thesis, Utrecht University}}
  (\bibinfo{date}{09} \bibinfo{year}{2020}).
\newblock
\newblock
\shownote{\url{https://dspace.library.uu.nl/handle/1874/400790}; Coq code:
  \url{https://github.com/crtschin/thesis}.}


\bibitem[\protect\citeauthoryear{de~Vilhena and Pottier}{de~Vilhena and
  Pottier}{2021}]%
        {ad-2021-verifying}
\bibfield{author}{\bibinfo{person}{Paulo~Em{\'\i}lio de Vilhena} {and}
  \bibinfo{person}{Fran{\c{c}}ois Pottier}.} \bibinfo{year}{2021}\natexlab{}.
\newblock \showarticletitle{Verifying a Minimalist Reverse-Mode AD Library}.
\newblock \bibinfo{journal}{\emph{arXiv preprint arXiv:2112.07292}}
  (\bibinfo{year}{2021}).
\newblock


\bibitem[\protect\citeauthoryear{Elliott}{Elliott}{2018}]%
        {adfp-2018-categories-ad}
\bibfield{author}{\bibinfo{person}{Conal Elliott}.}
  \bibinfo{year}{2018}\natexlab{}.
\newblock \showarticletitle{The simple essence of automatic differentiation}.
\newblock \bibinfo{journal}{\emph{Proc. {ACM} Program. Lang.}}
  \bibinfo{volume}{2}, \bibinfo{number}{{ICFP}} (\bibinfo{year}{2018}),
  \bibinfo{pages}{70:1--70:29}.
\newblock
\urldef\tempurl%
\url{https://doi.org/10.1145/3236765}
\showDOI{\tempurl}


\bibitem[\protect\citeauthoryear{Elsman, Henglein, Kaarsgaard, Mathiesen, and
  Schenck}{Elsman et~al\mbox{.}}{2022}]%
        {ad-2022-elsman-combinatorial}
\bibfield{author}{\bibinfo{person}{Martin Elsman}, \bibinfo{person}{Fritz
  Henglein}, \bibinfo{person}{Robin Kaarsgaard}, \bibinfo{person}{Mikkel~Kragh
  Mathiesen}, {and} \bibinfo{person}{Robert Schenck}.}
  \bibinfo{year}{2022}\natexlab{}.
\newblock \showarticletitle{Combinatory Adjoints and Differentiation}. In
  \bibinfo{booktitle}{\emph{Proceedings Ninth Workshop on Mathematically
  Structured Functional Programming, MSFP@ETAPS 2022, Munich, Germany, 2nd
  April 2022}} \emph{(\bibinfo{series}{{EPTCS}})},
  \bibfield{editor}{\bibinfo{person}{Jeremy Gibbons} {and}
  \bibinfo{person}{Max~S. New}} (Eds.), Vol.~\bibinfo{volume}{360}.
  \bibinfo{pages}{1--26}.
\newblock
\urldef\tempurl%
\url{https://doi.org/10.4204/EPTCS.360.1}
\showDOI{\tempurl}


\bibitem[\protect\citeauthoryear{Griewank and Walther}{Griewank and
  Walther}{2008}]%
        {adbook-2008-griewank-walther}
\bibfield{author}{\bibinfo{person}{Andreas Griewank} {and}
  \bibinfo{person}{Andrea Walther}.} \bibinfo{year}{2008}\natexlab{}.
\newblock \bibinfo{booktitle}{\emph{Evaluating derivatives - principles and
  techniques of algorithmic differentiation, Second Edition}}.
\newblock \bibinfo{publisher}{{SIAM}}.
\newblock
\showISBNx{978-0-89871-659-7}
\urldef\tempurl%
\url{https://doi.org/10.1137/1.9780898717761}
\showDOI{\tempurl}


\bibitem[\protect\citeauthoryear{Henriksen}{Henriksen}{2017}]%
        {futhark-2017-thesis}
\bibfield{author}{\bibinfo{person}{Troels Henriksen}.}
  \bibinfo{year}{2017}\natexlab{}.
\newblock \emph{\bibinfo{title}{Design and Implementation of the Futhark
  Programming Language}}.
\newblock \bibinfo{thesistype}{Ph.D. Dissertation}. \bibinfo{school}{University
  of Copenhagen}, \bibinfo{address}{Universitetsparken 5, 2100 København}.
\newblock


\bibitem[\protect\citeauthoryear{Henriksen, Serup, Elsman, Henglein, and
  Oancea}{Henriksen et~al\mbox{.}}{2017}]%
        {futhark-2017-pldi}
\bibfield{author}{\bibinfo{person}{Troels Henriksen}, \bibinfo{person}{Niels
  G.~W. Serup}, \bibinfo{person}{Martin Elsman}, \bibinfo{person}{Fritz
  Henglein}, {and} \bibinfo{person}{Cosmin~E. Oancea}.}
  \bibinfo{year}{2017}\natexlab{}.
\newblock \showarticletitle{Futhark: purely functional GPU-programming with
  nested parallelism and in-place array updates}. In
  \bibinfo{booktitle}{\emph{Proceedings of the 38th {ACM} {SIGPLAN} Conference
  on Programming Language Design and Implementation, {PLDI} 2017, Barcelona,
  Spain, June 18-23, 2017}}, \bibfield{editor}{\bibinfo{person}{Albert Cohen}
  {and} \bibinfo{person}{Martin~T. Vechev}} (Eds.). \bibinfo{publisher}{{ACM}},
  \bibinfo{pages}{556--571}.
\newblock
\urldef\tempurl%
\url{https://doi.org/10.1145/3062341.3062354}
\showDOI{\tempurl}


\bibitem[\protect\citeauthoryear{Hughes}{Hughes}{1986}]%
        {fp-1986-difference-lists}
\bibfield{author}{\bibinfo{person}{R.~John~M. Hughes}.}
  \bibinfo{year}{1986}\natexlab{}.
\newblock \showarticletitle{A Novel Representation of Lists and its Application
  to the Function "reverse"}.
\newblock \bibinfo{journal}{\emph{Inf. Process. Lett.}} \bibinfo{volume}{22},
  \bibinfo{number}{3} (\bibinfo{year}{1986}), \bibinfo{pages}{141--144}.
\newblock
\urldef\tempurl%
\url{https://doi.org/10.1016/0020-0190(86)90059-1}
\showDOI{\tempurl}


\bibitem[\protect\citeauthoryear{Huot, Staton, and V{\'{a}}k{\'{a}}r}{Huot
  et~al\mbox{.}}{2020}]%
        {ad-2020-sam-mathieu-matthijs}
\bibfield{author}{\bibinfo{person}{Mathieu Huot}, \bibinfo{person}{Sam Staton},
  {and} \bibinfo{person}{Matthijs V{\'{a}}k{\'{a}}r}.}
  \bibinfo{year}{2020}\natexlab{}.
\newblock \showarticletitle{Correctness of Automatic Differentiation via
  Diffeologies and Categorical Gluing}. In
  \bibinfo{booktitle}{\emph{Foundations of Software Science and Computation
  Structures - 23rd International Conference, {FOSSACS} 2020, Held as Part of
  the European Joint Conferences on Theory and Practice of Software, {ETAPS}
  2020, Dublin, Ireland, April 25-30, 2020, Proceedings}}
  \emph{(\bibinfo{series}{Lecture Notes in Computer Science})},
  \bibfield{editor}{\bibinfo{person}{Jean Goubault{-}Larrecq} {and}
  \bibinfo{person}{Barbara K{\"{o}}nig}} (Eds.), Vol.~\bibinfo{volume}{12077}.
  \bibinfo{publisher}{Springer}, \bibinfo{pages}{319--338}.
\newblock
\urldef\tempurl%
\url{https://doi.org/10.1007/978-3-030-45231-5\_17}
\showDOI{\tempurl}


\bibitem[\protect\citeauthoryear{Huot, Staton, and V{\'{a}}k{\'{a}}r}{Huot
  et~al\mbox{.}}{2021}]%
        {vakar-staton-huot-2021}
\bibfield{author}{\bibinfo{person}{Mathieu Huot}, \bibinfo{person}{Sam Staton},
  {and} \bibinfo{person}{Matthijs V{\'{a}}k{\'{a}}r}.}
  \bibinfo{year}{2021}\natexlab{}.
\newblock \showarticletitle{Higher Order Automatic Differentiation of Higher
  Order Functions}.
\newblock \bibinfo{journal}{\emph{CoRR}}  \bibinfo{volume}{abs/2101.06757}
  (\bibinfo{year}{2021}).
\newblock
\showeprint[arXiv]{2101.06757}
\urldef\tempurl%
\url{https://arxiv.org/abs/2101.06757}
\showURL{%
\tempurl}


\bibitem[\protect\citeauthoryear{Kerjean and P{\'e}drot}{Kerjean and
  P{\'e}drot}{2022}]%
        {ad-2022-kerjean:hal-03123968}
\bibfield{author}{\bibinfo{person}{Marie Kerjean} {and}
  \bibinfo{person}{Pierre-Marie P{\'e}drot}.} \bibinfo{year}{2022}\natexlab{}.
\newblock \bibinfo{title}{{$\partial$ is for Dialectica}}.
  (\bibinfo{date}{Jan.} \bibinfo{year}{2022}).
\newblock
\urldef\tempurl%
\url{https://hal.archives-ouvertes.fr/hal-03123968}
\showURL{%
\tempurl}
\newblock
\shownote{working paper or preprint.}


\bibitem[\protect\citeauthoryear{Krawiec, Jones, Krishnaswami, Ellis,
  Eisenberg, and Fitzgibbon}{Krawiec et~al\mbox{.}}{2022}]%
        {ad-2021-krawiec-kmett-ad}
\bibfield{author}{\bibinfo{person}{Faustyna Krawiec},
  \bibinfo{person}{Simon~Peyton Jones}, \bibinfo{person}{Neel Krishnaswami},
  \bibinfo{person}{Tom Ellis}, \bibinfo{person}{Richard~A. Eisenberg}, {and}
  \bibinfo{person}{Andrew~W. Fitzgibbon}.} \bibinfo{year}{2022}\natexlab{}.
\newblock \showarticletitle{Provably correct, asymptotically efficient,
  higher-order reverse-mode automatic differentiation}.
\newblock \bibinfo{journal}{\emph{Proc. {ACM} Program. Lang.}}
  \bibinfo{volume}{6}, \bibinfo{number}{{POPL}} (\bibinfo{year}{2022}),
  \bibinfo{pages}{1--30}.
\newblock
\urldef\tempurl%
\url{https://doi.org/10.1145/3498710}
\showDOI{\tempurl}


\bibitem[\protect\citeauthoryear{Krishnaswami, Pradic, and Benton}{Krishnaswami
  et~al\mbox{.}}{2015}]%
        {krishnaswami2015integrating}
\bibfield{author}{\bibinfo{person}{Neelakantan~R Krishnaswami},
  \bibinfo{person}{Pierre Pradic}, {and} \bibinfo{person}{Nick Benton}.}
  \bibinfo{year}{2015}\natexlab{}.
\newblock \showarticletitle{Integrating linear and dependent types}.
\newblock \bibinfo{journal}{\emph{ACM SIGPLAN Notices}} \bibinfo{volume}{50},
  \bibinfo{number}{1} (\bibinfo{year}{2015}), \bibinfo{pages}{17--30}.
\newblock


\bibitem[\protect\citeauthoryear{Launchbury and Jones}{Launchbury and
  Jones}{1994}]%
        {fp-1994-st-monad}
\bibfield{author}{\bibinfo{person}{John Launchbury} {and}
  \bibinfo{person}{Simon L.~Peyton Jones}.} \bibinfo{year}{1994}\natexlab{}.
\newblock \showarticletitle{Lazy Functional State Threads}. In
  \bibinfo{booktitle}{\emph{Proceedings of the {ACM} SIGPLAN'94 Conference on
  Programming Language Design and Implementation (PLDI), Orlando, Florida, USA,
  June 20-24, 1994}}, \bibfield{editor}{\bibinfo{person}{Vivek Sarkar},
  \bibinfo{person}{Barbara~G. Ryder}, {and} \bibinfo{person}{Mary~Lou Soffa}}
  (Eds.). \bibinfo{publisher}{{ACM}}, \bibinfo{pages}{24--35}.
\newblock
\urldef\tempurl%
\url{https://doi.org/10.1145/178243.178246}
\showDOI{\tempurl}


\bibitem[\protect\citeauthoryear{Leary and Wang}{Leary and Wang}{2017}]%
        {ad-2017-xla}
\bibfield{author}{\bibinfo{person}{Chris Leary} {and} \bibinfo{person}{Todd
  Wang}.} \bibinfo{year}{2017}\natexlab{}.
\newblock \showarticletitle{XLA: TensorFlow, compiled}.
\newblock \bibinfo{journal}{\emph{TensorFlow Dev Summit}}
  (\bibinfo{year}{2017}).
\newblock
\urldef\tempurl%
\url{https://developers.googleblog.com/2017/03/xla-tensorflow-compiled.html}
\showURL{%
\tempurl}


\bibitem[\protect\citeauthoryear{Linnainmaa}{Linnainmaa}{1976}]%
        {linnainmaa1976taylor}
\bibfield{author}{\bibinfo{person}{Seppo Linnainmaa}.}
  \bibinfo{year}{1976}\natexlab{}.
\newblock \showarticletitle{Taylor expansion of the accumulated rounding
  error}.
\newblock \bibinfo{journal}{\emph{BIT Numerical Mathematics}}
  \bibinfo{volume}{16}, \bibinfo{number}{2} (\bibinfo{year}{1976}),
  \bibinfo{pages}{146--160}.
\newblock


\bibitem[\protect\citeauthoryear{Margossian}{Margossian}{2019}]%
        {ad-2018-survey-ad-implementation}
\bibfield{author}{\bibinfo{person}{Charles~C. Margossian}.}
  \bibinfo{year}{2019}\natexlab{}.
\newblock \showarticletitle{A review of automatic differentiation and its
  efficient implementation}.
\newblock \bibinfo{journal}{\emph{Wiley Interdiscip. Rev. Data Min. Knowl.
  Discov.}} \bibinfo{volume}{9}, \bibinfo{number}{4} (\bibinfo{year}{2019}).
\newblock
\urldef\tempurl%
\url{https://doi.org/10.1002/widm.1305}
\showDOI{\tempurl}


\bibitem[\protect\citeauthoryear{McDonell, Chakravarty, Keller, and
  Lippmeier}{McDonell et~al\mbox{.}}{2013}]%
        {acc-2013-optim}
\bibfield{author}{\bibinfo{person}{Trevor~L. McDonell}, \bibinfo{person}{Manuel
  M.~T. Chakravarty}, \bibinfo{person}{Gabriele Keller}, {and}
  \bibinfo{person}{Ben Lippmeier}.} \bibinfo{year}{2013}\natexlab{}.
\newblock \showarticletitle{Optimising purely functional {GPU} programs}. In
  \bibinfo{booktitle}{\emph{{ACM} {SIGPLAN} International Conference on
  Functional Programming, ICFP'13, Boston, MA, {USA} - September 25 - 27,
  2013}}, \bibfield{editor}{\bibinfo{person}{Greg Morrisett} {and}
  \bibinfo{person}{Tarmo Uustalu}} (Eds.). \bibinfo{publisher}{{ACM}},
  \bibinfo{pages}{49--60}.
\newblock
\urldef\tempurl%
\url{https://doi.org/10.1145/2500365.2500595}
\showDOI{\tempurl}


\bibitem[\protect\citeauthoryear{Merrill and Garland}{Merrill and
  Garland}{2016}]%
        {array-2016-nvidia-decoupled-lookback-scan}
\bibfield{author}{\bibinfo{person}{Duane Merrill} {and}
  \bibinfo{person}{Michael Garland}.} \bibinfo{year}{2016}\natexlab{}.
\newblock \bibinfo{booktitle}{\emph{Single-pass Parallel Prefix Scan with
  Decoupled Look-back}}.
\newblock \bibinfo{type}{{T}echnical {R}eport} NVR-2016-002.
  \bibinfo{institution}{NVIDIA}.
\newblock


\bibitem[\protect\citeauthoryear{Minamide, Morrisett, and Harper}{Minamide
  et~al\mbox{.}}{1996}]%
        {DBLP:conf/popl/MinamideMH96}
\bibfield{author}{\bibinfo{person}{Yasuhiko Minamide},
  \bibinfo{person}{J.~Gregory Morrisett}, {and} \bibinfo{person}{Robert
  Harper}.} \bibinfo{year}{1996}\natexlab{}.
\newblock \showarticletitle{Typed Closure Conversion}. In
  \bibinfo{booktitle}{\emph{Conference Record of POPL'96: The 23rd {ACM}
  {SIGPLAN-SIGACT} Symposium on Principles of Programming Languages, Papers
  Presented at the Symposium, St. Petersburg Beach, Florida, USA, January
  21-24, 1996}}, \bibfield{editor}{\bibinfo{person}{Hans{-}Juergen Boehm} {and}
  \bibinfo{person}{Guy L.~Steele Jr.}} (Eds.). \bibinfo{publisher}{{ACM}
  Press}, \bibinfo{pages}{271--283}.
\newblock
\urldef\tempurl%
\url{https://doi.org/10.1145/237721.237791}
\showDOI{\tempurl}


\bibitem[\protect\citeauthoryear{Norell}{Norell}{2007}]%
        {agda-2007-norell}
\bibfield{author}{\bibinfo{person}{Ulf Norell}.}
  \bibinfo{year}{2007}\natexlab{}.
\newblock \bibinfo{booktitle}{\emph{Towards a practical programming language
  based on dependent type theory}}. Vol.~\bibinfo{volume}{32}.
\newblock \bibinfo{publisher}{Chalmers University of Technology}.
\newblock


\bibitem[\protect\citeauthoryear{Nunes and V{\'{a}}k{\'{a}}r}{Nunes and
  V{\'{a}}k{\'{a}}r}{2022}]%
        {nunes-2022-dual-numbers-long}
\bibfield{author}{\bibinfo{person}{Fernando~Lucatelli Nunes} {and}
  \bibinfo{person}{Matthijs V{\'{a}}k{\'{a}}r}.}
  \bibinfo{year}{2022}\natexlab{}.
\newblock \showarticletitle{{Automatic Differentiation for ML-family languages:
  correctness via logical relations}}.
\newblock \bibinfo{journal}{\emph{CoRR}}  \bibinfo{volume}{abs/2210.07724}
  (\bibinfo{year}{2022}).
\newblock
\showeprint[arXiv]{2210.07724}
\urldef\tempurl%
\url{https://arxiv.org/abs/2210.07724}
\showURL{%
\tempurl}


\bibitem[\protect\citeauthoryear{Nunes and Vákár}{Nunes and Vákár}{2023}]%
        {nunes-2022-chad-expressive}
\bibfield{author}{\bibinfo{person}{Fernando~Lucatelli Nunes} {and}
  \bibinfo{person}{Matthijs Vákár}.} \bibinfo{year}{2023}\natexlab{}.
\newblock \showarticletitle{CHAD for expressive total languages}.
\newblock \bibinfo{journal}{\emph{Mathematical Structures in Computer Science}}
  \bibinfo{volume}{33}, \bibinfo{number}{4-5} (\bibinfo{year}{2023}),
  \bibinfo{pages}{311–426}.
\newblock
\urldef\tempurl%
\url{https://doi.org/10.1017/S096012952300018X}
\showDOI{\tempurl}


\bibitem[\protect\citeauthoryear{Paszke, Gross, Chintala, Chanan, Yang, DeVito,
  Lin, Desmaison, Antiga, and Lerer}{Paszke et~al\mbox{.}}{2017}]%
        {ad-2017-pytorch}
\bibfield{author}{\bibinfo{person}{Adam Paszke}, \bibinfo{person}{Sam Gross},
  \bibinfo{person}{Soumith Chintala}, \bibinfo{person}{Gregory Chanan},
  \bibinfo{person}{Edward Yang}, \bibinfo{person}{Zachary DeVito},
  \bibinfo{person}{Zeming Lin}, \bibinfo{person}{Alban Desmaison},
  \bibinfo{person}{Luca Antiga}, {and} \bibinfo{person}{Adam Lerer}.}
  \bibinfo{year}{2017}\natexlab{}.
\newblock \showarticletitle{Automatic differentiation in {PyTorch}}. In
  \bibinfo{booktitle}{\emph{NIPS 2017 Autodiff Workshop: The future of
  gradient-based machine learning software and techniques}}.
  \bibinfo{publisher}{Curran Associates, Inc.}, \bibinfo{address}{Red Hook, NY,
  USA}.
\newblock


\bibitem[\protect\citeauthoryear{Paszke, Johnson, Duvenaud, Vytiniotis, Radul,
  Johnson, Ragan{-}Kelley, and Maclaurin}{Paszke et~al\mbox{.}}{2021a}]%
        {dex-2021-ad}
\bibfield{author}{\bibinfo{person}{Adam Paszke}, \bibinfo{person}{Daniel~D.
  Johnson}, \bibinfo{person}{David Duvenaud}, \bibinfo{person}{Dimitrios
  Vytiniotis}, \bibinfo{person}{Alexey Radul}, \bibinfo{person}{Matthew~J.
  Johnson}, \bibinfo{person}{Jonathan Ragan{-}Kelley}, {and}
  \bibinfo{person}{Dougal Maclaurin}.} \bibinfo{year}{2021}\natexlab{a}.
\newblock \showarticletitle{Getting to the point: index sets and
  parallelism-preserving autodiff for pointful array programming}.
\newblock \bibinfo{journal}{\emph{Proc. {ACM} Program. Lang.}}
  \bibinfo{volume}{5}, \bibinfo{number}{{ICFP}} (\bibinfo{year}{2021}),
  \bibinfo{pages}{1--29}.
\newblock
\urldef\tempurl%
\url{https://doi.org/10.1145/3473593}
\showDOI{\tempurl}


\bibitem[\protect\citeauthoryear{Paszke, Johnson, Frostig, and
  Maclaurin}{Paszke et~al\mbox{.}}{2021b}]%
        {ad-2021-diff-scan}
\bibfield{author}{\bibinfo{person}{Adam Paszke}, \bibinfo{person}{Matthew~J.
  Johnson}, \bibinfo{person}{Roy Frostig}, {and} \bibinfo{person}{Dougal
  Maclaurin}.} \bibinfo{year}{2021}\natexlab{b}.
\newblock \showarticletitle{Parallelism-preserving automatic differentiation
  for second-order array languages}. In \bibinfo{booktitle}{\emph{{FHPNC} 2021:
  Proceedings of the 9th {ACM} {SIGPLAN} International Workshop on Functional
  High-Performance and Numerical Computing, FHPNC@ICFP 2021, Virtual Event,
  Korea, August 22, 2021}}, \bibfield{editor}{\bibinfo{person}{Gabriele Keller}
  {and} \bibinfo{person}{Troels Henriksen}} (Eds.). \bibinfo{publisher}{{ACM}},
  \bibinfo{pages}{13--23}.
\newblock
\urldef\tempurl%
\url{https://doi.org/10.1145/3471873.3472975}
\showDOI{\tempurl}


\bibitem[\protect\citeauthoryear{Pearlmutter and Siskind}{Pearlmutter and
  Siskind}{2008}]%
        {ad-2008-reverse-functional-ad}
\bibfield{author}{\bibinfo{person}{Barak~A. Pearlmutter} {and}
  \bibinfo{person}{Jeffrey~Mark Siskind}.} \bibinfo{year}{2008}\natexlab{}.
\newblock \showarticletitle{Reverse-mode {AD} in a functional framework: Lambda
  the ultimate backpropagator}.
\newblock \bibinfo{journal}{\emph{{ACM} Trans. Program. Lang. Syst.}}
  \bibinfo{volume}{30}, \bibinfo{number}{2} (\bibinfo{year}{2008}),
  \bibinfo{pages}{7:1--7:36}.
\newblock
\urldef\tempurl%
\url{https://doi.org/10.1145/1330017.1330018}
\showDOI{\tempurl}


\bibitem[\protect\citeauthoryear{Plotkin and Power}{Plotkin and Power}{2002}]%
        {fp-2002-notions-computation-monads}
\bibfield{author}{\bibinfo{person}{Gordon~D. Plotkin} {and}
  \bibinfo{person}{John Power}.} \bibinfo{year}{2002}\natexlab{}.
\newblock \showarticletitle{Notions of Computation Determine Monads}. In
  \bibinfo{booktitle}{\emph{Foundations of Software Science and Computation
  Structures, 5th International Conference, {FOSSACS} 2002. Held as Part of the
  Joint European Conferences on Theory and Practice of Software, {ETAPS} 2002
  Grenoble, France, April 8-12, 2002, Proceedings}}
  \emph{(\bibinfo{series}{Lecture Notes in Computer Science})},
  \bibfield{editor}{\bibinfo{person}{Mogens Nielsen} {and}
  \bibinfo{person}{Uffe Engberg}} (Eds.), Vol.~\bibinfo{volume}{2303}.
  \bibinfo{publisher}{Springer}, \bibinfo{pages}{342--356}.
\newblock
\urldef\tempurl%
\url{https://doi.org/10.1007/3-540-45931-6\_24}
\showDOI{\tempurl}


\bibitem[\protect\citeauthoryear{Plotkin and Pretnar}{Plotkin and
  Pretnar}{2013}]%
        {fp-2013-effect-handlers}
\bibfield{author}{\bibinfo{person}{Gordon~D. Plotkin} {and}
  \bibinfo{person}{Matija Pretnar}.} \bibinfo{year}{2013}\natexlab{}.
\newblock \showarticletitle{Handling Algebraic Effects}.
\newblock \bibinfo{journal}{\emph{Log. Methods Comput. Sci.}}
  \bibinfo{volume}{9}, \bibinfo{number}{4} (\bibinfo{year}{2013}).
\newblock
\urldef\tempurl%
\url{https://doi.org/10.2168/LMCS-9(4:23)2013}
\showDOI{\tempurl}


\bibitem[\protect\citeauthoryear{Radul, Paszke, Frostig, Johnson, and
  Maclaurin}{Radul et~al\mbox{.}}{2023}]%
        {ad-2023-yolo}
\bibfield{author}{\bibinfo{person}{Alexey Radul}, \bibinfo{person}{Adam
  Paszke}, \bibinfo{person}{Roy Frostig}, \bibinfo{person}{Matthew~J. Johnson},
  {and} \bibinfo{person}{Dougal Maclaurin}.} \bibinfo{year}{2023}\natexlab{}.
\newblock \showarticletitle{You Only Linearize Once: Tangents Transpose to
  Gradients}.
\newblock \bibinfo{journal}{\emph{Proc. {ACM} Program. Lang.}}
  \bibinfo{volume}{7}, \bibinfo{number}{{POPL}} (\bibinfo{year}{2023}),
  \bibinfo{pages}{1246--1274}.
\newblock
\urldef\tempurl%
\url{https://doi.org/10.1145/3571236}
\showDOI{\tempurl}


\bibitem[\protect\citeauthoryear{Reynolds}{Reynolds}{1998}]%
        {fp-1998-defunctionalisation}
\bibfield{author}{\bibinfo{person}{John~C. Reynolds}.}
  \bibinfo{year}{1998}\natexlab{}.
\newblock \showarticletitle{Definitional Interpreters for Higher-Order
  Programming Languages}.
\newblock \bibinfo{journal}{\emph{High. Order Symb. Comput.}}
  \bibinfo{volume}{11}, \bibinfo{number}{4} (\bibinfo{year}{1998}),
  \bibinfo{pages}{363--397}.
\newblock
\urldef\tempurl%
\url{https://doi.org/10.1023/A:1010027404223}
\showDOI{\tempurl}


\bibitem[\protect\citeauthoryear{Schenck, R{\o}nning, Henriksen, and
  Oancea}{Schenck et~al\mbox{.}}{2022}]%
        {ad-2022-futhark-partial-recompute}
\bibfield{author}{\bibinfo{person}{Robert Schenck}, \bibinfo{person}{Ola
  R{\o}nning}, \bibinfo{person}{Troels Henriksen}, {and}
  \bibinfo{person}{Cosmin~E. Oancea}.} \bibinfo{year}{2022}\natexlab{}.
\newblock \showarticletitle{{AD} for an Array Language with Nested
  Parallelism}.
\newblock \bibinfo{journal}{\emph{CoRR}}  \bibinfo{volume}{abs/2202.10297}
  (\bibinfo{year}{2022}).
\newblock
\showeprint[arXiv]{2202.10297}
\urldef\tempurl%
\url{https://arxiv.org/abs/2202.10297}
\showURL{%
\tempurl}


\bibitem[\protect\citeauthoryear{Shaikhha, Fitzgibbon, Vytiniotis, and
  Jones}{Shaikhha et~al\mbox{.}}{2019}]%
        {ad-2019-fwd-ad-gradient-compiler-opts}
\bibfield{author}{\bibinfo{person}{Amir Shaikhha}, \bibinfo{person}{Andrew
  Fitzgibbon}, \bibinfo{person}{Dimitrios Vytiniotis}, {and}
  \bibinfo{person}{Simon~Peyton Jones}.} \bibinfo{year}{2019}\natexlab{}.
\newblock \showarticletitle{Efficient differentiable programming in a
  functional array-processing language}.
\newblock \bibinfo{journal}{\emph{Proc. {ACM} Program. Lang.}}
  \bibinfo{volume}{3}, \bibinfo{number}{{ICFP}} (\bibinfo{year}{2019}),
  \bibinfo{pages}{97:1--97:30}.
\newblock
\urldef\tempurl%
\url{https://doi.org/10.1145/3341701}
\showDOI{\tempurl}


\bibitem[\protect\citeauthoryear{Shaikhha, Huot, and Hashemian}{Shaikhha
  et~al\mbox{.}}{2023}]%
        {shaikhha2023nabla}
\bibfield{author}{\bibinfo{person}{Amir Shaikhha}, \bibinfo{person}{Mathieu
  Huot}, {and} \bibinfo{person}{Shideh Hashemian}.}
  \bibinfo{year}{2023}\natexlab{}.
\newblock \showarticletitle{$\nabla$SD: Differentiable Programming for Sparse
  Tensors}.
\newblock \bibinfo{journal}{\emph{arXiv preprint arXiv:2303.07030}}
  (\bibinfo{year}{2023}).
\newblock


\bibitem[\protect\citeauthoryear{Siskind and Pearlmutter}{Siskind and
  Pearlmutter}{2016}]%
        {ad-2016-scheme-higher-order-ad}
\bibfield{author}{\bibinfo{person}{Jeffrey~Mark Siskind} {and}
  \bibinfo{person}{Barak~A. Pearlmutter}.} \bibinfo{year}{2016}\natexlab{}.
\newblock \showarticletitle{Efficient Implementation of a Higher-Order Language
  with Built-In {AD}}.
\newblock \bibinfo{journal}{\emph{CoRR}}  \bibinfo{volume}{abs/1611.03416}
  (\bibinfo{year}{2016}).
\newblock
\showeprint[arxiv]{1611.03416}


\bibitem[\protect\citeauthoryear{Smeding and V{\'{a}}k{\'{a}}r}{Smeding and
  V{\'{a}}k{\'{a}}r}{2023a}]%
        {efficient-chad-artifact}
\bibfield{author}{\bibinfo{person}{Tom Smeding} {and} \bibinfo{person}{Matthijs
  V{\'{a}}k{\'{a}}r}.} \bibinfo{year}{2023}\natexlab{a}.
\newblock \bibinfo{title}{Artifact for Efficient CHAD}.
\newblock
\newblock
\urldef\tempurl%
\url{https://doi.org/10.5281/zenodo.10015321}
\showDOI{\tempurl}
\newblock
\shownote{Artifact for this publication.}


\bibitem[\protect\citeauthoryear{Smeding and V{\'{a}}k{\'{a}}r}{Smeding and
  V{\'{a}}k{\'{a}}r}{2023b}]%
        {ad-dualrev-th}
\bibfield{author}{\bibinfo{person}{Tom Smeding} {and} \bibinfo{person}{Matthijs
  V{\'{a}}k{\'{a}}r}.} \bibinfo{year}{2023}\natexlab{b}.
\newblock \showarticletitle{Efficient Dual-Numbers Reverse {AD} via Well-Known
  Program Transformations}.
\newblock \bibinfo{journal}{\emph{Proc. {ACM} Program. Lang.}}
  \bibinfo{volume}{7}, \bibinfo{number}{{POPL}} (\bibinfo{year}{2023}),
  \bibinfo{pages}{1573--1600}.
\newblock
\urldef\tempurl%
\url{https://doi.org/10.1145/3571247}
\showDOI{\tempurl}


\bibitem[\protect\citeauthoryear{Staton}{Staton}{2010}]%
        {fp-2010-local-state-staton}
\bibfield{author}{\bibinfo{person}{Sam Staton}.}
  \bibinfo{year}{2010}\natexlab{}.
\newblock \showarticletitle{Completeness for Algebraic Theories of Local
  State}. In \bibinfo{booktitle}{\emph{Foundations of Software Science and
  Computational Structures, 13th International Conference, {FOSSACS} 2010, Held
  as Part of the Joint European Conferences on Theory and Practice of Software,
  {ETAPS} 2010, Paphos, Cyprus, March 20-28, 2010. Proceedings}}
  \emph{(\bibinfo{series}{Lecture Notes in Computer Science})},
  \bibfield{editor}{\bibinfo{person}{C.{-}H.~Luke Ong}} (Ed.),
  Vol.~\bibinfo{volume}{6014}. \bibinfo{publisher}{Springer},
  \bibinfo{pages}{48--63}.
\newblock
\urldef\tempurl%
\url{https://doi.org/10.1007/978-3-642-12032-9\_5}
\showDOI{\tempurl}


\bibitem[\protect\citeauthoryear{V{\'{a}}k{\'{a}}r}{V{\'{a}}k{\'{a}}r}{2021}]%
        {vakar-2021-higher-order-reverse-ad}
\bibfield{author}{\bibinfo{person}{Matthijs V{\'{a}}k{\'{a}}r}.}
  \bibinfo{year}{2021}\natexlab{}.
\newblock \showarticletitle{Reverse {AD} at Higher Types: Pure, Principled and
  Denotationally Correct}. In \bibinfo{booktitle}{\emph{Programming Languages
  and Systems}} \emph{(\bibinfo{series}{Lecture Notes in Computer Science})},
  \bibfield{editor}{\bibinfo{person}{Nobuko Yoshida}} (Ed.),
  Vol.~\bibinfo{volume}{12648}. \bibinfo{publisher}{Springer},
  \bibinfo{pages}{607--634}.
\newblock
\urldef\tempurl%
\url{https://doi.org/10.1007/978-3-030-72019-3\_22}
\showDOI{\tempurl}


\bibitem[\protect\citeauthoryear{V{\'{a}}k{\'{a}}r and
  Smeding}{V{\'{a}}k{\'{a}}r and Smeding}{2022}]%
        {vakar-2022-chad}
\bibfield{author}{\bibinfo{person}{Matthijs V{\'{a}}k{\'{a}}r} {and}
  \bibinfo{person}{Tom Smeding}.} \bibinfo{year}{2022}\natexlab{}.
\newblock \showarticletitle{{CHAD:} Combinatory Homomorphic Automatic
  Differentiation}.
\newblock \bibinfo{journal}{\emph{{ACM} Trans. Program. Lang. Syst.}}
  \bibinfo{volume}{44}, \bibinfo{number}{3}, \bibinfo{pages}{20:1--20:49}.
\newblock
\urldef\tempurl%
\url{https://doi.org/10.1145/3527634}
\showDOI{\tempurl}


\bibitem[\protect\citeauthoryear{van~den Berg, Schrijvers, McKinna, and
  Vandenbroucke}{van~den Berg et~al\mbox{.}}{2024}]%
        {van2024forward}
\bibfield{author}{\bibinfo{person}{Birthe van~den Berg}, \bibinfo{person}{Tom
  Schrijvers}, \bibinfo{person}{James McKinna}, {and}
  \bibinfo{person}{Alexander Vandenbroucke}.} \bibinfo{year}{2024}\natexlab{}.
\newblock \showarticletitle{Forward-or reverse-mode automatic differentiation:
  What's the difference?}
\newblock \bibinfo{journal}{\emph{Science of Computer Programming}}
  \bibinfo{volume}{231} (\bibinfo{year}{2024}), \bibinfo{pages}{103010}.
\newblock


\bibitem[\protect\citeauthoryear{Vytiniotis, Belov, Wei, Plotkin, and
  Abadi}{Vytiniotis et~al\mbox{.}}{2019}]%
        {ad-2019-vytiniotisdifferentiable}
\bibfield{author}{\bibinfo{person}{Dimitrios Vytiniotis}, \bibinfo{person}{Dan
  Belov}, \bibinfo{person}{Richard Wei}, \bibinfo{person}{Gordon Plotkin},
  {and} \bibinfo{person}{Martin Abadi}.} \bibinfo{year}{2019}\natexlab{}.
\newblock \showarticletitle{The differentiable curry}.
\newblock \bibinfo{journal}{\emph{NeurIPS Workshop on Program Transformations}}
  (\bibinfo{year}{2019}).
\newblock


\bibitem[\protect\citeauthoryear{Vákár}{Vákár}{2015}]%
        {DBLP:conf/fossacs/Vakar15}
\bibfield{author}{\bibinfo{person}{Matthijs Vákár}.}
  \bibinfo{year}{2015}\natexlab{}.
\newblock \showarticletitle{A Categorical Semantics for Linear Logical
  Frameworks}, \bibfield{editor}{\bibinfo{person}{Andrew~M. Pitts}} (Ed.).
\newblock \bibinfo{journal}{\emph{Foundations of Software Science and
  Computation Structures - 18th International Conference, FoSSaCS 2015, Held as
  Part of the European Joint Conferences on Theory and Practice of Software,
  {ETAPS} 2015, London, UK, April 11-18, 2015. Proceedings}}
  \bibinfo{volume}{9034}, \bibinfo{pages}{102--116}.
\newblock
\urldef\tempurl%
\url{https://doi.org/10.1007/978-3-662-46678-0\_7}
\showDOI{\tempurl}


\bibitem[\protect\citeauthoryear{Vákár}{Vákár}{2020}]%
        {DBLP:journals/corr/abs-2007-05282}
\bibfield{author}{\bibinfo{person}{Matthijs Vákár}.}
  \bibinfo{year}{2020}\natexlab{}.
\newblock \showarticletitle{Denotational Correctness of Forward-Mode Automatic
  Differentiation for Iteration and Recursion}.
\newblock \bibinfo{journal}{\emph{CoRR}}  \bibinfo{volume}{abs/2007.05282}
  (\bibinfo{year}{2020}).
\newblock
\showeprint[arXiv]{2007.05282}
\urldef\tempurl%
\url{https://arxiv.org/abs/2007.05282}
\showURL{%
\tempurl}


\bibitem[\protect\citeauthoryear{Wang, Decker, Wu, Essertel, and Rompf}{Wang
  et~al\mbox{.}}{2018}]%
        {ad-2018-cps-by-callstack}
\bibfield{author}{\bibinfo{person}{Fei Wang}, \bibinfo{person}{James~M.
  Decker}, \bibinfo{person}{Xilun Wu}, \bibinfo{person}{Gr{\'{e}}gory~M.
  Essertel}, {and} \bibinfo{person}{Tiark Rompf}.}
  \bibinfo{year}{2018}\natexlab{}.
\newblock \showarticletitle{Backpropagation with Callbacks: Foundations for
  Efficient and Expressive Differentiable Programming}. In
  \bibinfo{booktitle}{\emph{Advances in Neural Information Processing Systems
  31: Annual Conference on Neural Information Processing Systems 2018, NeurIPS
  2018, 3-8 December 2018, Montr{\'{e}}al, Canada}},
  \bibfield{editor}{\bibinfo{person}{Samy Bengio}, \bibinfo{person}{Hanna~M.
  Wallach}, \bibinfo{person}{Hugo Larochelle}, \bibinfo{person}{Kristen
  Grauman}, \bibinfo{person}{Nicol{\`{o}} Cesa{-}Bianchi}, {and}
  \bibinfo{person}{Roman Garnett}} (Eds.). \bibinfo{pages}{10201--10212}.
\newblock
\urldef\tempurl%
\url{http://papers.nips.cc/paper/8221-backpropagation-with-callbacks-foundations-for-efficient-and-expressive-differentiable-programming}
\showURL{%
\tempurl}


\bibitem[\protect\citeauthoryear{Wang and Rompf}{Wang and Rompf}{2018}]%
        {ad-2018-rev-delimited-continuations}
\bibfield{author}{\bibinfo{person}{Fei Wang} {and} \bibinfo{person}{Tiark
  Rompf}.} \bibinfo{year}{2018}\natexlab{}.
\newblock \showarticletitle{A Language and Compiler View on Differentiable
  Programming}. In \bibinfo{booktitle}{\emph{6th International Conference on
  Learning Representations, {ICLR} 2018, Vancouver, BC, Canada, April 30 - May
  3, 2018, Workshop Track Proceedings}}. \bibinfo{publisher}{OpenReview.net}.
\newblock
\urldef\tempurl%
\url{https://openreview.net/forum?id=SJxJtYkPG}
\showURL{%
\tempurl}


\bibitem[\protect\citeauthoryear{Wang, Zheng, Decker, Wu, Essertel, and
  Rompf}{Wang et~al\mbox{.}}{2019}]%
        {ad-2019-delimited-continuations}
\bibfield{author}{\bibinfo{person}{Fei Wang}, \bibinfo{person}{Daniel Zheng},
  \bibinfo{person}{James~M. Decker}, \bibinfo{person}{Xilun Wu},
  \bibinfo{person}{Gr{\'{e}}gory~M. Essertel}, {and} \bibinfo{person}{Tiark
  Rompf}.} \bibinfo{year}{2019}\natexlab{}.
\newblock \showarticletitle{Demystifying differentiable programming:
  shift/reset the penultimate backpropagator}.
\newblock \bibinfo{journal}{\emph{Proc. {ACM} Program. Lang.}}
  \bibinfo{volume}{3}, \bibinfo{number}{{ICFP}} (\bibinfo{year}{2019}),
  \bibinfo{pages}{96:1--96:31}.
\newblock
\urldef\tempurl%
\url{https://doi.org/10.1145/3341700}
\showDOI{\tempurl}


\end{thebibliography}
